\documentclass{LMCS}

\usepackage{amsfonts}
\usepackage{enumerate}

\def\doi{7 (2:1) 2011}
\lmcsheading%
{\doi}
{1--55}
{}
{}
{Jul.~12, 2010}
{Apr.~21, 2011}
{}

\begin{document}


\newcommand{\emptyrun}{\langle\rangle} 
\newcommand{\oo}{\bot}            
\newcommand{\pp}{\top}            
\newcommand{\xx}{\wp}               
\newcommand{\legal}[2]{\mbox{\bf Lr}^{#1}_{#2}} 
\newcommand{\win}[2]{\mbox{\bf Wn}^{#1}_{#2}} 
\newcommand{\seq}[1]{\langle #1 \rangle}           


\newcommand{\ade}{\mbox{\Large $\sqcup$}\hspace{1pt}}      
\newcommand{\ada}{\mbox{\Large $\sqcap$}\hspace{1pt}}      
\newcommand{\sst}{\mbox{\raisebox{-0.07cm}{\scriptsize $-$}\hspace{-0.2cm}$\pst$}}
\newcommand{\scost}{\mbox{\raisebox{0.20cm}{\scriptsize $-$}\hspace{-0.2cm}$\pcost$}}
\newcommand{\sqc}{\mbox{\hspace{2pt}\small \raisebox{0.0cm}{$\bigtriangleup$}\hspace{2pt}}}
\newcommand{\sqd}{\mbox{\hspace{2pt}\small \raisebox{0.06cm}{$\bigtriangledown$}\hspace{2pt}}}
\newcommand{\sqe}{\mbox{\large \raisebox{0.07cm}{$\bigtriangledown$}}}
\newcommand{\sqa}{\mbox{\large \raisebox{0.0cm}{$\bigtriangleup$}}}
\newcommand{\mld}{\vee}    
\newcommand{\mlc}{\wedge}  
\newcommand{\tgd}{\mbox{\hspace{2pt}$\vee$\hspace{-1.29mm}\raisebox{0.1mm}{\rule{0.13mm}{2mm}}\hspace{5pt}}}    
\newcommand{\tgc}{\mbox{\hspace{2pt}$\wedge$\hspace{-1.29mm}\raisebox{0.02mm}{\rule{0.13mm}{2mm}}\hspace{5pt}}}    
\newcommand{\tge}{\hspace{1pt}\mbox{\Large $\vee$\hspace{-1.84mm}\raisebox{0.1mm}{\rule{0.13mm}{3.0mm}}\hspace{6pt}}}   
\newcommand{\tga}{\mbox{\hspace{1pt}\Large $\wedge$\hspace{-1.84mm}\raisebox{0.02mm}{\rule{0.13mm}{3.0mm}}\hspace{6pt}}}     
\newcommand{\tgpst}{\mbox{\raisebox{-0.01cm}{\scriptsize $\wedge$}\hspace{-4pt}\raisebox{0.06cm}{\small $\mid$}\hspace{2pt}}}
\newcommand{\tgpcost}{\mbox{\raisebox{0.12cm}{\scriptsize $\vee$}\hspace{-3.8pt}\raisebox{0.04cm}{\small $\mid$}\hspace{2pt}}}
\newcommand{\tgst}{\mbox{\raisebox{-0.05cm}{$\circ$}\hspace{-0.12cm}\raisebox{0.05cm}{\small $\mid$}\hspace{2pt}}} 
\newcommand{\tgcost}{\mbox{\raisebox{0.12cm}{$\circ$}\hspace{-0.12cm}\raisebox{0.04cm}{\small $\mid$}\hspace{2pt}}}
\newcommand{\mle}{\mbox{\hspace{1pt}\Large $\vee$}\hspace{1pt}}    
\newcommand{\mla}{\mbox{\hspace{1pt}\Large $\wedge$}\hspace{1pt}}  
\newcommand{\add}{\hspace{0pt}\sqcup}                      
\newcommand{\adc}{\hspace{0pt}\sqcap}                      
\newcommand{\gneg}{\neg}                  
\newcommand{\mli}{\rightarrow}                     
\newcommand{\tlg}{\bot}               
\newcommand{\twg}{\top}               

\newcommand{\pst}{\mbox{\raisebox{-0.01cm}{\scriptsize $\wedge$}\hspace{-4pt}\raisebox{0.16cm}{\tiny $\mid$}\hspace{2pt}}}
\newcommand{\cla}{\mbox{\large $\forall$}\hspace{1pt}}      
\newcommand{\cle}{\mbox{\large $\exists$}\hspace{1pt}}        
\newcommand{\pcost}{\mbox{\raisebox{0.12cm}{\scriptsize $\vee$}\hspace{-4pt}\raisebox{0.02cm}{\tiny $\mid$}\hspace{2pt}}}




\title{From formulas to cirquents in computability logic}
\author[G.~Japaridze]{Giorgi Japaridze}
\thanks{This material is based upon work supported by the National Science Foundation under Grant No. 0208816}
\address{School of Computer Science and Technology, Shandong
  University, China; and
  Department of Computing Sciences, Villanova University, USA}
\email{giorgi.japaridze@villanova.edu}

\begin{abstract} 
{\em Computability logic} (CoL) is a recently introduced semantical platform and  ambitious program for redeveloping logic as a formal theory of computability, as opposed to the formal theory of truth that logic has more traditionally been. Its expressions represent interactive computational tasks seen as games played by a machine against the environment, and  ``truth'' is understood as existence of an algorithmic winning strategy. With logical operators standing for operations on games, the formalism of CoL is open-ended, and has already undergone series of extensions.  
This article extends the expressive power of CoL in a qualitatively new way, generalizing {\em formulas} (to which the earlier languages  of CoL were  limited) to circuit-style structures termed {\em cirquents}. The latter, unlike formulas, are able to account for subgame/subtask sharing  between different parts of the overall game/task. Among the many advantages offered by this ability is that it allows us to capture, refine and generalize the well known   {\em independence-friendly logic} which, after the present leap forward,  naturally becomes a conservative fragment of CoL, just as classical logic had been known to be a conservative fragment of the formula-based version of CoL. 
Technically, this paper is self-contained, and can be read without any prior familiarity with CoL.
\end{abstract}



\maketitle
\vfill\eject
\tableofcontents
 
\section{Introduction}\label{sintr}

\noindent {\em Computability logic} (CoL), introduced in \cite{Jap03,Japic,Japfin}, is a semantical platform and  ambitious program for redeveloping logic as a formal theory of computability, as opposed to the formal theory of truth that logic has more traditionally been. Its expressions stand for interactive computational tasks seen as games played by a machine against its environment, and  ``truth''  is understood as existence of an effective solution, i.e., of an algorithmic winning strategy. 

With this semantics, CoL provides a systematic answer to the fundamental question ``{\em what can be computed?}\hspace{1pt}'', just as classical logic is a systematic tool for finding what is true. Furthermore, as it turns out, in positive cases ``{\em what} can be computed'' always allows itself to be replaced by ``{\em how} can be computed'', which makes CoL of potential interest in not only theoretical computer science, but many more applied areas as well, including interactive knowledge base systems, resource oriented systems for planning and action, or declarative programming languages. 
On the logical side, CoL promises to be an appealing, constructive and computationally meaningful alternative to classical logic as a basis for applied theories. The first concrete steps towards realizing this potential have been made very recently in
 \cite{Japtowards,cla4,cla5}, where  CoL-based versions of Peano arithmetic were elaborated. The system constructed in  \cite{Japtowards} is an axiomatic theory of {\em effectively solvable} number-theoretic {\em problems} (just as the ordinary Peano arithmetic is an axiomatic theory of {\em true} number-theoretic {\em facts}); the system constructed in  \cite{cla4} is an axiomatic theory of {\em efficiently solvable} (namely, solvable in polynomial time) number-theoretic {\em problems}; in the same style, \cite{cla5} constructs systems for polynomial space, elementary recursive, and primitive recursive computabilities. In all cases, a solution for a problem can be effectively --- in fact, efficiently --- extracted from a proof of it in the system, which reduces problem-solving to theorem-proving.

The formalism of CoL is open-ended. It has already undergone series of extensions (\cite{Japseq}-\cite{Japtogl}) through introducing logical operators for new, actually or potentially interesting, operations on games, and this process will probably still continue in the future. The present work is also devoted to expanding the expressive power of CoL, but in a very different way. Namely, while the earlier languages of CoL were limited to {\em formulas}, this paper makes a leap forward by generalizing formulas to circuit-style structures termed {\em cirquents}. These structures, in a very limited form (compared with their present form), were introduced earlier \cite{Cirq,Japdeep} in the context of the new proof-theoretic approach called {\em cirquent calculus}. Cirquent-based formalisms have  significant advantages over formula-based ones, including exponentially higher efficiency and substantially greater expressive power. Both \cite{Cirq} and \cite{Japdeep} pointed out the possibility and expediency of bringing cirquents into CoL. But a CoL-semantics for  cirquents 
had never really been set up until now. 

Unlike most of its predecessors, from the technical (albeit perhaps not philosophical or motivational) point of view, the present paper is written without assuming that the reader is already familiar with the basic concepts and techniques of computability logic. It is organized as follows.

Section \ref{s2} reintroduces the concept of games and interactive computability on which the semantics of CoL is based. A reader familiar with the basics of CoL may safely skip this section. 

Section \ref{s3} introduces the simplest kind of cirquents, containing only the traditional $\mld$ and $\mlc$ sorts of gates (negation, applied directly 
to inputs, is also present). These are nothing but (possibly infinite)  Boolean circuits in the usual sense, and the semantics of CoL for them coincides with the traditional, classical semantics. While such cirquents --- 
when finite --- do not offer higher expressiveness than formulas do, they {\em do} offer dramatically higher efficiency. This fact alone, in our days of logic being increasingly CS-oriented, provides sufficient motivation for considering a switch from formulas to cirquents in logic, even if we are only concerned with classical logic. The first steps in this direction have already been made in \cite{Japdeep}, where a cirquent-based sound and complete deductive system was set up for classical logic. That system was shown to provide an exponential speedup of proofs over its formula-based counterparts. 

Each of the subsequent Sections \ref{s4}-\ref{s8} conservatively generalizes the cirquents and the semantics of the preceding sections. 

Section \ref{s4} strengthens the expressiveness of cirquents by allowing new, so called {\em selectional}, sorts of gates, with the latter coming in three --- {\em choice} $\add,\adc$, {\em sequential} $\sqd\hspace{-2pt},\hspace{-2pt}\sqc$ and {\em toggling} $\tgd\hspace{-2pt},\hspace{-2pt}\tgc$ --- flavors of disjunction and conjunction. Unlike  $\mld$ and $\mlc$ which stand for parallel combinations of (sub)tasks,  selectional gates model decision steps in the process of interaction of the machine with its environment. Cirquents with such gates, allowing us to account for the possibility of {\em sharing} nontrivial subgames/subtasks, have never been considered in the past. They, even when finite, are  more expressive (let alone efficient) than formulas with selectional connectives. 

Section \ref{s5} introduces the idea of {\em clustering} selectional gates. Clusters are, in fact, generalized gates --- switch-style devices that permit to select one of several $n$-tuples of inputs and connect them, in a parallel fashion, with the $n$-tuple of outputs, with ordinary gates being nothing but the special cases of clusters where $n=1$. Clustering makes it possible to express a new kind of sharing, which can be characterized as ``sharing children without sharing grandchildren'' --- that is, sharing decisions (associated with child gates) without sharing the results of such decisions  (the children of those  gates). The ability to account for this sort of sharing yields a further qualitative increase in the expressiveness of the associated formalism.  

Sections \ref{s6} and \ref{s7} extend clustering from selectional gates to the traditional sorts $\mld,\mlc$ of gates. It turns out that the resulting cirquents --- even without selectional gates --- are expressive enough to fully capture the well known and extensively studied {\em independence-friendly (IF) logic} introduced by Hintikka and Sandu \cite{Hintikka96,HS97}.   At the same time, cirquents with clustered $\mld,\mlc$-gates yield substantially higher expressiveness than IF logic does. Due to this fact, they overcome a series of problems emerging within the earlier known approaches to IF logic. One of such problems is the inability of the traditional formalisms of IF logic to account for independence from conjunctions and disjunctions in the same systematic way as they account for independence from quantifiers. Correspondingly, attempts to develop IF logic at the purely propositional level have not been able to go beyond certain very limited and special syntactic fragments of the language.
In contrast, our approach saves classical logic's nice phenomenon of   quantifiers being nothing but ``long'' conjunctions and disjunctions, so that one can do with the latter everything that can be done with the former, and vice versa. As a result, we can now (at last)  meaningfully talk about {\em propositional IF logic} in the proper sense without any unsettling syntactic restrictions. Another problem arising with IF logic is the ``unplayability'' (cf. \cite{Stev09}) of the incomplete-information games traditionally associated with its formulas, as such games violate certain natural game-theoretic principles such as {\em perfect recall}. Attempts to associate reasonable game-theoretic intuitions with imperfect-information games typically have to resort to viewing the two parties not as individual players but rather as teams of cooperating but non-communicating players. This approach, however, may often get messy, and tends to give rise to series of new sorts of problems. V\"{a}\"{a}n\"{a}nen \cite{Vaa02} managed to construct a semantics for IF logic based on perfect-information games. This, however, made games unplayable for a new reason: moves in V\"{a}\"{a}n\"{a}nen's games are second-order objects and hence are ``unmakable''. Our approach  avoids the need to deal with imperfect-information, second-order-move, or multiple-player games altogether. 

Section \ref{s7} also introduces a further generalization of cirquents through what is  termed {\em ranking}. Cirquents with ranking (and with only $\mld,\mlc$ gates) allow us to further capture the so called {\em  extended IF logic} (cf. \cite{Tul09}), but are substantially more expressive than the latter. They also overcome one notable problem arising in extended IF logic, which is the (weak) negation's not being able to act as an ordinary connective that can be meaningfully applied anywhere within a formula. 

Section \ref{s8} extends the formalism of cirquents by allowing additional sorts of inputs termed {\em general}, as opposed to the {\em elementary} inputs to which the cirquents of the earlier sections are limited. Unlike elementary inputs that are interpreted just as $\twg$ (``true'') or $\tlg$(``false''), general inputs stand for any, possibly nontrivial, games. This way, cirquents become and can be viewed as (complex) {\em operations} on games, only very few of which are expressible through formulas. 

Section \ref{s9} sets up an alternative semantics for cirquents termed {\em abstract resource semantics} (ARS). It is a companion and faithful technical assistant of CoL. ARS also has good claims to be a materialization and generalization of the resource intuitions traditionally --- but somewhat wrongly ---  associated with linear logic and its variations. Unlike CoL, ARS has already met cirquents in the past, namely, in \cite{Cirq,Japdeep}. The latter, however, unlike the present paper, elaborated ARS only for a very limited class of cirquents --- cirquents with just $\mld,\mlc$ gates and without clustering or ranking.

Section \ref{s10} proves that CoL and ARS validate exactly the same cirquents. Among the expected applications of this theorem is facilitating soundness/completeness proofs for various deductive systems for (fragments of) CoL, as technically it is by an order of magnitude easier to deal with (the simple and ``naive'') ARS than to deal with  (the complex and ``serious'')  CoL directly. 

\section{Games}\label{s2}

\noindent As  noted, computability logic understands interactive computational problems as games played between two players:  {\em machine} and {\em environment}. The symbolic names   for these two players are   
$\twg$ and 
$\tlg$,
 respectively. $\pp$ is a deterministic mechanical device (thus) only capable of following algorithmic strategies, whereas there are no restrictions on the behavior of $\oo$, which represents a capricious user, the blind forces of nature, or the devil himself.  Our sympathies are with $\twg$, 
and by just saying ``won'' or ``lost'' without specifying a player, we always mean won or lost by $\twg$. $\xx$ is always  a variable ranging over $\{\twg,\tlg\}$, with 
\(\gneg \xx\) meaning $\xx$'s adversary, i.e. the player which is not $\xx$.

Before getting to a formal definition of games, we agree that a 
{\bf move} is always a finite string over the standard keyboard alphabet. 
A {\bf labeled move} ({\bf labmove}) is a move prefixed with $\pp$ or $\oo$, with its prefix ({\bf label}) indicating which player has made the move. 
A {\bf run} is a (finite or infinite) sequence\footnote{Throughout this paper, by a {\em sequence} we mean a $\kappa$-sequence for some $\kappa\leq\omega$.} of labmoves, and a 
{\bf position} is a finite run.
Runs will be often delimited by ``$\langle$'' and ``$\rangle$'', with $\emptyrun$ thus denoting the {\bf empty run}. 

 The following is a formal definition of the concept of a game, combined with some less formal conventions regarding the usage of certain terminology. It should be noted that the concept of a game considered in CoL is more general than the one defined below, with games in our present sense called {\em constant games}. Since we (for simplicity) only consider constant games in this paper, we omit the word ``constant'' and just say ``game''.

\begin{defi}\label{game}
 A {\bf game}  is a pair $A=(\legal{A}{},\win{A}{})$, where:\medskip

1. $\legal{A}{}$ is a set of runs satisfying the condition that a finite or infinite run is in $\legal{A}{}$ iff all of its nonempty finite --- not necessarily proper --- initial
segments are in $\legal{A}{}$ (notice that this implies $\emptyrun\in\legal{A}{}$). The elements of $\legal{A}{}$ are
said to be {\bf legal runs} of $A$, and all other runs are said to be {\bf illegal runs} of $A$. We say that $\alpha$ is a {\bf legal move} for $\xx$ in a position $\Phi$ of $A$ iff $\seq{\Phi,\xx\alpha}\in\legal{A}{}$; otherwise 
$\alpha$ is an {\bf illegal move}. When the last move of the shortest illegal initial segment of $\Gamma$  is $\xx$-labeled, we say that $\Gamma$ is a {\bf $\xx$-illegal run} of $A$.\medskip

2. $\win{A}{}$ is a function that sends every run $\Gamma$ to one of the players $\pp$ or $\oo$, satisfying the condition that if $\Gamma$ is a $\xx$-illegal run of $A$, then $\win{A}{}\seq{\Gamma}=\gneg\xx$.\footnote{We write $\win{A}{}\seq{\Gamma}$ for $\win{A}{}(\Gamma)$.} When $\win{A}{}\seq{\Gamma}=\xx$, we say that $\Gamma$ is a {\bf $\xx$-won} (or {\bf won by $\xx$}) run of $A$; otherwise $\Gamma$ is {\bf lost by $\xx$}. Thus, an illegal run is always lost by the player who has made the first illegal move in it.  
\end{defi}

It is clear from the above definition that, when defining a particular game $A$, it would be sufficient to specify what {\em positions}  (finite runs) are legal, and what {\em legal runs} are won by $\pp$. Such a definition will then uniquely extend to all --- including infinite and illegal --- runs. We will implicitly rely on this observation in the sequel. 

A game is said to be {\bf elementary} iff it has no legal runs other than (the always legal) empty run $\emptyrun$. That is, an elementary game is a ``game'' without any (legal) moves,  automatically won or lost. There are exactly two such games, for which we use the same symbols $\twg$ and $\tlg$ as for the two players: the game $\twg$ automatically won by player $\pp$, and the game $\tlg$ automatically won by player $\oo$.\footnote{Precisely, we have $\win{\twg}{}\emptyrun=\twg$ and $\win{\tlg}{}\emptyrun=\tlg$.} Computability logic is a conservative extension of classical logic, understanding classical propositions as elementary games. And, just as classical logic, it sees no difference between any two true propositions such as ``$0= 0$'' and ``{\em Snow is white}'', and identifies them with the elementary game $\pp$; similarly, it treats false propositions such as ``$0= 1$'' or ``{\em Snow is black}'' as the elementary game $\tlg$. 

The {\em negation} $\gneg A$ of a game $A$ is understood as the game obtained from $A$ by interchanging the roles of the two players, i.e., making $\pp$'s (legal) moves and wins $\oo$'s moves and wins, and vice versa. Precisely, let us agree that, for a run $\Gamma$, $\gneg \Gamma$ means the result of changing in $\Gamma$ each label $\pp$ to $\oo$ and vice versa. Then:  

\begin{defi}\label{negdef}
The {\bf negation} $\gneg A$ of a game $A$ is defined by stipulating that, for any run $\Gamma$,
\begin{enumerate}[$\bullet$]
\item $\Gamma\in\legal{\gneg A}{}$ iff $\gneg \Gamma\in \legal{A}{}$; 
\item $\win{\gneg A}{}\seq{\Gamma}=\pp$ iff $\win{A}{}\seq{\gneg \Gamma}=\oo$. 
\end{enumerate} 
\end{defi}\medskip  

\noindent Obviously the negation of an elementary game is also elementary. Generally, when applied to elementary games, the meaning of $\gneg$ fully coincides with its classical meaning. So, $\gneg$ is a conservative generalization of classical negation from elementary games to all games. 

Note the relaxed nature of our games. They do not impose any regulations on when either player can or should move. This is entirely up to the players. Even if we assume that illegal moves physically cannot be made, it is still possible that in certain (or all) positions  both of the players have legal moves, and then the next move will be made (if made at all) by the player who wants or can act sooner. This brings us to the next question to clarify: how are our games really played, and what does a  {\em strategy}  mean here?

In traditional game-semantical approaches, including those of Lorenzen \cite{Lor61}, Hintikka \cite{Hintikka73} or Blass \cite{Bla92},  player's strategies are understood as {\em functions} --- typically as functions from interaction histories (positions) to moves, or sometimes (Abramsky and Jagadeesan \cite{Abr94}) as functions that only look at the latest move of the history. This {\em strategies-as-functions} approach, however, is inapplicable in the context of computability logic, whose relaxed semantics, in striving to get rid of any ``bureaucratic pollutants'' and only deal with the remaining true essence of games, has no structural rules and thus does not regulate 
the order of moves. As noted,  here often either player may have (legal) moves, and then it is unclear whether the next move should be the one prescribed by $\pp$'s strategy function or the one prescribed by the strategy function of $\oo$. In fact, for a game semantics whose ambition is to provide a comprehensive, natural and direct tool for modeling interactive computations, the strategies-as-functions approach would be less than adequate, even if technically possible. This is so for the simple reason that  the strategies that real computers follow are not functions. If the strategy of your personal computer was a function from the history of interaction with you, then its performance would keep noticeably worsening due to the need to read the continuously lengthening --- and, in fact, practically infinite --- interaction history every time before responding. Fully ignoring that history and looking only at your latest keystroke in the spirit of \cite{Abr94} is also certainly not what your computer does, either. The inadequacy of the strategies-as-functions approach becomes especially evident when one tries to bring computational complexity issues into interactive computation, the next natural target towards which CoL has already started making its first steps (\cite{lbcs,cla4,cla5}).  

In computability logic, ($\pp$'s effective) strategies are defined in terms of interactive machines, where computation is one continuous process interspersed with --- and influenced by --- multiple ``input'' (environment's moves) and ``output'' (machine's moves) events. Of several, seemingly rather different yet equivalent,  machine models of interactive computation studied in CoL, here we will consider the most basic, {\bf HPM} (``Hard-Play Machine'') model. 

An HPM is nothing but a Turing machine with the additional capability of making moves. The adversary can also move at any time, with such moves being the only nondeterministic events from the machine's perspective. Along with the ordinary  read/write {\em work tape}, the machine also has an additional  tape\footnote{An HPM often also has a third tape called the {\em valuation tape}. Its function is to provide values for the variables on which a game may depend. However, as we remember, in this paper we only consider constant games --- games that do not depend on any variables. This makes it possible to safely remove the valuation tape from the HPM model (or leave it there but fully ignore), as this tape is no longer relevant.} called  the  {\em run tape}. The latter, at any time,  spells the ``current position'' of the play. The role of this tape is to make the interaction history fully visible to the machine.  It is read-only, and its content is automatically updated every time either player makes a move.

In these terms,  an  {\bf algorithmic solution} ($\pp$'s {\bf winning strategy}) for a given  game $A$ is understood as an HPM $\mathcal M$ such that,  no matter how the environment acts during its interaction with $\mathcal M$ (what moves it makes and when), the run incrementally spelled on the run tape is a $\pp$-won run of $A$. When this is the case, we say that ${\mathcal M}$ {\bf wins}, or {\bf solves}, $A$, and that $A$ is a {\bf computable}, or {\bf algorithmically solvable}, game.   

As for $\oo$'s strategies, there is no need to define them: all possible behaviors by $\oo$ are accounted for by the different possible nondeterministic updates  of the run tape of an HPM. 

In the above outline, we described HPMs in a relaxed fashion, without being specific about technical details such as, say, how, exactly, moves are made by the machine, how many moves either player can make at once, what happens if both players attempt to move ``simultaneously'', etc. As it turns out (cf. \cite{Jap03}), all reasonable design choices yield the same class of winnable games as long as we consider a certain natural subclass of games called {\em static}. 

Intuitively, static games are interactive tasks where the relative speeds of the players are irrelevant, as it never hurts a player to postpone making moves. In other words, they are games that are contests of intellect rather than contests of speed. And one of the theses that computability logic philosophically relies on is that static games present an adequate formal counterpart of our intuitive concept of ``pure'', speed-independent interactive computational problems. Correspondingly, computability logic restricts its attention (more specifically, possible interpretations of the atoms of its formal language) to static games.  Below comes a formal definition of this concept.

For either player $\xx$, we say that a run $\Upsilon$ is a {\bf $\xx$-delay} of a run $\Gamma$ iff:
\begin{enumerate}[$\bullet$]
\item for both players $\xx'\in\{\pp,\oo\}$, the subsequence of $\xx'$-labeled moves of $\Upsilon$ is the same as that of $\Gamma$, and
\item for any $n,k\geq 1$, if the $n$th $\xx$-labeled move is made later than (is to the right of) the $k$th $\gneg\xx$-labeled move in $\Gamma$, then so is it in $\Upsilon$.
\end{enumerate}\medskip

\noindent The above conditions mean that in  $\Upsilon$  each player has made the same sequence of moves as in $\Gamma$, only, in $\Upsilon$, $\xx$ might have been acting with some delay.

Let us say that a run is {\bf $\xx$-legal} iff it is not $\xx$-illegal. That is, a $\xx$-legal run is either simply legal, or the player responsible for (first) making it illegal is $\gneg \xx$ rather than $\xx$. 
 
Now, we say that a game  $A$ is {\bf static} iff, whenever a run $\Upsilon$ is a $\xx$-delay of 
a run $\Gamma$, we have:
\begin{enumerate}[$\bullet$]
\item if $\Gamma$ is a $\xx$-legal run of $A$, then so is $\Upsilon$;\footnote{In most papers on CoL, the concept of static games is defined without this (first) condition. In such cases, however, the existence of an always-illegal move $\spadesuit$ is stipulated in the definition of games. The first condition of our present definition of static games turns out to be simply derivable from that stipulation. 
This and a couple of other minor technical differences between our present formulations
 from those given in other pieces of literature on CoL only signify presentational and by no means conceptual variations.}
\item if $\Gamma$ is a $\xx$-won run of $A$, then so is $\Upsilon$.
\end{enumerate}\medskip

\noindent The class of static games is closed under all game operations studied in CoL, and all games that we shall see in this paper are static. 
Throughout this paper, we use the term ``{\bf computational problem}", or simply ``{\bf problem}", is a synonym of ``static game''.

A simplest example of a non-static game would be the game where all moves are legal, and which is won by the player who moves first. The chances of a player to succeed only depend on its relative speed, that is. Such a game hardly represents any meaningful computational problem.

\section{The simplest kind of cirquents}\label{s3}

\noindent We fix some  infinite collection of finite alphanumeric expressions called  {\bf atoms}, and use  $p$, $q$, $r$, $s$, $p_1$, $p_2$, $p(3,6)$, $q_7(1,1,8)$, \ldots as metavariables for them. If $p$ is an atom, then the expressions $p$ and $\gneg p$ are said to be {\bf literals}.

Let us agree that, in this section, by a  {\bf graph} we mean a directed acyclic multigraph with countably many nodes and edges,  where the outgoing edges of each node are arranged in a fixed left-to-right order (edges $\#1$, $\#2$, etc.), and where  each node is labeled  with either a literal or $\mld$ or $\mlc$. 
Since the sets of nodes and edges are countable, we assume that they are always subsets of $\{1,2,3,\ldots\}$. For a node $n$ of the graph, the string representing $n$ in the standard decimal notation is said to be the {\bf ID number}, or just the {\bf ID}, of the node.  Similarly for edges.

The nodes labeled with $\mlc$ or $\mld$  we call {\bf gates}, and the nodes labeled with literals we call {\bf ports}. Specifically, a node labeled with a literal $L$ is said to be a an {\bf $L$-port}; a $\mlc$-labeled node is said to be a {\bf $\mlc$-gate}; and a 
$\mld$-labeled node is said to be a {\bf $\mld$-gate}. When there is an edge from a node $a$ to a node $b$, we say that $b$ is a {\bf child} of $a$ and $a$ is a {\bf parent} of $b$. 
The relations ``{\bf descendant}'' and ``{\bf ancestor}'' are the transitive closures of the relations ``child'' and ``parent'', respectively. The meanings of some other standard relations such as ``grandchild'', ``grandparent'', etc. should also be clear. 

The {\bf outdegree} of a node of a graph is the quantity of outgoing edges of that node, which can be finite or infinite. Since there are only countably many edges, any two infinite outdegrees are equal.  Similarly, the {\bf indegree} of a node is the quantity of the incoming edges of that node.

We say that a  graph is  {\bf well-founded} iff there is no infinite sequence $a_1,a_2,a_3,\ldots$ of nodes where each $a_i$ is a predecessor of $a_{i+1}$. 
Of course, any (directed acyclic) graph with finitely many nodes is well-founded. 

We say that a graph is {\bf effective} iff the following basic predicates and partial functions characterizing it are recursive:  ``{\em $x$ is a node}'', ``{\em $x$ is an edge}'', ``{\em the label of node $x$}'', ``{\em the outdegree of node $x$}'', ``{\em the $y$'th outgoing edge of node $x$}'', ``{\em the origin of edge $x$}'', ``{\em the destination of edge $x$}''.   

\begin{defi}\label{may5a}
A {\bf cirquent} is an effective, well-founded graph  satisfying the following two  conditions:
\begin{enumerate}[(1)]
\item Ports have no children.
\item There is a node, called the {\bf root}, which is an ancestor of all other nodes in the graph.
\end{enumerate}
\end{defi}\medskip

\noindent We say that a cirquent is {\bf finite} iff it has only finitely many edges (and hence nodes); otherwise it is {\bf infinite}.

We say that a cirquent is {\bf tree-like} iff the indegree of each of its non-root node is $1$.

Graphically, we represent ports through the corresponding literals,  $\mld$-gates through  $\mld$-inscribed circles, and 
$\mlc$-gates through  $\mlc$-inscribed circles.   We agree that the direction of an edge is always upward, which allows us to draw lines rather than arrows for edges. It is understood that the official order of the outgoing edges of a gate coincides with the (left to right) order in which the edges are drawn.  Also, typically we do not indicate the IDs of nodes unless necessary. In most cases, what particular IDs are assigned to the nodes of a cirquent is irrelevant and such an assignment can be chosen arbitrarily. Similarly, edge  IDs are usually irrelevant, and they will never be indicated. 

Below are a few examples of cirquents. Note that cirquents may contain parallel edges, as we agreed that ``graph'' means ``multigraph''. Note also that not only ports but also gates can be childless. 

\begin{center} \begin{picture}(285,112)
\put(33,89){$\gneg p$}
\put(64,89){$p$}
\put(53,71){\line(-1,1){12}}
\put(53,71){\line(1,1){12}}
\put(53,65){\circle{12}}
\put(50,62){$\mld$}

\put(91,89){$\gneg p$}
\put(122,89){$p$}
\put(111,71){\line(-1,1){12}}
\put(111,71){\line(1,1){12}}
\put(111,65){\circle{12}}
\put(108,62){$\mld$}

\put(0,65){$\gneg p$}
\put(157,65){$p$}
\put(83,40){\line(-3,2){29}}
\put(83,40){\line(3,2){29}}
\put(83,40){\line(-4,1){75}}
\put(83,40){\line(4,1){75}}
\put(83,34){\circle{12}}
\put(80,31){$\mlc$}

\put(239,89){$\gneg p$}
\put(279,89){$p$}
\put(263,66){\line(-1,1){16}}
\put(263,66){\line(1,1){16}}
\put(263,60){\circle{12}}
\put(260,57){$\mld$}

\put(263,34){\circle{12}}
\put(260,31){$\mlc$}
\put(261,40){\line(0,1){14}}
\put(265,40){\line(0,1){14}}
\put(269,38){\line(1,1){10}}
\put(279,48){\line(0,1){35}}
\put(257,38){\line(-1,1){10}}
\put(247,48){\line(0,1){35}}

\put(82,10){{\bf Figure 1:} Finite cirquents}
\end{picture}\end{center}

\begin{center} \begin{picture}(390,122)
\put(0,99){$p(1,1)$}
\put(35,99){$p(1,2)$}
\put(70,99){$\ldots$}

\put(3,81){\line(0,1){12}}
\put(3,81){\line(3,1){36}}
\put(3,81){\line(6,1){26}}
\put(40,84){$\ldots$}
\put(3,75){\circle{12}}
\put(0,72){$\mld$}

\put(113,99){$p(2,1)$}
\put(148,99){$p(2,2)$}
\put(183,99){$\ldots$}

\put(116,81){\line(0,1){12}}
\put(116,81){\line(3,1){36}}
\put(116,81){\line(6,1){26}}
\put(153,84){$\ldots$}
\put(116,75){\circle{12}}
\put(113,72){$\mld$}

\put(3,40){\line(0,1){29}}
\put(3,40){\line(4,1){114}}
\put(3,40){\line(6,1){89}}
\put(3,34){\circle{12}}
\put(0,31){$\mlc$}
\put(110,51){\huge $\ldots$}
\put(200,74){\huge $\ldots$}

\put(303,100){\circle{12}}
\put(300,97){$\mld$}

\put(321,81){\circle{12}}
\put(318,78){$\mld$}
\put(316,85){\line(-1,1){10}}

\put(341,67){\circle{12}}
\put(338,63){$\mld$}
\put(336,71){\line(-5,3){10}}

\put(363,61){\circle{12}}
\put(360,57){$\mld$}
\put(357,63){\line(-4,1){10}}

\put(379,59){\line(-6,1){10}}

\put(303,40){\line(0,1){54}}
\put(303,40){\line(1,2){17}}
\put(303,40){\line(5,3){36}}
\put(303,40){\line(3,1){55}}
\put(303,40){\line(6,1){35}}
\put(303,34){\circle{12}}
\put(300,31){$\mld$}

\put(381,57){$\ldots$}
\put(342,44){$\ldots$}

\put(318,100){$p_1$}
\put(338,100){$p_2$}
\put(360,100){$p_3$}
\put(380,100){$\ldots$}
\put(380,80){$\ldots$}
\put(321,96){\line(0,-1){10}}
\put(341,96){\line(0,-1){23}}
\put(363,96){\line(0,-1){29}}

\put(131,10){{\bf Figure 2:} Infinite cirquents}
\end{picture}\end{center}

\noindent By an {\bf interpretation}\footnote{This \label{ftn1} concepts is termed  ``perfect interpretation'' in the other pieces of literature on CoL, where the word ``interpretation'' is reserved for a slightly more general concept. Since we only deal with perfect interpretations in this paper, we omit the word ``perfect'' and just say ``interpretation''.} in this and the following few sections we mean a function $^*$ that sends each atom $p$ to one of the values (elementary games) $p^*\in\{\twg,\tlg\}$. It immediately extends to a mapping from all literals to $\{\twg,\tlg\}$ by stipulating that $(\gneg p)^*=\gneg (p^*)$; that is, $^*$ sends $\gneg p$ to $\twg$ iff it sends $p$ to $\tlg$. 

Each interpretation $^*$ induces the predicate of {\em truth} with respect to (w.r.t.) $^*$ for cirquents and their nodes, as defined below. This definition, as well as similar definitions given later, silently but essentially relies on the fact that the graphs that we consider are well-founded. 

\begin{defi}\label{may14b}
Let $C$ be a cirquent and $^*$ an interpretation. With ``port'' and ``gate'' below meaning those of $C$, and ``true'' to be read as ``{\bf true w.r.t. $^*$}'',  we say that: 
\begin{enumerate}[$\bullet$]
\item An $L$-port
is true iff $L^*=\twg$ (any literal $L$).
\item A $\mld$-gate is true iff so is at least one of its children (thus, a childless $\mld$-gate is never true).
\item A $\mlc$-gate is true iff so are all of its children (thus, a childless $\mlc$-gate is always true).
\end{enumerate}
Finally, we say that the cirquent $C$ is true iff so is its root. 
\end{defi}

\begin{defi}\label{may14bb}
Let $C$ be a cirquent and $^*$ an interpretation. We define $C^*$ to be the elementary game the (only) legal run $\emptyrun$ of which is won by $\pp$ iff $C$ is true w.r.t. $^*$. 
\end{defi}

Thus, every interpretation $^*$ ``interprets'' a cirquent $C$ as
 one of the elementary games $\twg$ or $\tlg$. This will not necessarily be the case for the more general sorts of cirquents introduced in later sections though.

We say that two cirquents $C_1$ and $C_2$ are {\bf extensionally identical} iff, for every interpretation $^*$, $C_{1}^{*}=C_{2}^{*}$.  For instance, the two cirquents of Figure 1  are extensionally identical. Occasionally we may say ``{\em equivalent}'' instead of ``extensionally identical'', even though one should remember that equivalence often (in other pieces of literature) may mean something weaker than extensional identity. 

It should  be pointed out that the above definition of extensional identity applies not only to cirquents in the sense of the present section. It extends, without any changes in phrasing, to cirquents in the more general sense of any of the subsequent sections of this paper as well.  

Finally, we say that a cirquent $C$  is {\bf valid} iff, for any interpretation $^*$, $C^*=\twg$. 

Obviously the semantics that we have defined in this section is nothing but the kind old semantics of classical logic. Computability logic fully agrees with and adopts the latter. This is exactly what makes computability logic a conservative extension of classical logic. 

Let us agree that, whenever we speak about formulas of classical logic, they are assumed to be written in {\em negation normal form}, that is, in the form where the negation symbol $\gneg$ is only applied to atoms. If an expression violates this condition, it is to be understood just as a standard abbreviation. Similarly, if we write $E\mli F$, it is to be understood as an abbreviation of $\gneg E\mld E$. Also, slightly deviating from the tradition, we allow any finite numbers of arguments for conjunctions and disjunctions in classical formulas. The symbol $\twg$ will be understood as an abbreviation of the empty conjunction, $\tlg$ as an abbreviation of the empty disjunction,  
the expression $\mlc\{E\}$ will be used for the conjunction whose single conjunct is $E$, and similarly for $\mld\{E\}$. More generally, for any $n\geq 0$, $\mlc\{E_1,\ldots,E_n\}$ can (but not necessarily will)  be written instead of $E_1\mlc\ldots\mlc E_n$, and similarly for $\mld\{E_1,\ldots,E_n\}$. 

Every formula of classical propositional logic then can and will be  seen as a finite tree-like cirquent, namely, the cirquent which is nothing but  the parse tree for that formula. For instance, we shall understand the formula $\gneg p\mlc (\gneg p\mld p)\mlc (\gneg p\mld p)\mlc p$ as the left cirquent of Figure 1. 

Every finite --- not necessarily tree-like --- cirquent can also be translated into an equivalent formula of classical propositional logic. This can be done by first turning the cirquent into an extensionally identical  tree-like cirquent by duplicating and separating shared nodes, and then writing the formula whose parse tree such a cirquent is. For instance, applying  this procedure to the right cirquent of Figure 1 turns it into the left cirquent of the same figure and thus the formula $\gneg p\mlc (\gneg p\mld p)\mlc (\gneg p\mld p)\mlc p$. 

Without loss of generality, we assume that, unless otherwise specified, the universe of discourse in all cases that we consider --- i.e., the set over which the variables of classical first order logic range --- is $\{1,2,3,\ldots\}$. Then, from the perspective of classical first order logic, the universal quantification  of $E(x)$ is nothing but the ``long conjunction'' $E(1)\mlc E(2)\mlc E(3)\mlc\ldots$, and the existential quantification of $E(x)$ is nothing but the ``long disjunction'' $E(1)\mld E(2)\mld E(3)\mld\ldots$. To emphasize this connection, let us agree to use the expression $\mla\hspace{-1pt} xE(x)$ for the former and $\mle\hspace{-2pt} xE(x)$ for the latter, instead of the more usual $\cla xE(x)$ and $\cle x E(x)$ (computability logic reserves $\cla x$ and $\cle x$ for another, so called {\em blind}, sort of quantifiers; semantically they, just like $\mla\hspace{-1pt} x$ and $\mle \hspace{-2pt}x$, are conservative generalizations of the classical quantifiers).

Since quantifiers  are conjunctions or disjunctions,  it is obvious that all formulas of classical first order logic can also be seen as tree-like (albeit no longer finite) cirquents. For instance, the formula $\mla\hspace{-1pt} x\hspace{-1pt}\mle\hspace{-2pt} y \hspace{3pt} p(x,y)$ is nothing but the left cirquent of Figure 2. 

On the other hand, unlike the case with finite cirquents, obviously not all infinite cirquents can be directly and adequately translated into formulas of classical logic.
Of course, the great expressive power achieved by infinite cirquents, by itself, does not mean much, because such cirquents are generally ``unwritable'' or ``undrawable''. The cirquents of Figure 2 are among the not many lucky exceptions, but even there, the usage of the ellipsis is very informal, relying on a human reader's ability to see patterns and generalize. In order to take advantage of the expressiveness of cirquents, one needs to introduce notational conventions allowing to represent certain infinite patterns and (sub)cirquents through finite means. The quantifiers $\mla\hspace{-1pt} x$ and $\mle\hspace{-2pt} x$  are among such means. Defining new, ever more expressive means (not reducible to $\mla \hspace{-1pt}x,\mle \hspace{-2pt}x$) is certainly possible. Among the advantages of considering all --- rather than only finitely represented --- cirquents as we do in this paper is that a semantics for them has to be set up only once. If and when various abbreviations and finite means of expression are introduced, one will only need to explain what kinds of cirquents or subcirquents they stand for, without otherwise redefining or extending the already agreed-upon general semantics for cirquents. 

But higher expressiveness is not the only advantage of cirquents over formulas. Another very serious advantage is the higher efficiency of cirquents. Let us for now talk only about finite cirquents. As we know, all such cirquents can be written as  formulas of classical propositional logic. So, they do not offer any additional expressive power. But they {\em do} offer dramatically improved efficiency in representing Boolean functions. In Section 8 of \cite{Japdeep} one can find examples of naturally emerging sets of Boolean functions which can be represented through polynomial size cirquents but require exponential sizes if represented through formulas. The higher efficiency of cirquents is achieved due to the possibility to {\em share} children between different parents --- the mechanism absent in formulas, because of which an exponential explosion may easily occur when translating a (non-tree-like) cirquent into an equivalent formula.   
Imagine where the computer industry would be at present if, for some strange reason, the computer engineers had insisted on tree-like (rather than graph-like) circuitries! 

Cirquents offer not only improved efficiency of expression, but also improved efficiency of proofs in deductive systems. In \cite{Japdeep}, a cirquent-based analytic deductive system {\bf CL8} for classical propositional logic is set up which is shown to yield an exponential improvement (over formula-based analytic deductive systems) in proof efficiency. For instance, the notorious propositional Pigeonhole principle, which is known to have no polynomial size analytic proofs  in traditional systems, has been shown to have such proofs in  {\bf CL8}.

\section{Selectional gates}\label{s4}

\noindent We now start a series of generalizations for cirquents and their semantics. We agree that, throughout the rest of this paper, unless otherwise specified, ``cirquent'' and all related terms are always to be understood in their latest, most general, sense.    

In this section we extend the concept of {\bf cirquents} defined in the previous section by allowing, along with $\mld$ and $\mlc$, the following six additional possible labels for gates: $\tgd\hspace{-2pt},\hspace{-2pt}\tgc\hspace{-2pt},\hspace{-2pt}\sqd\hspace{-2pt},\hspace{-2pt}\sqc\hspace{-2pt},\add,\adc$. Gates labeled with any of these new symbols we call {\bf selectional} gates. And gates labeled with the old $\mld$ or $\mlc$ we call {\bf parallel} gates. 

Selectional gates, in turn, are subdivided into three groups: 
\begin{enumerate}[$\bullet$]
\item $\{\hspace{-2pt}\tgd\hspace{-2pt},\hspace{-2pt}\tgc\hspace{-2pt}\}$, referred to as {\bf toggling} gates; 
\item $\{\hspace{-2pt}\sqd\hspace{-2pt},\hspace{-2pt}\sqc\hspace{-2pt}\}$, referred to as {\bf sequential} gates;
\item $\{\add,\adc\}$, referred to as {\bf choice} gates.
\end{enumerate}

 The eight kinds of gates are also divided into the following two groups: 
\begin{enumerate}[$\bullet$]
\item $\{\mld,\hspace{-2pt}\tgd\hspace{-2pt},\hspace{-2pt}\sqd\hspace{-2pt},\add\}$, termed {\bf disjunctive} gates (or simply {\bf disjunctions});
\item   $\{\mlc,\hspace{-2pt}\tgc\hspace{-2pt},\hspace{-2pt}\sqc\hspace{-2pt},\adc\}$, termed {\bf conjunctive} gates (or simply {\bf conjunctions}). 
\end{enumerate}\medskip

\noindent Thus, $\mlc$ is to be referred to as ``parallel conjunction'', $\add$ as ``choice disjunction'', etc.
 
The same eight symbols can be used to construct formulas in the  standard way. Of course, in the context of formulas, these symbols will be referred to as {\bf operators} or {\bf connectives} rather than as gates. Formulas of computability logic accordingly may also use eight sorts of {\bf quantifiers}. They are written as symbols of the same shape as the corresponding connectives, but in a larger size. Two of such quantifiers --- $\mla \hspace{-1pt}x$ and $\mle \hspace{-2pt}x$ --- have already been explained in the previous section. The remaining quantifiers are understood in the same way: $\ada xE(x)$ is the infinite choice conjunction $E(1)\adc E(2)\adc\ldots$, \ $\ade xE(x)$ is the infinite choice disjunction $E(1)\add E(2)\add\ldots$, \ and similarly for $\tga\hspace{-1pt},\hspace{-1pt}\tge\hspace{-1pt},\hspace{-1pt}\sqa,\sqe$. Since quantifiers are again nothing but ``long'' conjunctions and disjunctions, as pointed out in the previous section, there is no necessity to have special gates for them, as our approach that permits  gates with infinite outdegrees covers them all. For similar reasons, there is no necessity to have special gates for what computability logic calls {\em (co)recurrence operators}. The {\em parallel recurrence} $\pst E$ of $E$ is defined as the infinite parallel conjunction $E\mlc E\mlc\ldots$, the {\em parallel corecurrence} $\pcost E$ of $E$ is defined as the infinite parallel disjunction $E\mld E\mld\ldots$, and similarly for {\em toggling recurrence} $\tgpst$, {\em toggling corecurrence} $\tgpcost$, {\em sequential recurrence} $\sst$ and {\em sequential corecurrence} $\scost$ (choice recurrence and corecurrence are not considered because, semantically, both $E\adc E\adc\ldots$ and $E\add E\add\ldots$ are simply equivalent to $E$).  

All of the operators $\mld,\mlc,\hspace{-2pt}\tgd\hspace{-2pt},\hspace{-2pt}\tgc\hspace{-2pt},\hspace{-2pt}\sqd\hspace{-2pt},\hspace{-2pt}\sqc\hspace{-2pt},\add,\adc$, including their quantifier and recurrence versions, have been already motivated, defined and studied in computability logic. This, however, has been done only in the context of formulas. In this paper we extend the earlier approach and concepts of computability logic from formulas to cirquents as generalized formulas. 

As before, an  interpretation is an assignment $^*$ of $\twg$ or $\tlg$ to each atom, extended to all literals by commuting with $\gneg$. And, as before, such an assignment induces a mapping that sends each cirquent $C$ to a game $C^*$.  When $C$ is a cirquent in the sense of the previous section, $C^*$ is always an elementary game ($\twg$ or $\tlg$). When, however, $C$ contains selectional gates, the game $C^*$ is no longer elementary. 

To define such games $C^*$, let us agree that, throughout this paper, positive integers are identified with their decimal representations, so that, when we say ``number $n$'', we may mean either the {\em number} $n$ as an abstract object, or the {\em string} that represents $n$ in the decimal notation. Among the benefits of  this convention is that it allows us to identify nodes of a cirquent with their IDs. 

\begin{defi}\label{may14a}
Let $C$ be a cirquent, $^*$ an interpretation, and $\Phi$ a position. $\Phi$ is a {\bf legal position} of the game $C^*$ iff, with a ``gate'' below meaning a gate of $C$,  the following conditions are satisfied:
\begin{enumerate}[(1)]
\item Each labmove of $\Phi$ has one of the following forms:
\begin{enumerate}
\item $\pp g.i$, where $g$ is a $\tgd$\hspace{-2pt}-,\hspace{-2pt} $\sqd$\hspace{-2pt}-\hspace{-2pt} or $\add$-gate  and $i$ is a positive integer not exceeding 
the outdegree of $g$; 
\item $\oo g.i$, where $g$ is a $\tgc$\hspace{-2pt}-,\hspace{-2pt} $\sqc$\hspace{-2pt}-\hspace{-2pt} or $\adc$-gate  and $i$ is a positive integer not exceeding 
the outdegree of $g$. 
\end{enumerate} 
\item Whenever $g$ is a choice gate, $\Phi$ contains at most one occurrence of a labmove of the form $\xx g.i$.   
\item Whenever $g$ is a sequential gate and $\Phi$ is of the form $\seq{\ldots , \xx g.i ,\ldots, \xx g.j ,\ldots}$, we have $i<j$. 
\end{enumerate}
\end{defi}\medskip

\noindent Note that the set of legal runs of $C^*$ does not depend on $^*$. Hence, in the sequel,  we can unambiguously omit ``$^*$'' and simply say ``legal run of $C$''. 

The intuitive meaning of a move of the form $g.i$ is {\bf selecting} the $i$th outgoing edge --- together with the child pointed at by such an edge --- of (in) the selectional gate $g$. In a disjunctive selectional gate, a selection is always made by player $\pp$; and in a conjunctive selectional gate, a selection is always made by player $\oo$. The difference between the three types of selectional gates is only in how many selections and in what order are allowed to be made in the same gate. In a choice gate, a selection can be made only once. In a sequential gate, selections can be reconsidered any number of times, but only in the left-to-right order (once an edge $\#i$ is selected, no edge $\# j$ with $j\leq i$ can be (re)selected afterwards). In a toggling gate, selections can be reconsidered any number of times and in any order. This includes the possibility to select the same edge over and over again. 

\begin{defi}\label{mm20}
In the context of a given cirquent $C$ and a  legal run $\Gamma$ of $C$, we will say that a selectional gate $g$ is {\bf unresolved} iff either no moves of the form $g.j$ have been made in $\Gamma$, or infinitely many such moves  have been made.\footnote{The latter, of course, may not be the case if $g$ is a choice gate, or a sequential gate with a finite outdegree.} Otherwise $g$ is {\bf resolved} and, where $g.i$ is the last move of the form $g.j$ made in $\Gamma$, the child pointed at by the $i$th outgoing edge of $g$ is said to be the {\bf resolvent} of $g$. 
\end{defi}

Intuitively, the resolvent is the ``final'', or ``ultimate'' selection of a child made in gate $g$ by the corresponding player. There is no such ``ultimate'' selection in unresolved gates.

The following definition conservatively generalizes Definition \ref{may14b} of truth. Now we have a legal run $\Gamma$ of the cirquent as an additional parameter, which was trivial (namely, $\Gamma=\emptyrun$) in Definition \ref{may14b} and hence not mentioned there.

\begin{defi}\label{may14c}
Let $C$ be a cirquent, $^*$ an interpretation, and $\Gamma$ a legal run of $C$. In this context, with ``true'' to be read as ``{\bf true w.r.t. $(^*,\Gamma)$}'', we say that: 
\begin{enumerate}[$\bullet$]
\item An $L$-port  is true iff $L^*=\twg$.
\item A $\mld$-gate is true iff so is at least one of its children.
\item A $\mlc$-gate is true iff so are all of its children.
\item A resolved selectional gate is true iff so is its resolvent.
\item No unresolved disjunctive selectional gate is true.
\item Each unresolved conjunctive selectional gate is true. 
\end{enumerate}
Finally, we say that $C$ is true iff so is its root.
\end{defi}

As we just did in the above definition, when $^*$ and $\Gamma$ are fixed in a context or are otherwise irrelevant, we may omit ``w.r.t. $(^*,\Gamma)$'' and just say ``true''.
 
\begin{defi}\label{may14aa}
Let $C$ be a cirquent, $^*$ an interpretation, and $\Gamma$ a legal run of $C$. Then $\Gamma$ is a $\pp$-won run of the game $C^*$ iff $C$ is true 
w.r.t. $(^*,\Gamma)$. 
\end{defi}

Definitions \ref{may14a} and \ref{may14aa}, together, provide a definition of the game $C^*$, for any cirquent $C$ and interpretation $^*$. To such a game $C^*$ we may refer  as ``the game represented by $C$ under interpretation $^*$'', or as ``$C$ under interpretation $^*$'', or --- when $^*$ is fixed or irrelevant --- as ``the game represented by $C$''.
We may also say that ``$^*$  interprets $C$ as $C^*$''. Similarly for atoms and literals instead of cirquents. Also, in informal contexts we may identify a cirquent $C$ or a literal $L$ with a (the) game represented by it, and write $C$ or $L$ where, strictly speaking, $C^*$ or $L^*$ is meant.  These and similar terminological conventions apply not only to the present section, but the rest of the paper as well. 

We now need to generalize the definition of  validity given in the previous section to  cirquents in the sense of the present section. As it turns out, such a generalization can be made in two, equally natural, ways:

\begin{defi}\label{may20}
Let $C$ be a cirquent in the sense of the present or any of the subsequent sections. We say that:\\medskip

1. $C$ is {\bf weakly valid} iff, for any interpretation $^*$, there is an HPM ${\mathcal M}$ such that $\mathcal M$ wins the game $C^*$.

2. $C$ is {\bf strongly valid}
iff there is an HPM ${\mathcal M}$ such that, for any interpretation $^*$,  ${\mathcal M}$  wins the game $C^*$. 
\end{defi}  

When $\mathcal M$ and $C$ satisfy the second clause of the above definition, we say that $\mathcal M$ is a {\bf uniform solution} for $C$. Intuitively, a uniform solution $\mathcal M$ for a cirquent $C$ is an interpretation-independent winning strategy: since the ``intended'' or ``actual'' interpretation $^*$ is not visible to the machine,  $\mathcal M$  has to play in some standard, uniform way that would be successful 
for any possible interpretation of $C$. To put it in other words, a uniform solution is a {\em purely logical solution}, which is based only on the {\em form} of a cirquent rather than any extra-logical {\em meaning} (interpretation) we have or could have associated with it. 

It is obvious that for cirquents of  the previous section, the weak and strong forms of validity coincide with what we simply called ``validity'' there. In general, however, not every weakly valid cirquent would also be strongly valid.   A simplest example of such a weakly valid cirquent is $\gneg p\add p$.\footnote{If, however, we do not limit our considerations to perfect interpretations (see the footnote on page \pageref{ftn1}) and allow all interpretations in the definition of weak validity, this cirquent will no longer be weakly valid. In fact, for all 
well studied fragments of CoL, when interpretations are not required to be perfect, the weak and the strong versions of validity have been shown to yield the same classes of formulas. Strong validity in the CoL literature is usually referred to as {\em uniform validity}, and weak validity as (simply) {\em validity}.}
 Under any interpretation $^*$, the game 
represented by this cirquent is won by one of the two HPMs ${\mathcal M}_1$ or ${\mathcal M}_2$, where ${\mathcal M}_1$ is the machine that selects the left disjunct and rests, while  ${\mathcal M}_2$ is the machine that selects the right disjunct and rests. However, which of these two machines wins the game depends on whether $^*$ interprets $p$ as $\tlg$ or $\twg$. In general, there is no {\em one} machine that wins the game for {\em any} possible interpretation. That is, the cirquent $\gneg p\add p$ has no uniform solution, and thus it is not strongly valid.

Which of the two versions of validity is more interesting depends on the motivational standpoint. 
Weak validity tells us what can be computed in principle. So, a 
com\-putability-theoretician would focus on this concept.  
On the other hand, it is strong rather than weak validity that would be  of interest in all application areas of CoL.
There we want a logic on which a universal problem-solving machine can be based. Such a machine would or should be able to solve problems represented by logical expressions  without any specific knowledge of the meanings of 
their atoms, i.e. without knowledge of the actual interpretation, which may vary from situation to situation or from application to application. Strong validity is exactly what fits the bill in this case. Throughout this paper, our primary focus will be on strong rather than weak validity.

To appreciate the difference between the parallel and choice groups of gates or connectives, let us compare the two cirquents of Figure 3. 

\begin{center} \begin{picture}(390,142)
\put(0,119){$\gneg p(1)$}
\put(33,119){$p(1)$}
\put(25,101){\line(-1,1){12}}
\put(25,101){\line(1,1){12}}
\put(25,95){\circle{12}}
\put(22,92){$\mld$}

\put(64,119){$\gneg p(2)$}
\put(97,119){$p(2)$}
\put(89,101){\line(-1,1){12}}
\put(89,101){\line(1,1){12}}
\put(89,95){\circle{12}}
\put(86,92){$\mld$}

\put(128,119){$\gneg p(3)$}
\put(161,119){$p(3)$}
\put(153,101){\line(-1,1){12}}
\put(153,101){\line(1,1){12}}
\put(153,95){\circle{12}}
\put(150,92){$\mld$}

\put(25,60){\line(0,1){29}}
\put(25,60){\line(2,1){60}}
\put(25,60){\line(4,1){124}}
\put(25,60){\line(6,1){124}}
\put(25,54){\circle{12}}
\put(22,51){$\mlc$}
\put(165,80){\large $\ldots$}
\put(19,31){$\mla x\bigl(\gneg p(x)\mld p(x)\bigr)$}

\put(220,119){$\gneg p(1)$}
\put(253,119){$p(1)$}
\put(245,101){\line(-1,1){12}}
\put(245,101){\line(1,1){12}}
\put(245,95){\circle{12}}
\put(242,92){$\add$}

\put(284,119){$\gneg p(2)$}
\put(317,119){$p(2)$}
\put(309,101){\line(-1,1){12}}
\put(309,101){\line(1,1){12}}
\put(309,95){\circle{12}}
\put(306,92){$\add$}

\put(348,119){$\gneg p(3)$}
\put(381,119){$p(3)$}
\put(373,101){\line(-1,1){12}}
\put(373,101){\line(1,1){12}}
\put(373,95){\circle{12}}
\put(370,92){$\add$}

\put(245,60){\line(0,1){29}}
\put(245,60){\line(2,1){60}}
\put(245,60){\line(4,1){124}}
\put(245,60){\line(6,1){124}}
\put(245,54){\circle{12}}
\put(242,51){$\adc$}
\put(385,80){\large $\ldots$}
\put(240,31){$\ada x\bigl(\gneg p(x)\add p(x)\bigr)$}

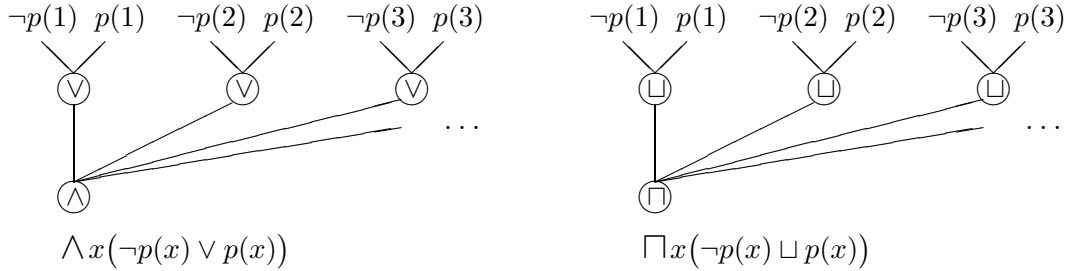
\put(119,10){{\bf Figure 3:} Parallel versus choice gates}
\end{picture}\end{center}\medskip
 
\noindent The game represented by the left cirquent of Figure 3 is elementary,  where no moves can or should be made by either player. It is also easy to see that this game (the only legal run $\emptyrun$ of it, that is) is automatically won by $\pp$, no matter what interpretation $^*$ is applied, so, it is both weakly and strongly valid. On the other hand, the game represented by the right cirquent of the same figure is not elementary. And it is neither strongly valid nor weakly valid.  A legal move by $\oo$ in this game consists in selecting one of the infinitely many outgoing edges (and hence children) of the root, intuitively corresponding to choosing a value for $x$ in the formula $\ada x\bigl(\gneg p(x)\add p(x)\bigr)$. And a(ny) legal move by $\pp$ consists in selecting one of the two outgoing edges  (and hence children) of one of the $\add$-gates. Making more than one selection  in the same choice (unlike toggling or sequential) gate is not allowed, so that a selection automatically also is the resolvent of the gate. The overall game is won by $\pp$ iff either $\oo$ failed to make a selection in the root, or else, 
where $i$ is the outgoing edge of the root selected by $\oo$ and $a_i$ is the corresponding ($i$th, that is) child of the root, either (1) $\pp$ has selected the left outgoing edge of $a_i$ and $\gneg p(i)$ is true, or  (2) $\pp$ has selected the right outgoing edge of $a_i$ and $p(i)$ is true. There are no conditions on when the available moves should be made, and generally they can be made by either player at any time and in any order. So, in the present example, $\pp$ can legally make selections in several or even all $\add$-gates. But, of course, a reasonable strategy for $\pp$ is to first wait till $\oo$ resolves the root (otherwise $\pp$ wins), and then focus only on the resolvent of the root (what happens in the other $\add$-gates no longer matters), trying to select the true child of it.  

From the above explanation it should be clear that the right cirquent of Figure 3 expresses the problem of deciding (in the traditional sense) the predicate $p(x)$. That is, under any given interpretation $^*$, the game represented by that cirquent has an algorithmic winning strategy  by $\pp$ if and only if the predicate $\bigl(p(x)\bigr)^*$ --- which we simply write  as 
 $p(x)$ --- is decidable. As not all predicates are decidable, the cirquent  is not weakly valid, let alone being strongly valid. 

To get a feel for sequential and toggling gates, let us look at Figure 4.

\begin{center} \begin{picture}(390,142)
\put(0,119){$\gneg p(1)$}
\put(33,119){$p(1)$}
\put(25,101){\line(-1,1){12}}
\put(25,101){\line(1,1){12}}
\put(25,95){\circle{12}}
\put(19,92){$\sqd$}

\put(64,119){$\gneg p(2)$}
\put(97,119){$p(2)$}
\put(89,101){\line(-1,1){12}}
\put(89,101){\line(1,1){12}}
\put(89,95){\circle{12}}
\put(83,92){$\sqd$}

\put(128,119){$\gneg p(3)$}
\put(161,119){$p(3)$}
\put(153,101){\line(-1,1){12}}
\put(153,101){\line(1,1){12}}
\put(153,95){\circle{12}}
\put(147,92){$\sqd$}

\put(25,60){\line(0,1){29}}
\put(25,60){\line(2,1){60}}
\put(25,60){\line(4,1){124}}
\put(25,60){\line(6,1){124}}
\put(25,54){\circle{12}}
\put(22,51){$\adc$}
\put(165,80){\large $\ldots$}
\put(19,31){$\ada x\bigl(\gneg p(x)\sqd p(x)\bigr)$}

\put(220,119){$\gneg p(1)$}
\put(253,119){$p(1)$}
\put(245,101){\line(-1,1){12}}
\put(245,101){\line(1,1){12}}
\put(245,95){\circle{12}}
\put(240,92){$\tgd$}

\put(284,119){$\gneg p(2)$}
\put(317,119){$p(2)$}
\put(309,101){\line(-1,1){12}}
\put(309,101){\line(1,1){12}}
\put(309,95){\circle{12}}
\put(304,92){$\tgd$}

\put(348,119){$\gneg p(3)$}
\put(381,119){$p(3)$}
\put(373,101){\line(-1,1){12}}
\put(373,101){\line(1,1){12}}
\put(373,95){\circle{12}}
\put(368,92){$\tgd$}

\put(245,60){\line(0,1){29}}
\put(245,60){\line(2,1){60}}
\put(245,60){\line(4,1){124}}
\put(245,60){\line(6,1){124}}
\put(245,54){\circle{12}}
\put(242,51){$\adc$}
\put(385,80){\large $\ldots$}
\put(240,31){$\ada x\bigl(\gneg p(x)\tgd p(x)\bigr)$}

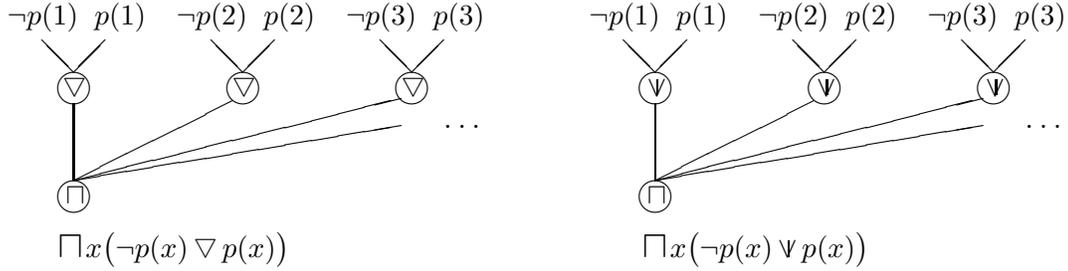
\put(105,10){{\bf Figure 4:} Sequential versus toggling gates}
\end{picture}\end{center}
 
\noindent The cirquents of Figure 4 look similar to the right cirquent of Figure 3. And the latter, as we know, represents the problem of {\em deciding} $p(x)$. Then what are the
problems  represented by the cirquents of Figure 4?

The left cirquent of Figure 4 represents the problem of {\em semideciding} $p(x)$. That is, under any given interpretation, the game represented by this cirquent is computable if and only if the predicate $p(x)$ is semidecidable (recursively enumerable). Indeed, suppose $p(x)$ is semidecidable. Then an algorithmic winning strategy for the game represented by the cirquent goes like this. Wait till the environment selects the $i$th child $a_i$ of the root for some $i$. Then select the left child of $a_i$, after which start looking for a certificate of the truth of $p(i)$. If and when such a certificate is found, select the  right child of $a_i$, and rest your case. It is obvious that this strategy indeed wins. For the opposite direction, suppose $\mathcal M$ is an HPM that wins the game represented by the  cirquent. Then a semidecision procedure for the predicate $p(x)$ goes like this. After receiving an input $i$, simulate the work of $\mathcal M$ in the scenario where, at the beginning of the play, the environment selected the $i$th child $a_i$ of the root. If and when you see in this simulation that $\mathcal M$ selected the right child of $a_i$, accept the input. 

As for the right cirquent of Figure 4, it also represents a decision-style problem, which is further weaker than the problem of semideciding $p(x)$.
 This problem is known in the literature as {\em recursive approximation} (cf. \cite{Hinman}, Definition 8.3.9). Recursively approximating $p(x)$ means telling whether $p(x)$ is true or not, but doing so in the same style as semideciding does in negative cases: by correctly saying ``Yes'' or ``No'' at some point (after perhaps taking back previous answers several times) and never reconsidering this answer afterwards.  Observe that semideciding $p(x)$ can be seen as always saying ``No'' at the beginning and then, if this answer is incorrect, changing it to ``Yes'' at some later time; so, when the answer is negative, this will be expressed by saying ``No'' and never taking back this answer, yet without ever indicating that the answer is final and will not change.\footnote{Unless, of course, the procedure halts by good luck. Halting without saying ``Yes'' can then be seen as an explicit indication that the original answer ``No'' was final.} Thus, the difference between semideciding and recursively approximating is that, unlike a semidecision procedure, a recursive approximation procedure can reconsider {\em both} negative and positive answers, and  do so several times rather than only once. 
In perhaps more familiar terms, according to Shoenfield's Limit Lemma (Cf. \cite{Hinman}, Lemma 8.3.12), a predicate $p(x)$ is recursively approximable (the problem of its recursive approximation has an algorithmic solution) iff $p(x)$ is of arithmetical complexity $\Delta_2$, i.e.,  both $p(x)$ and its negation can be written in the form $\cle z\cla y\hspace{1pt}s(z,y,x)$, where $s(z,y,x)$ is a decidable predicate.

We could go on and on illustrating how our formalism --- even at the formula level --- makes it possible to express various known and unknown natural and interesting properties, relations and operations on predicates or games as generalized predicates, but this can take us too far. Plenty of examples and discussions in that style can be found in, say, \cite{Jap03,Japseq,Japfin,Japtogl}.  Here we only want to 
point out the difference between our treatment of $\mld,\mlc$ (including quantifiers as infinite versions of $\mld,\mlc$) and the more traditional game-semantical approaches, most notably that of Hintikka's \cite{Hintikka73} game-theoretic semantics. The latter essentially treats $\mld,\mlc$ as we treat $\add,\adc$ --- namely, associates $\pp$'s moves/choices with $\mld$ and $\oo$'s moves/choices with $\mlc$. Computability logic, on the other hand, in the style used by Blass \cite{Bla92} for the multiplicatives of linear logic, treats $A\mld B$ and $A\mlc B$ as parallel combinations of games: these are simultaneous plays on ``two boards'' (within the two components). In order to win $A\mlc B$, $\pp$ needs to win in both components, while in order to win $A\mld B$, it is sufficient for $\pp$ to win in just one component. No choice between $A$ and $B$ is expected to be made at any time by either player. 
Note that otherwise strong validity would not at all be an interesting concept: there would be no strongly valid cirquents except for some pathological cases such as the cirquent whose root is a childless conjunctive gate. 

Another crucial difference between our approach and that of Hintikka, as well as the approach of Blass, is that we insist on the effectiveness of strategies while the latter allow any strategies. It is not hard to see that, if we allowed any (rather than only algorithmic) strategies, the system of our gates would semantically collapse to merely the parallel group. That is, a cirquent $C$ (under whatever interpretation) would have a winning strategy by $\pp$ if and only if  $C'$ does, where $C'$ is the result of replacing in $C$ every disjunctive selectional gate by $\mld$ and every conjunctive selectional gate by $\mlc$. 

Anyway, an important issue for the present paper is that of the advantages of cirquents over formulas. As we remember, finite cirquents without selectional gates are more efficient tools of expression than formulas of classical logic are, but otherwise their expressive power is the same as that of formulas. How about finite cirquents in the more general sense of this section --- ones containing selectional gates? In this case, finite (let alone infinite) cirquents are not only more efficient but also more expressive than formulas. To get some insights, let us look at Figure 5. 

\begin{center} \begin{picture}(324,162)

\put(23,142){$p$}
\put(50,142){$q$}
\put(39,125){\line(-1,1){12}}
\put(39,125){\line(1,1){12}}
\put(39,119){\circle{12}}
\put(36,116){$\add$}

\put(3,118){$\gneg p$}
\put(28,101){\line(-1,1){12}}
\put(28,101){\line(1,1){12}}
\put(28,95){\circle{12}}
\put(25,92){$\mld$}

\put(102,142){$q$}
\put(75,142){$p$}
\put(91,125){\line(-1,1){12}}
\put(91,125){\line(1,1){12}}
\put(91,119){\circle{12}}
\put(88,116){$\add$}

\put(112,118){$\gneg q$}
\put(103,101){\line(-1,1){12}}
\put(103,101){\line(1,1){12}}
\put(103,95){\circle{12}}
\put(100,92){$\mld$}

\put(66,64){\line(-3,2){37}}
\put(66,64){\line(3,2){37}}
\put(66,58){\circle{12}}
\put(63,55){$\mlc$}

\put(0,31){$\bigl(\gneg p\mld (p\add q)\bigr)\mlc \bigl((p\add q)\mld \gneg q\bigr)$}

\put(249,142){$p$}
\put(276,142){$q$}
\put(265,125){\line(-1,1){12}}
\put(265,125){\line(1,1){12}}
\put(265,119){\circle{12}}
\put(262,116){$\add$}

\put(203,118){$\gneg p$}
\put(228,101){\line(-1,1){12}}
\put(228,101){\line(3,1){36}}
\put(228,95){\circle{12}}
\put(225,92){$\mld$}

\put(312,118){$\gneg q$}
\put(303,101){\line(-3,1){36}}
\put(303,101){\line(1,1){12}}
\put(303,95){\circle{12}}
\put(300,92){$\mld$}

\put(266,64){\line(-3,2){37}}
\put(266,64){\line(3,2){37}}
\put(266,58){\circle{12}}
\put(263,55){$\mlc$}

\put(238,31){\em No formula}

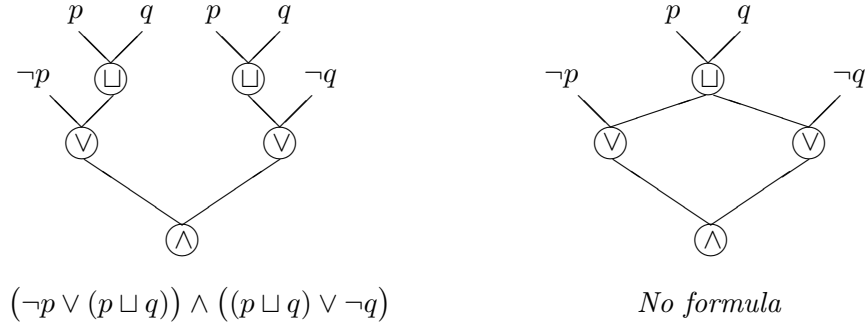
\put(77,10){{\bf Figure 5:} Unshared versus shared gates}
\end{picture}\end{center}
  
\noindent On the left of Figure 5 we see a tree-like cirquent obtained from the non-tree-like cirquent on the right by duplicating and separating shared descendants in the same style as we obtained the left cirquent of Figure 1 from the right cirquent.  The two cirquents of Figure 5, however, are not extensionally identical --- after all, every legal run of the right cirquent contains at most one move while legal runs of the left cirquent may contain two moves. The two cirquents are different in a much stronger sense than extensional non-identity though. A machine that selects the left outgoing edge of the left $\add$-gate and the right outgoing edge of the right $\add$-gate wins the game represented by the left cirquent of Figure  5 under any interpretation, so that the cirquent is strongly valid. On the other hand, the right cirquent of Figure 5 is obviously not strongly valid. Morally, the difference between the two cirquents is that, in the right cirquent, unlike the left cirquent, the subcirquent (``resource'') $p\add q$  is {\em shared} between the two $\mlc$-conjuncted parents. Sharing means that only one resolution can be made in the $\add$-gate --- a resolution that ``works well'' for both parents. There is no such resolution though, because the two parents ``want'' two different choices of a $\add$-disjunct. In contrast, in the left cirquent, both of these conflicting ``desires'' can be satisfied as each parent has its own $\add$-child, so that there is no need to coordinate resolutions.  
  
In layman's terms, a shared resource (subcirquent) can be explained using the following metaphor. Imagine Victor and his wife Peggy own a joint bank account, with a balance of \$ 20,000, reserved for family needs. This resource can be seen as a shared choice combination of {\em truck}, {\em furniture}, and anything of family use that \$20,000 can buy.\footnote{If Victor and Peggy may change their mind several times, and the sellers' return policies are flexible enough, then this is a toggling combination rather than a choice one.} However, if Victor prefers a truck while Peggy prefers furniture (and both cost \$20,000), then they will have to sit down and decide together whether furniture is more important for the family or a truck, as their budget does not allow to get both. The situation would be very different if both Victor and Peggy had their own accounts, each worth \$20,000. After all, the total balance in this case would be \$40,000 rather than \$20,000.

As we saw, the right cirquent of Figure 5 cannot be adequately translated into a formula using the standard way of turning non-trees into trees. However, using some non-standard and creative ways, a formula which is extensionally identical to that cirquent can still be found. For instance, one can easily see that 
 $(p\add q)\mld (\gneg p\mlc\gneg q)$ is such a formula (as long as its $\add$-gate is given the same ID as the ID of the original cirquent's $\add$-gate, of course). Well, we just got lucky in this particular case. ``Luck'' would not have been on our side if the cirquent under question was slightly more evolved, such as the one of Figure 6.

\begin{center} \begin{picture}(180,125)

\put(23,105){$p_1$}
\put(50,105){$p_2$}
\put(39,88){\line(-1,1){12}}
\put(39,88){\line(1,1){12}}
\put(39,82){\circle{12}}
\put(36,79){$\add$}

\put(91,66){\line(-5,1){51}}
\put(91,66){\line(5,1){51}}
\put(91,60){\circle{12}}
\put(88,57){$\mlc$}

\put(102,105){$p_4$}
\put(75,105){$p_3$}
\put(91,88){\line(-1,1){12}}
\put(91,88){\line(1,1){12}}
\put(91,82){\circle{12}}
\put(88,79){$\add$}

\put(154,105){$p_6$}
\put(127,105){$p_5$}
\put(143,88){\line(-1,1){12}}
\put(143,88){\line(1,1){12}}
\put(143,82){\circle{12}}
\put(140,79){$\add$}

\put(39,66){\line(0,1){10}}
\put(39,66){\line(5,1){51}}
\put(39,60){\circle{12}}
\put(36,57){$\mlc$}

\put(143,66){\line(0,1){10}}
\put(143,66){\line(-5,1){51}}
\put(143,60){\circle{12}}
\put(140,57){$\mlc$}

\put(91,44){\line(0,1){10}}
\put(91,44){\line(-5,1){51}}
\put(91,44){\line(5,1){51}}
\put(91,38){\circle{12}}
\put(88,35){$\mld$}

\put(-13,10){{\bf Figure 6:} A more evolved example of sharing}
\end{picture}\end{center}

\section{Clustering selectional gates}\label{s5}

\noindent We now further generalize {\bf cirquents} by adding an extra parameter to them,  called {\bf clustering} (for selectional gates). The latter is nothing but a partition of the set of all selectional gates into subsets, called {\bf clusters}, satisfying the condition that all gates within any given cluster have the same label (all are $\add$-gates, or all are $\tgc$-gates, or \ldots) and the same outdegree. Due to this condition, we can talk about the {\bf outdegree} of a cluster meaning the common outdegree of its elements, or the {\bf type}  of a cluster meaning the common type (label) of its elements. An additional condition that we require to be satisfied by all cirquents is that the question on whether any two given selectional gates are in the same cluster be decidable.  

Just like nodes do, each cluster also has its own {\bf ID}. For clarity, we assume that the ID of a cluster is the same as the smallest of the IDs  of the elements of the cluster. 
The {\em extended ID}  of a selectional gate is the expression $n_k$, where $n$ is the ID of the gate and the subscript $k$ is the ID  of the cluster to which the gate belongs. When representing cirquents graphically, one could require to show the extended ID of each selectional gate and (just) the ID of any other node. More often, however, we draw cirquents in a lazy yet unambiguous way, where only cluster IDs are indicated; furthermore, such IDs can be (though not always will be) omitted for 
singleton clusters. Figure 7 shows the same cirquent, on the left represented fully and on the right represented in a lazy way, with the identical cluster ID ``$7$'' attached to two $\add$-gates indicating that they are in the same cluster, while the absence of a cluster ID for the middle $\add$-gate indicating that it is in a different, singleton cluster (and so would be any other selectional gate if drawn without a cluster ID). Thus, altogether there are two clusters here, one --- cluster $8$ --- containing gate $8$, and the other --- cluster $7$ containing gates $7$ and $9$. 

\begin{center} \begin{picture}(362,165)

\put(23,137){$p_1$}
\put(24,145){$1$}
\put(50,137){$p_2$}
\put(51,145){$2$}
\put(39,120){\line(-1,1){12}}
\put(39,120){\line(1,1){12}}
\put(39,114){\circle{12}}
\put(36,111){$\add$}

\put(21,110){$7_7$}
\put(152,110){$9_7$}
\put(87,97){$8_8$}
\put(20,70){$10$}
\put(152,70){$11$}
\put(86,50){$12$}

\put(102,137){$p_4$}
\put(103,145){$4$}
\put(75,137){$p_3$}
\put(76,145){$3$}
\put(91,120){\line(-1,1){12}}
\put(91,120){\line(1,1){12}}
\put(91,114){\circle{12}}
\put(88,111){$\add$}

\put(154,137){$p_6$}
\put(155,145){$6$}
\put(127,137){$p_5$}
\put(128,145){$5$}
\put(143,120){\line(-1,1){12}}
\put(143,120){\line(1,1){12}}
\put(143,114){\circle{12}}
\put(140,111){$\add$}

\put(39,82){\line(0,1){26}}
\put(39,82){\line(2,1){51}}
\put(39,76){\circle{12}}
\put(36,73){$\mlc$}

\put(143,82){\line(0,1){26}}
\put(143,82){\line(-2,1){51}}
\put(143,76){\circle{12}}
\put(140,73){$\mlc$}

\put(91,44){\line(-2,1){51}}
\put(91,44){\line(2,1){51}}
\put(91,38){\circle{12}}
\put(88,35){$\mld$}

\put(223,137){$p_1$}
\put(250,137){$p_2$}
\put(239,120){\line(-1,1){12}}
\put(239,120){\line(1,1){12}}
\put(239,114){\circle{12}}
\put(236,111){$\add$}

\put(228,110){$_7$}
\put(350,110){$_7$}

\put(302,137){$p_4$}
\put(275,137){$p_3$}
\put(291,120){\line(-1,1){12}}
\put(291,120){\line(1,1){12}}
\put(291,114){\circle{12}}
\put(288,111){$\add$}

\put(354,137){$p_6$}
\put(327,137){$p_5$}
\put(343,120){\line(-1,1){12}}
\put(343,120){\line(1,1){12}}
\put(343,114){\circle{12}}
\put(340,111){$\add$}

\put(239,82){\line(0,1){26}}
\put(239,82){\line(2,1){51}}
\put(239,76){\circle{12}}
\put(236,73){$\mlc$}

\put(343,82){\line(0,1){26}}
\put(343,82){\line(-2,1){51}}
\put(343,76){\circle{12}}
\put(340,73){$\mlc$}

\put(291,44){\line(-2,1){51}}
\put(291,44){\line(2,1){51}}
\put(291,38){\circle{12}}
\put(288,35){$\mld$}

\put(69,10){{\bf Figure 7:} A  cirquent with clustered selectional gates}
\end{picture}\end{center}

\noindent It may be helpful for one's intuition to think of each cluster as a single gate-style physical device rather than a collection of individual gates.  Namely, a cluster consisting of $n$ gates of outdegree $m$, as a single device, would have $n$ outputs  
and $m$ \ $n$-tuples of inputs.
Figure 8 depicts this new kind of a ``gate'' for the case when $n=3$, $m=2$ and the type of the cluster is $\add$. 

\begin{center} \begin{picture}(200,125)

\put(30,101){\vector(0,-1){10}}
\put(50,101){\vector(0,-1){10}}
\put(70,101){\vector(0,-1){10}}

\put(130,101){\vector(0,-1){10}}
\put(150,101){\vector(0,-1){10}}
\put(170,101){\vector(0,-1){10}}

\put(20,55){\line(0,1){30}}
\put(80,75){\line(0,1){10}}
\put(120,75){\line(0,1){10}}
\put(180,55){\line(0,1){30}}
\put(20,85){\line(1,0){60}}
\put(120,85){\line(1,0){60}}
\put(80,75){\line(1,0){40}}
\put(20,55){\line(1,0){50}}
\put(130,55){\line(1,0){50}}
\put(70,55){\line(1,0){60}}

\put(97,78){$\add$}

\put(76,59){\small $a_0$}
\put(96,59){\small $b_0$}
\put(116,59){\small $c_0$}

\put(27,77){\small $a_1$}
\put(47,77){\small $b_1$}
\put(67,77){\small $c_1$}

\put(127,77){\small $a_2$}
\put(147,77){\small $b_2$}
\put(167,77){\small $c_2$}

\put(37,106){\em inputs}
\put(137,106){\em inputs}

\put(80,49){\vector(0,-1){10}}
\put(100,49){\vector(0,-1){10}}
\put(120,49){\vector(0,-1){10}}
\put(84,30){\em outputs}

\put(80,53){\circle*{5}}
\put(100,53){\circle*{5}}
\put(120,53){\circle*{5}}

\put(30,87){\circle*{5}}
\put(50,87){\circle*{5}}
\put(70,87){\circle*{5}}

\put(130,87){\circle*{5}}
\put(150,87){\circle*{5}}
\put(170,87){\circle*{5}}

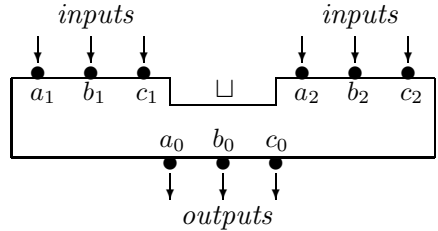
\put(11,10){{\bf Figure 8:} Clusters as generalized gates}
\end{picture}\end{center}

\noindent The device shown in Figure 8 should be thought of as a switch that can be set to one of the positions $1$ or $2$ (otherwise no signals will pass through it). Setting it to $1$ simultaneously connects the three input lines $a_1$, $b_1$ and $c_1$ to the output lines $a_0$, $b_0$ and $c_0$, respectively (the three lines are parallel, isolated from each other, so that no signal can jump from one line to another).  Similarly, setting the switch to $2$ connects $a_2$, $b_2$ and $c_2$ to $a_0$, $b_0$ and $c_0$, respectively. This is thus an ``{\em either $(a_1,b_1,c_1)$ or $(a_2,b_2,c_2)$}'' kind of a switch; combinations such as, say, $(a_1,b_2,c_1)$, are not available. 

Figure 9 shows the cirquent of Figure 7 with its clusters re-drawn in the style of Figure 8.

\begin{center} \begin{picture}(175,173)

\put(14,147){\small $p_1$}
\put(29,147){\small $p_5$}
\put(59,147){\small $p_2$}
\put(74,147){\small $p_6$}
\put(126,147){\small $p_3$}
\put(156,147){\small $p_4$}

\put(17,131){\line(0,1){10}}
\put(32,131){\line(0,1){10}}

\put(62,131){\line(0,1){10}}
\put(77,131){\line(0,1){10}}

\put(129,131){\line(0,1){10}}
\put(159,131){\line(0,1){10}}

\put(39,120){\line(1,0){15}}
\put(9,115){\line(1,0){75}}

\put(43,123){$\add$}

\put(39,114){\circle*{3}}
\put(54,114){\circle*{3}}
\put(54,114){\line(5,-1){99}}

\put(9,115){\line(0,1){15}}
\put(39,120){\line(0,1){10}}
\put(9,130){\line(1,0){30}}
\put(17,131){\circle*{3}}
\put(32,131){\circle*{3}}

\put(54,120){\line(0,1){10}}
\put(84,115){\line(0,1){15}}
\put(54,130){\line(1,0){30}}
\put(62,131){\circle*{3}}
\put(77,131){\circle*{3}}

\put(153,87){\line(0,1){7}}

\put(136,120){\line(1,0){15}}
\put(121,115){\line(1,0){45}}

\put(140,123){$\add$}

\put(143,114){\circle*{3}}

\put(121,115){\line(0,1){15}}
\put(136,120){\line(0,1){10}}
\put(121,130){\line(1,0){15}}
\put(129,131){\circle*{3}}

\put(166,115){\line(0,1){15}}
\put(151,120){\line(0,1){10}}
\put(151,130){\line(1,0){15}}
\put(159,131){\circle*{3}}

\put(39,82){\line(0,1){32}}
\put(114,107){\line(4,1){30}}
\put(39,82){\line(3,1){75}}
\put(39,76){\circle{12}}
\put(36,73){$\mlc$}

\put(143,82){\line(0,1){32}}
\put(143,82){\line(2,1){10}}
\put(143,76){\circle{12}}
\put(140,73){$\mlc$}

\put(91,44){\line(-2,1){51}}
\put(91,44){\line(2,1){51}}
\put(91,38){\circle{12}}
\put(88,35){$\mld$}

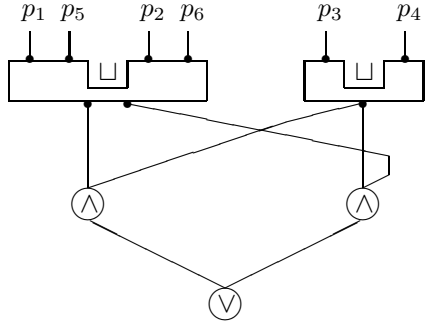
\put(-61,10){{\bf Figure 9:} An alternative representation of the cirquent of Figure 7}
\end{picture}\end{center}

\noindent Representing clusters in the way we have just done illustrates that they are nothing but generalized gates. In fact, this is a very natural generalization. Namely, a gate in the ordinary sense is the special case of this general sort of a gate where the number of output lines (as well as input lines in each group of inputs) equals $1$. Technically, however, we prefer to continue seeing clusters as sets of ordinary gates as they were officially defined at the beginning of this section. So, drawings in the style of Figures 8 or 9 will never reemerge in this paper, and whatever was said about clusters as individual ``gates'' of a new type can be safely forgotten.

Cirquents of the previous section should be viewed as special cases of cirquents in the new sense of this section. Namely, they are the cases where each selectional gate forms its own, single-element cluster. With this view, the semantical concepts that we reintroduce in this section conservatively generalize those of the previous section.  

The definition of a legal run of the game represented by a cirquent $C$ is the same as before (Definition \ref{may14a}), with the difference that now moves are made within clusters rather than individual gates. That is, each move of a legal run of $C$ looks like $c.i$, where $c$ is the ID of a cluster (rather than of a gate), and $i$ is a positive integer not exceeding the outdegree of that cluster. The intuitive meaning of such a move $c.i$ is selecting the $i$th outgoing edge (together with the corresponding child)  simultaneously in {\em each} gate belonging to cluster $c$.
All other conditions on the legality of runs remain literally the same as before. 
Anyway, let us not be lazy to fully (re)produce such a definition:

\begin{defi}\label{maya}
Let $C$ be a cirquent, $^*$ an interpretation, and $\Phi$ a position. $\Phi$ is a {\bf legal position} of the game $C^*$ iff, with a ``cluster'' below meaning a cluster of $C$,  the following conditions are satisfied:
\begin{enumerate}[(1)]
\item Each labmove of $\Phi$ has one of the following forms:
\begin{enumerate} 
\item $\pp c.i$, where $c$ is a $\tgd$\hspace{-2pt}-,\hspace{-2pt} $\sqd$\hspace{-2pt}-\hspace{-2pt} or $\add$-cluster  and $i$ is a positive integer not exceeding 
the outdegree of $c$; 
\item $\oo c.i$, where $c$ is a $\tgc$\hspace{-2pt}-,\hspace{-2pt} $\sqc$\hspace{-2pt}-\hspace{-2pt} or $\adc$-cluster  and $i$ is a positive integer not exceeding 
the outdegree of $c$. 
\end{enumerate} 
\item Whenever $c$ is a choice cluster, $\Phi$ contains at most one occurrence of a labmove of the form $\xx c.i$.   
\item Whenever $c$ is a sequential cluster and $\Phi$ is of the form $\seq{\ldots , \xx c.i ,\ldots, \xx c.j ,\ldots}$, we have $i<j$. 
\end{enumerate}
\end{defi}\medskip

\noindent So, for instance, $\seq{\pp 8.1,\pp 7.2}$ is a legal run of the cirquent of Figure 7, but $\seq{\pp 8.1,\pp 9.2}$ is not, because $9$ is (a gate ID but) not a cluster ID. To summarize once again, selections (moves) in individual {\em gates} are no longer available. Rather, they should be made in {\em clusters}.

\begin{defi}\label{m20}
In the context of a given cirquent $C$ and a  legal run $\Gamma$ of $C$, we will say that a selectional gate $g$ of a cluster $c$ is {\bf unresolved} iff either no moves of the form $c.j$ have been made in $\Gamma$, or infinitely many such moves  have been made. Otherwise $g$ is {\bf resolved} and, where $c.i$ is the last move of the form $c.j$ made in $\Gamma$,  the child pointed at by the $i$th outgoing edge of $g$ is said to be the {\bf resolvent} of $g$.
\end{defi} 

With the terms ``unresolved'', ``resolved'' and ``resolvent'' conservatively redefined this way, the definition of {\bf truth} for a cirquent and its nodes 
is 
{\em literally}  the same\footnote{Unlike Definition \ref{maya} which, at least, changed the word ``gate'' to the word ``cluster'' when reproducing the corresponding Definition \ref{may14a}.} in our present case as in the case of the cirquents of the previous section (Definition \ref{may14c}), so we do not reproduce it here. The same applies to the definition of the {\bf Wn} components of the games represented by cirquents (Definition \ref{may14aa}).

Let us look at the game represented by the cirquents of Figure 7 once again. The meaning of the move ``$7.2$'' in this game is   selecting outgoing edge $\#2$ in {\em both} gates of cluster $7$. Intuitively, the effect of such a move is connecting gate $10$ directly to node $2$ and gate $11$ directly to node $6$. Thus, the move ``$7.2$'' can be seen as a choice (choice $\#2$) {\em shared} between gates $10$ and $11$; that shared choice, however, yields different, unshared results for the two gates: result $2$ for gate $10$ while result $6$ for gate $11$. This sort of sharing is very different from the sort of sharing represented by gate $8$: the effect of the move ``$8.1$'' is sharing both the choice $\#1$ as well as the result $3$ of that choice. 

Back to the  world of Victor and Peggy, imagine they are in their family car on a road between two cities $A$ and $B$. Victor likes sports but never goes to theaters. Peggy likes theaters but never attends games. There is a basketball game and a ballet show  tonight in city $A$. And there is  a football game and an opera show in city $B$. The shared choice/move in this situation is a choice between ``{\em drive to $A$}'' and ``{\em drive to $B$}'' (they only have  one car!). The outcomes of either choice, however, are not shared. For instance, the outcome of the choice ``{\em drive to $A$}'' is ``{\em see the basketball match}'' for Victor while ``{\em see the ballet}'' for Peggy. Victor and Peggy can negotiate and decide between the two pairs  $(\mbox{\em Basketball},\mbox{\em Ballet})$ or  $(\mbox{\em Football},\mbox{\em Opera})$. But the pairs $(\mbox{\em Basketball},\mbox{\em Opera})$ and 
$(\mbox{\em Football},\mbox{\em Ballet})$ are not available. 

As for the stronger type of sharing corresponding to the middle $\add$-gate of Figure 7, it can be explained by the following modification of the situation: Both Victor and Peggy are fond of Impressionism. There is a Monet exhibition in city $A$, and a Pissarro  exhibition in city $B$. The action/choice --- driving to $A$ or driving to $B$ --- is again shared. But now so is the corresponding outcome ``{\em see Monet's paintings}'' or ``{\em see Pissarro's paintings}''. 

While we could continue elaborating on independent philosophical and mathematical/technical motivations for introducing clustering, here we just want to point out  
the very direct connections of our approach to the already well motivated {\em IF} ({\em independence-friendly}) {\em logic}. We assume that the reader is familiar with the basics of the latter, or else he or she may take a quick look at the concise yet complete (for our purposes) overview of the subject given in \cite{Tul09}. It should be noted that we use the term ``IF logic'' in a  generic sense, making no terminological distinction between Hintikka and Sandu's \cite{HS97} original version of IF logic and Hodges's \cite{Hodges97} generalization of it termed in \cite{Hodges07}  {\em slash logic}. In fact, when talking about the (``traditional'') syntax or semantics of IF logic, what we have in mind are those of slash logic. 
It is assumed that all formulas of IF logic are  written in negation normal form, and that different occurrences of quantifiers in them always bind different variables. 
As pointed out in Section 3.3 of \cite{Tul09}, ``in the literature the interest almost always pertains to the truth of a sentence'' (as opposed to falsity which, in IF logic, is not the same as ``not true''). 
And it is known that slashing universal quantifiers or conjunctions has no effect on the truth (as opposed to falsity) status of formulas. Hence, in this section, we further assume that only existential quantifiers and disjunctions are slashed in the formulas of IF logic.  Finally, without much (if any) loss of generality, we assume that the semantics of IF logic exclusively deals with models with countable domains; namely, such a domain is always $\{1,2,3,\ldots\}$ or some nonempty finite initial portion of it. 

Computability logic insists on algorithmicity of $\pp$'s strategies, while IF logic, in its game semantics, allows any strategies. Let us, for now, consider the version of IF logic which differs from its canonical version only in that it, like CoL, requires $\pp$'s (which there is usually  called $\cle$-Player, or Verifier, or Eloise) strategies to be effective, while imposing no restrictions on $\oo$'s ($\cla$-Player's, Falsifier's, Abelard's) strategies. Call this version of IF logic {\bf effective IF logic}. 

Remember that the outgoing edges of each node of a cirquent come in a fixed left-to-right order: edge $\#1$, edge $\#2$, edge $\#3$, etc. Let us call these numbers $1$, $2$, $3$, etc. the {\bf order numbers} of the corresponding edges. 
 
We now claim that the fragment of our cirquent logic where cirquents are allowed to have only two --- $\mlc$ and $\add$ --- sorts of gates is sufficient to cover effective IF logic, and far beyond. A verification of this claim --- perhaps at the philosophical rather than technical level --- is left to the reader. 

 Namely, each formula $E$ of effective IF logic can be understood as the cirquent obtained from it through performing  the following operations:

\begin{desc}\label{dsc1} \ 
\begin{enumerate}[(1)]
\item Ignoring slashes, write $E$ in the form of a tree-like cirquent, understanding $\cla$ as a ``long''  $\mlc$-conjunction and $\cle$ as a ``long''  $\mld$-disjunction. This way, each occurrence $O$ of a quantifier, conjunction or disjunction  gives rise to one (if $O$ is not in the scope of a quantifier) or many (if $O$ is in the scope of a quantifier)  gates. We say that each such gate, as well as each outgoing edge of it, {\bf originates} from $O$. Also, since we assume that different occurrences of quantifiers in IF formulas always bind different variables, instead of saying ``\ldots originates from the occurrence of $\cla y$ (or $\cle y$)'' we can unambiguously simply say ``\ldots originates from $y$''. 
\item Change the label of each $\mld$-gate to $\add$. 
\item Cluster the disjunctive gates so that, any two such gates $a$ and $b$ belong to the same cluster if and only if they originate from the same occurrence of $\cle x/y_1,\ldots,y_n$ or $\mld/y_1,\ldots,y_n$ in $E$, and the following condition is satisfied:
\begin{enumerate}[$\bullet$] 
\item Let $e_{1}^{a},\ldots,e_{k}^{a}$ and $e_{1}^{b},\ldots,e_{k}^{b}$ be the paths (sequences of edges) from the root of the tree to $a$ and $b$, respectively. Then, for any $i\in\{1,\ldots,k\}$, the order numbers of the edges $e_{i}^{a}$ and $e_{i}^{b}$ are identical unless these edges originate from one of the variables $y_1,\ldots,y_n$.   
\end{enumerate}
\end{enumerate}
\end{desc}\medskip

\noindent Let us see an example to visualize how our construction works.  
A traditional  starting point of introductory or survey papers on IF logic is the formula 
\begin{equation}\label{may21}
\cla x\cle y\cla z\cle t/x,\hspace{-2pt}y \ p(x,y,z,t).
\end{equation} Technically, its meaning can be expressed by the
second-order formula 
\[\cle f\cle g\cla x\cla z\ p\bigl(x,f(x),z,g(z)\bigr)\footnote{As
  long as we deal with effective IF logic, one should require here
  that $f$ and $g$ range over recursive functions.}
\]
  or the following formula with Henkin's branching quantifiers:
\[\begin{array}{r}
\cla x\cle y\\
\cla z\cle t
\end{array} \raisebox{2pt}{$p(x,y,z,t)$},\]
with the shape of the quantifier array indicating that the two quantifier blocks $\cla x\cle y$ and $\cla z\cle t$ are independent of each other even though, in (\ref{may21}), one occurs in the scope of the other.  We agreed earlier to write $\mla\hspace{-1pt},\mle$ instead of $\cla,\cle$. So,  we rewrite (\ref{may21})  as $\mla \hspace{-1pt}x\hspace{-1pt}\mle\hspace{-2pt} y\hspace{-1pt}\mla\hspace{-1pt} z\hspace{-1pt}\mle\hspace{-1pt} t/x,\hspace{-2pt}y \hspace{3pt} p(x,y,z,t)$. This formula, however, is not yet adequate. The philosophy of IF logic associates the intuitions of {\em finding} (rather than just {\em existence}) with existential quantifiers or disjunctions; and, 
as we know, it is the operator/gate $\add$ whose precise meaning is actually {\em finding} things (rather than $\mld$, which is  merely about {\em existence} of things). So, $\mle$ should be replaced with $\ade$,  and we get 
 the formula  $\mla \hspace{-1pt}x\ade y\mla\hspace{-1pt} z\ade t/x,\hspace{-2pt}y \hspace{2pt} p(x,y,z,t)$ which, ignoring the slash for now, can be seen as a tree-like cirquent in the sense of the previous section. It now remains to account for the slash by adequately clustering the cirquent. Namely, the clustering should make sure that, whenever $a$ and $b$ are two upper-level (those originating from $t$) $\add$-gates such that the paths from the root to them --- seen not as sequences of edges but as sequences of the corresponding order numbers ---  only differ in their first two elements (the ones originating from $x$ and $y$, of which $t$ should be independent), $a$ and $b$ are in the same cluster. Figure 10 illustrates this arrangement. For compactness considerations, in this figure we have assumed that the universe of discourse (the set over which $x,y,z,t$ range) is just $\{1,2\}$; also, we have written $p_{1111}$, $p_{1112}$, etc. instead of $p(1,1,1,1)$, $p(1,1,1,2)$, etc.
 
\begin{center} \begin{picture}(360,165)

\put(3,128){\scriptsize $3$}
\put(53,128){\scriptsize $4$}
\put(103,128){\scriptsize $3$}
\put(153,128){\scriptsize $4$}
\put(203,128){\scriptsize $3$}
\put(253,128){\scriptsize $4$}
\put(303,128){\scriptsize $3$}
\put(353,128){\scriptsize $4$}
\put(78,76){\scriptsize $1$}
\put(278,76){\scriptsize $2$}

\put(-17,142){\scriptsize $p_{1111}$}
\put(8,142){\scriptsize $p_{1112}$}
\put(5,122){\line(-2,3){11}}
\put(5,122){\line(2,3){11}}
\put(5,116){\circle{12}}
\put(2,113){$\add$}
\put(33,142){\scriptsize $p_{1121}$}
\put(58,142){\scriptsize $p_{1122}$}
\put(55,122){\line(-2,3){11}}
\put(55,122){\line(2,3){11}}
\put(55,116){\circle{12}}
\put(52,113){$\add$}
\put(30,97){\line(-2,1){25}}
\put(30,97){\line(2,1){25}}
\put(30,91){\circle{12}}
\put(27,88){$\mlc$}
\put(83,142){\scriptsize $p_{1211}$}
\put(108,142){\scriptsize $p_{1212}$}
\put(105,122){\line(-2,3){11}}
\put(105,122){\line(2,3){11}}
\put(105,116){\circle{12}}
\put(102,113){$\add$}
\put(133,142){\scriptsize $p_{1221}$}
\put(158,142){\scriptsize $p_{1222}$}
\put(155,122){\line(-2,3){11}}
\put(155,122){\line(2,3){11}}
\put(155,116){\circle{12}}
\put(152,113){$\add$}
\put(130,97){\line(-2,1){25}}
\put(130,97){\line(2,1){25}}
\put(130,91){\circle{12}}
\put(127,88){$\mlc$}
\put(80,72){\line(-4,1){51}}
\put(80,72){\line(4,1){51}}
\put(80,66){\circle{12}}
\put(77,63){$\add$}

\put(184,142){\scriptsize $p_{2111}$}
\put(208,142){\scriptsize $p_{2112}$}
\put(205,122){\line(-2,3){11}}
\put(205,122){\line(2,3){11}}
\put(205,116){\circle{12}}
\put(202,113){$\add$}
\put(233,142){\scriptsize $p_{2121}$}
\put(258,142){\scriptsize $p_{2122}$}
\put(255,122){\line(-2,3){11}}
\put(255,122){\line(2,3){11}}
\put(255,116){\circle{12}}
\put(252,113){$\add$}
\put(230,97){\line(-2,1){25}}
\put(230,97){\line(2,1){25}}
\put(230,91){\circle{12}}
\put(227,88){$\mlc$}
\put(283,142){\scriptsize $p_{2211}$}
\put(308,142){\scriptsize $p_{2212}$}
\put(305,122){\line(-2,3){11}}
\put(305,122){\line(2,3){11}}
\put(305,116){\circle{12}}
\put(302,113){$\add$}
\put(333,142){\scriptsize $p_{2221}$}
\put(358,142){\scriptsize $p_{2222}$}
\put(355,122){\line(-2,3){11}}
\put(355,122){\line(2,3){11}}
\put(355,116){\circle{12}}
\put(352,113){$\add$}
\put(330,97){\line(-2,1){25}}
\put(330,97){\line(2,1){25}}
\put(330,91){\circle{12}}
\put(327,88){$\mlc$}
\put(280,72){\line(-4,1){51}}
\put(280,72){\line(4,1){51}}
\put(280,66){\circle{12}}
\put(277,63){$\add$}

\put(181,43){\line(-6,1){100}}
\put(181,43){\line(6,1){100}}
\put(181,37){\circle{12}}
\put(178,34){$\mlc$}

\put(5,10){{\bf Figure 10:} Mimicking $\cla x\cle y\cla z\cle t/x,\hspace{-2pt}y \ p(x,y,z,t)$ (when the universe is $\{1,2\}$)}
\end{picture}\end{center}

\noindent To feel the difference created by clustering, let us consider the interpretation that sends the four atoms $p_{1111}$, $p_{1122}$, $p_{2212}$, $p_{2221}$ to $\twg$  and sends all other atoms to $\tlg$. Then the game represented by the cirquent of Figure 10 cannot be won (by $\pp$). On the other hand, it would be winnable if there was no clustering. Further, it is also winnable if clustering is made finer  (yet not trivial) as done in the cirquent of  Figure 11. The latter expresses the ``slightly'' modified form  $\cla x\cle y\cla z\cle t/x \hspace{3pt} p(x,y,z,t)$ of (\ref{may21}), and is won by an  HPM that generates the run \[\seq{\pp 1.1,\ \pp 3.1,\ \pp 4.2,\ \pp 2.2,\ \pp 5.2,\ \pp 6.1}\] (or any permutation of the above). Note that the same run is simply not a legal run of the cirquent of Figure 10.

\begin{center} \begin{picture}(360,165)

\put(3,128){\scriptsize $3$}
\put(53,128){\scriptsize $4$}
\put(103,128){\scriptsize $5$}
\put(153,128){\scriptsize $6$}
\put(203,128){\scriptsize $3$}
\put(253,128){\scriptsize $4$}
\put(303,128){\scriptsize $5$}
\put(353,128){\scriptsize $6$}

\put(78,76){\scriptsize $1$}
\put(278,76){\scriptsize $2$}

\put(-17,142){\scriptsize $p_{1111}$}
\put(8,142){\scriptsize $p_{1112}$}
\put(5,122){\line(-2,3){11}}
\put(5,122){\line(2,3){11}}
\put(5,116){\circle{12}}
\put(2,113){$\add$}
\put(33,142){\scriptsize $p_{1121}$}
\put(58,142){\scriptsize $p_{1122}$}
\put(55,122){\line(-2,3){11}}
\put(55,122){\line(2,3){11}}
\put(55,116){\circle{12}}
\put(52,113){$\add$}
\put(30,97){\line(-2,1){25}}
\put(30,97){\line(2,1){25}}
\put(30,91){\circle{12}}
\put(27,88){$\mlc$}
\put(83,142){\scriptsize $p_{1211}$}
\put(108,142){\scriptsize $p_{1212}$}
\put(105,122){\line(-2,3){11}}
\put(105,122){\line(2,3){11}}
\put(105,116){\circle{12}}
\put(102,113){$\add$}
\put(133,142){\scriptsize $p_{1221}$}
\put(158,142){\scriptsize $p_{1222}$}
\put(155,122){\line(-2,3){11}}
\put(155,122){\line(2,3){11}}
\put(155,116){\circle{12}}
\put(152,113){$\add$}
\put(130,97){\line(-2,1){25}}
\put(130,97){\line(2,1){25}}
\put(130,91){\circle{12}}
\put(127,88){$\mlc$}
\put(80,72){\line(-4,1){51}}
\put(80,72){\line(4,1){51}}
\put(80,66){\circle{12}}
\put(77,63){$\add$}

\put(184,142){\scriptsize $p_{2111}$}
\put(208,142){\scriptsize $p_{2112}$}
\put(205,122){\line(-2,3){11}}
\put(205,122){\line(2,3){11}}
\put(205,116){\circle{12}}
\put(202,113){$\add$}
\put(233,142){\scriptsize $p_{2121}$}
\put(258,142){\scriptsize $p_{2122}$}
\put(255,122){\line(-2,3){11}}
\put(255,122){\line(2,3){11}}
\put(255,116){\circle{12}}
\put(252,113){$\add$}
\put(230,97){\line(-2,1){25}}
\put(230,97){\line(2,1){25}}
\put(230,91){\circle{12}}
\put(227,88){$\mlc$}
\put(283,142){\scriptsize $p_{2211}$}
\put(308,142){\scriptsize $p_{2212}$}
\put(305,122){\line(-2,3){11}}
\put(305,122){\line(2,3){11}}
\put(305,116){\circle{12}}
\put(302,113){$\add$}
\put(333,142){\scriptsize $p_{2221}$}
\put(358,142){\scriptsize $p_{2222}$}
\put(355,122){\line(-2,3){11}}
\put(355,122){\line(2,3){11}}
\put(355,116){\circle{12}}
\put(352,113){$\add$}
\put(330,97){\line(-2,1){25}}
\put(330,97){\line(2,1){25}}
\put(330,91){\circle{12}}
\put(327,88){$\mlc$}
\put(280,72){\line(-4,1){51}}
\put(280,72){\line(4,1){51}}
\put(280,66){\circle{12}}
\put(277,63){$\add$}

\put(181,43){\line(-6,1){100}}
\put(181,43){\line(6,1){100}}
\put(181,37){\circle{12}}
\put(178,34){$\mlc$}

\put(5,10){{\bf Figure 11:} Mimicking $\cla x\cle y\cla z\cle t/x \hspace{3pt} p(x,y,z,t)$ (when the universe is $\{1,2\}$)}
\end{picture}\end{center}

\noindent While  $(\mlc,\add)$-cirquents allow us to fully capture effective IF logic, they, at the same time,  are significantly more expressive than the latter is, even if --- or, especially if --- we limit ourselves to finite cirquents, and correspondingly limit IF formulas to propositional ones. As mentioned earlier, IF logic  has no smooth approach at the purely propositional level, and is forced to severely limit the forms of (meaningful) propositional-level formulas as, for instance, done in \cite{SandPiet01}. In particular, it encounters serious difficulties in 
allowing independence from conjunctions or disjunctions (rather than quantifiers). These problems are automatically neutralized under our approach. We do not  introduce any restrictions whatsoever on the forms of cirquents or allowable clusterings in them; yet, all such expressions are semantically meaningful. 

Of course, a penalty for the higher expressive power of cirquents is the awkwardness 
associated with the necessity to draw graphs instead of writing formulas. But, again, various syntactic shortcuts can be introduced to make life easier, with recurrence operators, quantifiers or the slash notation being among such shortcuts. It should also be remembered that drawing cirquents may be some annoyance for humans (when writing papers) but not for computers (when using logic in their work); the latter, in fact, would {\em much} prefer cirquents, as they are exponentially more efficient means of expression than formulas are. In any case,  a more significant achievement than expressiveness is probably avoiding the necessity to deal with the unplayable and troublemaking imperfect-information games on which the traditional semantics for IF logic are based. We owe this effect to the fact that clustering enforces at the {\em game} level what IF logic calls {\em uniformity} and tries to enforce at the {\em strategy} level. Rather than relying on players' integrity in expecting that they --- to their disadvantage --- will conscientiously forget  the moves they have already seen but of which their actions should be independent, clustering simply makes ``cheating'' physically impossible. Perhaps this point is important enough to be repeated in an empasised form:
\begin{quote} {\em  While IF logic traditionally accounts for independence on the level of strategies (by imposing uniformity conditions on them), our approach does so directly on the level of game rules: independence is enforced through making ``dependent'' moves simply impossible to make --- such moves are ``physically'' unavailable.  } 
\end{quote} 

\noindent But it should be remembered that ``effective IF logic'',
while both mathematically and philosophically reasonable, is not at all the same as IF logic in its canonical form.\footnote{The idea of what we here call ``effective IF logic'' can be clearly found in  \cite{Hintikka96}, where Hintikka argued that restricting $\exists$'s ($\top$'s) strategies to effective ones could be a formulation of constructivism in the philosophy of mathematics. To the present author's best knowledge, however, no subsequent technical attempts have been made to correspondingly reconstruct the semantics of IF logic. The general idea of basing game semantics on effective strategies and this way realizing the philosophy of constructivism can be found in the even earlier work \cite{Jap93} by the present author (this paper was later refined and published in the form of \cite{Jap97}). \cite{Jap93,Jap97} also made the first, ``experimental'' steps towards technically realizing this line of thought. As for CoL, it takes pride not in putting forward the idea of considering effective strategies (which is so natural that it could hardly have escaped the thoughts of anyone working on game semantics), but in making it  actually work and generate non-trivial logics.
In contrast, restricting strategies to effective ones within Hintikka's framework leaves us with an essentially empty logic:  there are no valid (as opposed to true) formulas there other than trivialities such as $\top$ or $p(x)\rightarrow \top$ (but not $p(x)\rightarrow p(x)$: this principle, i.e. $\neg p(x)\vee p(x)$, fails when $p(x)$ is undecidable).  This is probably one of the reasons why the IF-logic community has remained focused on not-necessarily-effective strategies.}
So, a true merger between CoL and IF logic would seemingly require  a compromise from one of them: either IF logic should adopt the requirement of effectiveness of strategies, or computability logic should drop this (central to its philosophy) requirement. Probably neither camp  would be willing to make a compromise.

Fortunately, there is no real need for any compromises. The following section further generalizes the concept of cirquents in a conservative way. The new cirquents,  unlike the cirquents of the present section, are powerful enough to express anything that the traditional (``non-effective'') IF logic can. This is achieved through extending the idea of clustering from selectional gates to $\mld$-gates, yet without associating any moves with such gates or clusters.
 
\section{Clustering \texorpdfstring{$\mld$}{V}-gates}\label{s6}

\noindent A {\bf cirquent} in the sense of the present section is the same as one in the sense of the previous section, with the difference that now not only the set of 
selectional gates is partitioned into clusters, but also the set of $\mld$-gates (but not the set of $\mlc$-gates --- not yet, at least). The condition on clustering is the same as before: all gates within a given cluster are required to have the same type (label) and same outdegree. Cirquents in the sense of the previous section are seen as special cases of cirquents in the present sense, namely, the cases where all $\mld$-clusters are singletons.

The {\bf legal positions} of the game represented by a cirquent in this new sense are defined in literally the same way as before (Definition \ref{maya}). So, clustering $\mld$-gates  in a cirquent does not affect the set of its legal runs. 

By a {\bf metaselection} for a cirquent $C$ we will mean a (not necessarily effective) partial function $f$ from $\mld$-clusters of $C$ to the set of positive integers, such that, for any $\mld$-cluster $c$, whenever defined, $f(c)$ does not exceed the outdegree of $c$.\footnote{An equivalent approach would be letting $f$ be a {\em total} function from the set of $\mld$-clusters to the set $\{0,1,2,3,\ldots\}$ of {\em natural numbers} (rather than positive integers); then, instead of saying that $f$ is undefined at $c$, we could simply say that $f(c)=0$.}

In the context of a given cirquent $C$, a  legal run $\Gamma$ of $C$ and a metaselection $f$ for $C$, we will say that a $\mld$-gate $g$ of a cluster $c$ is {\bf unresolved} iff $f$  is undefined at $c$ (note that a childless $\mld$-gate will always be unresolved).   Otherwise $g$ is {\bf resolved}, and the child of it pointed at by the $f(c)$th  outgoing edge  is said to be the {\bf resolvent} of $g$. As for the  same-name concepts ``unresolved'', ``resolved'' and ``resolvent'' for selectional gates, they are defined literally as before (Definition \ref{m20}). Note that these three concepts depend on $f$ but not $\Gamma$ when $g$ is a $\mld$-gate, while they depend on $\Gamma$ but not $f$ when $g$ is a selectional gate. The function $f$ thus acts as a ``metaextension'' of $\Gamma$. Intuitively, it can be thought of as selections in $\mld$-clusters  made by the guardian angel of $\pp$  in favor of $\pp$ {\em after} the actual play took place (rather than {\em during} it), even if the latter lasted infinitely long; unlike $\pp$, its guardian angel has magic --- nonalgorithmic --- intellectual powers to make best possible selections. Technically, however, selections by the ``angel'', unlike selections made by either player, are not  moves of the game.

The following definition refines the earlier definitions of truth by relativizing this concept --- renamed into {\em metatruth} --- with respect to a metaselection as an additional parameter.
 
\begin{defi}\label{may14cc}
Let $C$ be a cirquent, $^*$ an interpretation,   $\Gamma$ a legal run of $C$, and $f$ a metaselection for $C$. In this context, with ``metatrue'' to be read as ``{\bf metatrue w.r.t. $(^*,\Gamma,f)$}'', we say that: 
\begin{enumerate}[$\bullet$]
\item An $L$-port is metatrue  iff $L^*=\twg$. 
\item A resolved selectional gate is metatrue iff so is its resolvent.
\item No unresolved disjunctive selectional gate is metatrue.
\item Each unresolved conjunctive selectional gate is metatrue.
\item A $\mld$-gate is metatrue iff it is resolved and its resolvent is metatrue.
\item A $\mlc$-gate is metatrue iff so are all of its children. 
\end{enumerate}
Finally, we say that $C$ is metatrue  iff so is its root.
\end{defi}

The following definition brings us from metatruth back to truth.

\begin{defi}\label{ma14cc}
Let $C$ be a cirquent, $^*$ an interpretation, and $\Gamma$ a legal run of $C$. We say that  
$C$ is {\bf true} w.r.t. $(^*,\Gamma)$ iff there is a metaselection $f$ for $C$ such that $C$ is metatrue w.r.t. 
$(^*,\Gamma,f)$. 
\end{defi}

It is left to the reader to see why the new concept of truth is a conservative generalization of its earlier counterparts. The same  applies to the following definition, which completes our definition of the game $C^*$ represented by any given cirquent $C$  under any given interpretation $^*$.  

\begin{defi}\label{may14aaa}
Let $C$ be a cirquent, $^*$ an interpretation, and $\Gamma$ a legal run of $C$. Then $\Gamma$ is a $\pp$-won run of the game $C^*$ iff $C$ is true w.r.t. $(^*,\Gamma)$. 
\end{defi}

We now claim that any formula $E$ of (this time ordinary, ``non-effective'') IF logic can be adequately written as a tree-like cirquent $C$ with only $\mld$ and $\mlc$ gates; namely, such a cirquent $C$ is obtained from $E$ through applying the steps 1 and 3 of Description \ref{dsc1}, with step 2 omitted. For any interpretation $^*$, we will then have $C^*=\twg$ iff $E$ is true (under the same interpretation of atoms) in IF logic. Figure 12 shows an example. As an easy 
exercise, the reader may want to verify that the cirquent of that figure is $\oo$ under the interpretation which sends $p_{1111}$, $p_{1122}$, $p_{2212}$, $p_{2221}$ to $\twg$ 
and sends all other atoms to $\tlg$. Note that this cirquent represents an elementary game, unlike its counterpart from Figure 10.  

\begin{center} \begin{picture}(360,165)

\put(3,128){\scriptsize $3$}
\put(53,128){\scriptsize $4$}
\put(103,128){\scriptsize $3$}
\put(153,128){\scriptsize $4$}
\put(203,128){\scriptsize $3$}
\put(253,128){\scriptsize $4$}
\put(303,128){\scriptsize $3$}
\put(353,128){\scriptsize $4$}
\put(78,76){\scriptsize $1$}
\put(278,76){\scriptsize $2$}

\put(-17,142){\scriptsize $p_{1111}$}
\put(8,142){\scriptsize $p_{1112}$}
\put(5,122){\line(-2,3){11}}
\put(5,122){\line(2,3){11}}
\put(5,116){\circle{12}}
\put(2,113){$\mld$}
\put(33,142){\scriptsize $p_{1121}$}
\put(58,142){\scriptsize $p_{1122}$}
\put(55,122){\line(-2,3){11}}
\put(55,122){\line(2,3){11}}
\put(55,116){\circle{12}}
\put(52,113){$\mld$}
\put(30,97){\line(-2,1){25}}
\put(30,97){\line(2,1){25}}
\put(30,91){\circle{12}}
\put(27,88){$\mlc$}
\put(83,142){\scriptsize $p_{1211}$}
\put(108,142){\scriptsize $p_{1212}$}
\put(105,122){\line(-2,3){11}}
\put(105,122){\line(2,3){11}}
\put(105,116){\circle{12}}
\put(102,113){$\mld$}
\put(133,142){\scriptsize $p_{1221}$}
\put(158,142){\scriptsize $p_{1222}$}
\put(155,122){\line(-2,3){11}}
\put(155,122){\line(2,3){11}}
\put(155,116){\circle{12}}
\put(152,113){$\mld$}
\put(130,97){\line(-2,1){25}}
\put(130,97){\line(2,1){25}}
\put(130,91){\circle{12}}
\put(127,88){$\mlc$}
\put(80,72){\line(-4,1){51}}
\put(80,72){\line(4,1){51}}
\put(80,66){\circle{12}}
\put(77,63){$\mld$}

\put(184,142){\scriptsize $p_{2111}$}
\put(208,142){\scriptsize $p_{2112}$}
\put(205,122){\line(-2,3){11}}
\put(205,122){\line(2,3){11}}
\put(205,116){\circle{12}}
\put(202,113){$\mld$}
\put(233,142){\scriptsize $p_{2121}$}
\put(258,142){\scriptsize $p_{2122}$}
\put(255,122){\line(-2,3){11}}
\put(255,122){\line(2,3){11}}
\put(255,116){\circle{12}}
\put(252,113){$\mld$}
\put(230,97){\line(-2,1){25}}
\put(230,97){\line(2,1){25}}
\put(230,91){\circle{12}}
\put(227,88){$\mlc$}
\put(283,142){\scriptsize $p_{2211}$}
\put(308,142){\scriptsize $p_{2212}$}
\put(305,122){\line(-2,3){11}}
\put(305,122){\line(2,3){11}}
\put(305,116){\circle{12}}
\put(302,113){$\mld$}
\put(333,142){\scriptsize $p_{2221}$}
\put(358,142){\scriptsize $p_{2222}$}
\put(355,122){\line(-2,3){11}}
\put(355,122){\line(2,3){11}}
\put(355,116){\circle{12}}
\put(352,113){$\mld$}
\put(330,97){\line(-2,1){25}}
\put(330,97){\line(2,1){25}}
\put(330,91){\circle{12}}
\put(327,88){$\mlc$}
\put(280,72){\line(-4,1){51}}
\put(280,72){\line(4,1){51}}
\put(280,66){\circle{12}}
\put(277,63){$\mld$}

\put(181,43){\line(-6,1){100}}
\put(181,43){\line(6,1){100}}
\put(181,37){\circle{12}}
\put(178,34){$\mlc$}

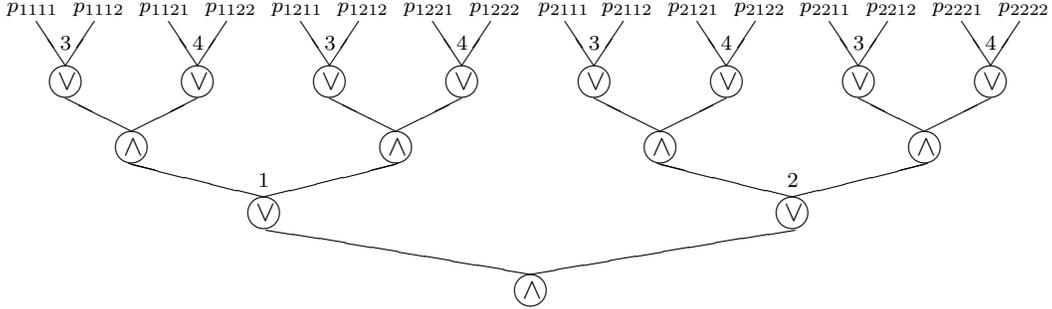
\put(0,10){{\bf Figure 12:} Representing $\cla x\cle y\cla z\cle t/x,\hspace{-2pt}y \hspace{3pt} p(x,y,z,t)$ (when the universe is $\{1,2\}$)}
\end{picture}\end{center}

\noindent Just as for any other claims made in this paper regarding connections to IF logic, we are not attempting to provide a proof of our present claim. Such a proof would require reproducing and analyzing one of the semantics accepted/recognized in the IF-logic community, which could take us too far --- the present paper is on computability   
logic rather than IF logic after all and, even if a known semantics of IF logic and the corresponding fragment of the semantics of CoL turned out to be not exactly equivalent, a question would arise about which one is a more adequate materialization of the original philosophy of IF logic. 

\section{Clustering all gates; ranking}\label{s7}

\noindent Other than the claimed fact that the cirquents of the previous section achieve the full expressive power of IF logic, there are no good reasons to stop at those cirquents. Indeed, if we clustered selectional and $\mld$-gates, why not do the same with the remaining $\mlc$ type of gates?  Naturally, the semantics of clustered $\mlc$ gates would have to be symmetric to that of clustered $\mld$-gates. Namely, a universally quantified metaselection $h$ should be associated with them, as we associated an existentially quantified metaselection $f$ with $\mld$-clusters in the previous section. Such an $h$ can be thought of as a guardian angel of $\oo$ that makes selections in $\mlc$-clusters in favor of $\oo$ after the game has been played by the two players. One can show that then, no matter how the $\mlc$-gates are clustered, truth in the sense of the previous section is equivalent to an assertion that sounds like the right side of Definition \ref{ma14cc} but starts with the words ``{\em there is an $f$ such that, for all $h$,\ldots}'' instead of just ``{\em there is an $f$ such that\ldots}''. But then the question comes: why this quantification order  and not ``{\em for all $h$ there is an $f$ such that\ldots}'', which would obviously yield a different yet meaningful concept of truth?\footnote{This predicate of truth, in contrast to the previous one, would depend on how $\mlc$-gates are clustered but not on how $\mld$-gates are clustered.}   Again, there is no good answer, and here we see the need for a yet more general approach that would be flexible enough to handle the semantical concepts induced by either quantification order. This brings us to the idea of introducing one more parameter into cirquents, which we call {\em ranking}. The latter is an indication of in what order selections by the ``guardian angels'' should be made. Furthermore, we allow not just one but several ``guardian angels'' for either player, with each ``angel'' responsible for a certain subset of clusters rather than all clusters of a given type. Then, again, ranking fixes the order in which these multiple ``angels'' should make their selections. Let us get down to formal definitions to make these intuitions precise and more clear. 

A {\bf cirquent} in the sense of the present section is the same as one in the sense of the previous section, with the following two changes. Firstly, now the set of {\em all gates} (including $\mlc$-gates) is partitioned into clusters, with each cluster, as before, satisfying the condition that all gates in it have the same type and the same outdegree. Secondly, there is an additional parameter called {\bf ranking}. The latter is a partition of the set of all  parallel ($\mld$ and $\mlc$) clusters into a finite number of subsets, called {\bf ranks}, arranged in a linear order, with each rank satisfying the condition that all clusters in it have the same type (but not necessarily the same outdegree). A rank containing $\mlc$-clusters is said to be {\bf conjunctive}, and a rank containing $\mld$-clusters {\bf disjunctive}. Since the ranks are linearly ordered, we can refer to them as the $1$st rank, the $2$nd rank, etc. or rank $1$, rank $2$, etc.  Also, instead of ``cluster $c$ is in the $i$th rank'', we may say ``$c$ is of (or has) rank $i$''.  

Cirquents in the sense of the previous section are understood as special cases of cirquents in the present sense, namely, as cirquents  where all $\mlc$-clusters are singletons of the highest rank, and  all $\mld$-clusters (which are not necessarily singletons) are of the lowest rank. Here we say ``highest rank'' and ``lowest rank'' instead of ``rank $2$'' and ``rank $1$'' just for safety, because, if one of the two types of parallel gates is absent, then rank $1$ would be both highest and lowest; and, if there are no parallel gates at all, the cirquent would not even have rank $1$.     

Let $C$ be a cirquent with $k$ ranks, and let $1\leq i\leq k$. An {\bf $i$-metaselection} for $C$ is a (not necessarily effective) partial function from the $i$th rank of $C$ to the set of positive integers, satisfying the condition that, for any given cluster $c$ of the $i$th rank, whenever $f_i(c)$ is defined, its value does not exceed the outdegree of $c$. And a (simply) {\bf metaselection} for $C$ is a $k$-tuple $(f_1,\ldots,f_k)$ where, for each $1\leq i\leq k$, $f_i$ is an $i$-metaselection for $C$. 

Clustering parallel gates and ranking has no effect on the set of {\bf legal runs} of the game represented by a cirquent, so the definition of legal positions for cirquents of Section \ref{s5} (Definition \ref{maya}) transfers to the present case without any changes.

\begin{defi}\label{jun4}
In the context of a given cirquent $C$ with $k$ ranks, a legal run $\Gamma$ of $C$ and a metaselection $\vec{f}=(f_1,\ldots,f_k)$ for $C$, we will say that a $\mld$-gate $g$ of a cluster $c$   is {\bf unresolved} iff, where  $i$ is the rank of $c$, the function $f_i$  is undefined at $c$.   Otherwise $g$ is {\bf resolved}, and the child of it pointed at by the $f_i(c)$th  outgoing edge  is said to be the {\bf resolvent} of $g$. As for the  same-name concepts ``unresolved'', ``resolved'' and ``resolvent'' for selectional gates, they are defined literally as before (Definition \ref{m20}). 
\end{defi}

The following definition of metatruth can be seen to conservatively generalize its predecessor, Definition \ref{may14cc}: 
 
\begin{defi}\label{may14ccc}
Let $C$ be a cirquent, $^*$ an interpretation,   $\Gamma$ a legal run of $C$, and $\vec{f}$ a metaselection for $C$. In this context, with ``metatrue'' to be read as ``{\bf metatrue w.r.t. $(^*,\Gamma,\vec{f})$}'', we say that: 
\begin{enumerate}[$\bullet$]
\item An $L$-port is metatrue iff $L^*=\twg$.
\item A resolved gate (of any of the eight types) is metatrue iff so is its resolvent.
\item No unresolved disjunctive gate (of any of the four types) is metatrue.
\item Every unresolved conjunctive gate (of any of the four types) is metatrue.
\end{enumerate}
Finally, we say that $C$ is metatrue iff so is its root.
\end{defi}

The following definition brings us from metatruth back to truth. Again, it can be seen to conservatively generalize its predecessor, Definition \ref{ma14cc}: 

\begin{defi}\label{may14k}
Let $C$ be a cirquent with $k$ ranks,   $^*$ an interpretation, and $\Gamma$ a legal run of $C$. We say that  
$C$ is {\bf true} w.r.t. $(^*,\Gamma)$  iff 
\[\mbox{\em ${\mathcal Q}_1 f_1\ldots {\mathcal Q}_k f_k$ such that  $C$ is metatrue w.r.t. $(^*,\Gamma,(f_1,\ldots,f_k))$.}\]
Here each ${\mathcal Q}_i f_i$ abbreviates the phrase ``for every $i$-metaselection  $f_i$ for $C$'' if the $i$th rank is conjunctive, and ``there is an  $i$-metaselection $f_i$ for $C$'' if the $i$th rank is disjunctive. 
\end{defi} 

With truth redefined this way, the (remaining) {\bf Wn} component of the game $C^*$ represented by a cirquent $C$ under an interpretation $^*$ is defined as before. Namely, a legal run $\Gamma$ of $C^*$ is considered $\pp$-won   
iff $C$ is true w.r.t. $(^*,\Gamma)$. 

To understand what we have achieved by introducing ranking and why such a generalization of cirquents was naturally called for, let us, for simplicity, limit our attention to selectional-gate-free cirquents. This fragment of our logic can be seen to be sufficient to  capture and naturally generalize the conservative  extension of IF logic known as {\em extended IF logic} (cf. \cite{Tul09}). The latter, in addition to what IF logic calls {\em strong negation} $\sim$, also considers {\em weak negation} $\gneg$. While $\sim \hspace{-3pt}E$ simply means the result of changing in $E$ each operator and atom to its dual ($p$ to $\gneg p$ and vice versa, $\cla$ to $\cle$ and vice versa, $\mlc$ to $\mld$ and vice versa) and hence there is no real need to have $\sim$ as a primitive,  
weak negation $\gneg $ in (extended) IF logic is quite problematic. Namely, the latter does not act like an ordinary operator that can be applied anywhere in a formula; rather, extended IF logic (essentially) only allows $\gneg$ to be applied to entire IF formulas, thus deeming meaningless and illegal  expressions such as, say, $\cle u\gneg\cla x\cle y\cla z\cle t/x \hspace{3pt} p(x,y,z,t,u)$.
 This is so because the traditional approaches to IF logic are not general enough to directly provide a semantics for $\gneg$. This odd situation makes it evident that more general approaches are necessary. Our approach can claim to be one that fits the bill.\footnote{It should be noted that there are other approaches closely related to or extending IF logic --- namely, Hodges's \cite{Hodges97} trump semantics and 
V\"{a}\"{a}n\"{a}nen's \cite{Vaa07} team logic --- in which one can form the weak negation of any formula. Those approaches are however quite different from ours, and we will not attempt a comparison.}

As noted earlier, the reader is expected to be familiar with the basic concepts and ideas of IF logic and, specifically, the two concepts of negation that we are discussing. So, we only explain the present point through particular examples.   

Our earlier assumption was that only existential quantifiers and disjunctions were slashed in formulas of non-extended IF logic, as slashing universal quantifiers or conjunctions had no effect on the truth conditions of formulas there. The same is no longer the case if one deals with negations though. Accordingly, from now on, we depart from the above assumption and allow slashing any operators in what we consider to be legitimate formulas of IF logic. 

Earlier we saw how to translate an IF formula $E$ into an equivalent cirquent. Now we conservatively refine that translation method in a way that accounts for the possibility that $E$ contains slashed conjunctions and/or universal quantifiers. Namely, $E$ should be understood as the cirquent $E^\circ$ defined below:  

\begin{desc}\label{dsc2} 
Let $E$ be any formula of (non-extended) IF logic. We define $E^\circ$ as the cirquent obtained from $E$ through performing the following steps:
\begin{enumerate}[(1)]
\item Ignoring slashes, write $E$ in the form of a tree-like cirquent, understanding $\cla$ as a ``long''  $\mlc$-conjunction and $\cle$ as a ``long''  $\mld$-disjunction. This way, each occurrence $O$ of a quantifier, conjunction or disjunction  gives rise to one (if $O$ is not in the scope of a quantifier) or many (if $O$ is in the scope of a quantifier)  gates. We say that each such gate, as well as each outgoing edge of it, {\bf originates} from $O$. Also, since we assume that different occurrences of quantifiers in IF formulas always bind different variables, instead of saying ``\ldots originates from the occurrence of $\cla y$ (or $\cle y$)'' we can unambiguously simply say ``\ldots originates from $y$''. 
\item Cluster the gates so that, any two gates $a$ and $b$ belong to the same cluster if and only if they originate from the same occurrence of $\cle x/y_1,\ldots,y_n$, \ $\mld/y_1,\ldots,y_n$, \ $\cla x/y_1,\ldots,y_n$ \ or \ $\mlc/y_1,\ldots,y_n$ \ in $E$, and the following condition is satisfied:
\begin{enumerate}[$\bullet$] 
\item Let $e_{1}^{a},\ldots,e_{k}^{a}$ and $e_{1}^{b},\ldots,e_{k}^{b}$ be the paths (sequences of edges) from the root of the tree to $a$ and $b$, respectively. Then, for any $i\in\{1,\ldots,k\}$, the order numbers of the edges $e_{i}^{a}$ and $e_{i}^{b}$ are identical unless these edges originate from one of the variables $y_1,\ldots,y_n$.   
\end{enumerate}
\item Impose ranking on the resulting cirquent, putting all $\mld$-clusters into the lowest rank and all $\mlc$-clusters into the highest rank. 
\end{enumerate}
\end{desc}

We claim that such a translation of IF formulas $E$ to cirquents $E^\circ$ is adequate. That is, for any interpretation/model $^*$, $E$ is true in $^*$ in the sense of IF logic if and only if  $(E^\circ)^*=\twg$ in our sense.

We further claim that any formula $\gneg E$ of extended IF logic adequately translates into the cirquent $(\gneg E)^\circ$ defined below:  

 \begin{desc}\label{dsc3}
Let $E$ be any formula of (non-extended) IF logic. We define $(\gneg E)^\circ$ as the cirquent obtained from $E$ through performing the following steps:
\begin{enumerate}[(1)]
\item Turn $E$ into $E^\circ$ according to Description \ref{dsc2}.
\item Change in $E^\circ$ every port label (literal) $p$ to $\gneg p$  and vice versa, and also change every gate label $\mld$ to $\mlc$ and vice versa.
\end{enumerate}
\end{desc}\medskip

\noindent Finally, we claim that any formula $\sim\hspace{-3pt}E$ of IF logic adequately translates into the cirquent $(\sim\hspace{-3pt} E)^\circ$ defined below:  

 \begin{desc}\label{dsc4}
Let $E$ be any formula of (non-extended) IF logic. We define $(\sim\hspace{-3pt} E)^\circ$ as the cirquent obtained from $E$ through performing the following steps:
\begin{enumerate}[(1)]
\item Turn $E$ into $(\gneg E)^\circ$ according to Description \ref{dsc3}.
\item Swap the ranks in the resulting cirquent. That is, make all elements of the formerly lowest rank now belong to the highest rank, and vice versa. 
\end{enumerate}
\end{desc}\medskip

\noindent Figures 13, 14 and 15 illustrate applications of our translations  to the IF-logic formula \[\cla x\cle y\cla z/x,\hspace{-2pt}y\cle t/x,\hspace{-2pt}y\hspace{3pt} p(x,y,z,t)\]
 and two forms of its negation. As before, the universe of discourse here is assumed to be $\{1,2\}$. For compactness considerations, we have written $\overline{p_{1111}}$,   $\overline{p_{1112}}$, etc. instead of $\gneg{p_{1111}}$,   $\gneg{p_{1112}}$, etc. To each gate in those figures we have attached an expression of the form $n^m$. It should be understood as an indication that the gate belongs to cluster $n$, and that such a cluster $n$ is of rank $m$.

\begin{center} \begin{picture}(360,165)

\put(2,128){\scriptsize $3^1$}
\put(52,128){\scriptsize $4^1$}
\put(102,128){\scriptsize $3^1$}
\put(152,128){\scriptsize $4^1$}
\put(202,128){\scriptsize $3^1$}
\put(252,128){\scriptsize $4^1$}
\put(302,128){\scriptsize $3^1$}
\put(352,128){\scriptsize $4^1$}
\put(77,76){\scriptsize $1^1$}
\put(277,76){\scriptsize $2^1$}
\put(178,46){\scriptsize $5^2$}
\put(27,101){\scriptsize $6^2$}
\put(127,101){\scriptsize $6^2$}
\put(227,101){\scriptsize $6^2$}
\put(327,101){\scriptsize $6^2$}

\put(-17,142){\scriptsize $p_{1111}$}
\put(8,142){\scriptsize $p_{1112}$}
\put(5,122){\line(-2,3){11}}
\put(5,122){\line(2,3){11}}
\put(5,116){\circle{12}}
\put(2,113){$\mld$}
\put(33,142){\scriptsize $p_{1121}$}
\put(58,142){\scriptsize $p_{1122}$}
\put(55,122){\line(-2,3){11}}
\put(55,122){\line(2,3){11}}
\put(55,116){\circle{12}}
\put(52,113){$\mld$}
\put(30,97){\line(-2,1){25}}
\put(30,97){\line(2,1){25}}
\put(30,91){\circle{12}}
\put(27,88){$\mlc$}
\put(83,142){\scriptsize $p_{1211}$}
\put(108,142){\scriptsize $p_{1212}$}
\put(105,122){\line(-2,3){11}}
\put(105,122){\line(2,3){11}}
\put(105,116){\circle{12}}
\put(102,113){$\mld$}
\put(133,142){\scriptsize $p_{1221}$}
\put(158,142){\scriptsize $p_{1222}$}
\put(155,122){\line(-2,3){11}}
\put(155,122){\line(2,3){11}}
\put(155,116){\circle{12}}
\put(152,113){$\mld$}
\put(130,97){\line(-2,1){25}}
\put(130,97){\line(2,1){25}}
\put(130,91){\circle{12}}
\put(127,88){$\mlc$}
\put(80,72){\line(-4,1){51}}
\put(80,72){\line(4,1){51}}
\put(80,66){\circle{12}}
\put(77,63){$\mld$}

\put(184,142){\scriptsize $p_{2111}$}
\put(208,142){\scriptsize $p_{2112}$}
\put(205,122){\line(-2,3){11}}
\put(205,122){\line(2,3){11}}
\put(205,116){\circle{12}}
\put(202,113){$\mld$}
\put(233,142){\scriptsize $p_{2121}$}
\put(258,142){\scriptsize $p_{2122}$}
\put(255,122){\line(-2,3){11}}
\put(255,122){\line(2,3){11}}
\put(255,116){\circle{12}}
\put(252,113){$\mld$}
\put(230,97){\line(-2,1){25}}
\put(230,97){\line(2,1){25}}
\put(230,91){\circle{12}}
\put(227,88){$\mlc$}
\put(283,142){\scriptsize $p_{2211}$}
\put(308,142){\scriptsize $p_{2212}$}
\put(305,122){\line(-2,3){11}}
\put(305,122){\line(2,3){11}}
\put(305,116){\circle{12}}
\put(302,113){$\mld$}
\put(333,142){\scriptsize $p_{2221}$}
\put(358,142){\scriptsize $p_{2222}$}
\put(355,122){\line(-2,3){11}}
\put(355,122){\line(2,3){11}}
\put(355,116){\circle{12}}
\put(352,113){$\mld$}
\put(330,97){\line(-2,1){25}}
\put(330,97){\line(2,1){25}}
\put(330,91){\circle{12}}
\put(327,88){$\mlc$}
\put(280,72){\line(-4,1){51}}
\put(280,72){\line(4,1){51}}
\put(280,66){\circle{12}}
\put(277,63){$\mld$}

\put(181,43){\line(-6,1){100}}
\put(181,43){\line(6,1){100}}
\put(181,37){\circle{12}}
\put(178,34){$\mlc$}

\put(15,10){{\bf Figure 13:} $\cla x\cle y\cla z/x,\hspace{-2pt}y \cle t/x,\hspace{-2pt}y \hspace{3pt} p(x,y,z,t)$ (when the universe is $\{1,2\}$)}
\end{picture}\end{center}

\begin{center} \begin{picture}(360,165)

\put(2,128){\scriptsize $3^1$}
\put(52,128){\scriptsize $4^1$}
\put(102,128){\scriptsize $3^1$}
\put(152,128){\scriptsize $4^1$}
\put(202,128){\scriptsize $3^1$}
\put(252,128){\scriptsize $4^1$}
\put(302,128){\scriptsize $3^1$}
\put(352,128){\scriptsize $4^1$}
\put(77,76){\scriptsize $1^1$}
\put(277,76){\scriptsize $2^1$}
\put(178,46){\scriptsize $5^2$}
\put(27,101){\scriptsize $6^2$}
\put(127,101){\scriptsize $6^2$}
\put(227,101){\scriptsize $6^2$}
\put(327,101){\scriptsize $6^2$}

\put(-17,142){\scriptsize $\overline{ p_{1111}}$}
\put(8,142){\scriptsize $\overline{ p_{1112}}$}
\put(5,122){\line(-2,3){11}}
\put(5,122){\line(2,3){11}}
\put(5,116){\circle{12}}
\put(2,113){$\mlc$}
\put(33,142){\scriptsize $\overline{ p_{1121}}$}
\put(58,142){\scriptsize $\overline{ p_{1122}}$}
\put(55,122){\line(-2,3){11}}
\put(55,122){\line(2,3){11}}
\put(55,116){\circle{12}}
\put(52,113){$\mlc$}
\put(30,97){\line(-2,1){25}}
\put(30,97){\line(2,1){25}}
\put(30,91){\circle{12}}
\put(27,88){$\mld$}
\put(83,142){\scriptsize $\overline{ p_{1211}}$}
\put(108,142){\scriptsize $\overline{ p_{1212}}$}
\put(105,122){\line(-2,3){11}}
\put(105,122){\line(2,3){11}}
\put(105,116){\circle{12}}
\put(102,113){$\mlc$}
\put(133,142){\scriptsize $\overline{ p_{1221}}$}
\put(158,142){\scriptsize $\overline{ p_{1222}}$}
\put(155,122){\line(-2,3){11}}
\put(155,122){\line(2,3){11}}
\put(155,116){\circle{12}}
\put(152,113){$\mlc$}
\put(130,97){\line(-2,1){25}}
\put(130,97){\line(2,1){25}}
\put(130,91){\circle{12}}
\put(127,88){$\mld$}
\put(80,72){\line(-4,1){51}}
\put(80,72){\line(4,1){51}}
\put(80,66){\circle{12}}
\put(77,63){$\mlc$}

\put(184,142){\scriptsize $\overline{ p_{2111}}$}
\put(208,142){\scriptsize $\overline{ p_{2112}}$}
\put(205,122){\line(-2,3){11}}
\put(205,122){\line(2,3){11}}
\put(205,116){\circle{12}}
\put(202,113){$\mlc$}
\put(233,142){\scriptsize $\overline{ p_{2121}}$}
\put(258,142){\scriptsize $\overline{ p_{2122}}$}
\put(255,122){\line(-2,3){11}}
\put(255,122){\line(2,3){11}}
\put(255,116){\circle{12}}
\put(252,113){$\mlc$}
\put(230,97){\line(-2,1){25}}
\put(230,97){\line(2,1){25}}
\put(230,91){\circle{12}}
\put(227,88){$\mld$}
\put(283,142){\scriptsize $\overline{ p_{2211}}$}
\put(308,142){\scriptsize $\overline{ p_{2212}}$}
\put(305,122){\line(-2,3){11}}
\put(305,122){\line(2,3){11}}
\put(305,116){\circle{12}}
\put(302,113){$\mlc$}
\put(333,142){\scriptsize $\overline{ p_{2221}}$}
\put(358,142){\scriptsize $\overline{ p_{2222}}$}
\put(355,122){\line(-2,3){11}}
\put(355,122){\line(2,3){11}}
\put(355,116){\circle{12}}
\put(352,113){$\mlc$}
\put(330,97){\line(-2,1){25}}
\put(330,97){\line(2,1){25}}
\put(330,91){\circle{12}}
\put(327,88){$\mld$}
\put(280,72){\line(-4,1){51}}
\put(280,72){\line(4,1){51}}
\put(280,66){\circle{12}}
\put(277,63){$\mlc$}

\put(181,43){\line(-6,1){100}}
\put(181,43){\line(6,1){100}}
\put(181,37){\circle{12}}
\put(178,34){$\mld$}

\put(15,10){{\bf Figure 14:} $\gneg \cla x\cle y\cla z/x,\hspace{-2pt}y \cle t/x,\hspace{-2pt}y \hspace{3pt} p(x,y,z,t)$ (when the universe is $\{1,2\}$)}
\end{picture}\end{center}

\begin{center} \begin{picture}(360,165)

\put(2,128){\scriptsize $3^2$}
\put(52,128){\scriptsize $4^2$}
\put(102,128){\scriptsize $3^2$}
\put(152,128){\scriptsize $4^2$}
\put(202,128){\scriptsize $3^2$}
\put(252,128){\scriptsize $4^2$}
\put(302,128){\scriptsize $3^2$}
\put(352,128){\scriptsize $4^2$}
\put(77,76){\scriptsize $1^2$}
\put(277,76){\scriptsize $2^2$}
\put(178,46){\scriptsize $5^1$}
\put(27,101){\scriptsize $6^1$}
\put(127,101){\scriptsize $6^1$}
\put(227,101){\scriptsize $6^1$}
\put(327,101){\scriptsize $6^1$}

\put(-17,142){\scriptsize $\overline{ p_{1111}}$}
\put(8,142){\scriptsize $\overline{ p_{1112}}$}
\put(5,122){\line(-2,3){11}}
\put(5,122){\line(2,3){11}}
\put(5,116){\circle{12}}
\put(2,113){$\mlc$}
\put(33,142){\scriptsize $\overline{ p_{1121}}$}
\put(58,142){\scriptsize $\overline{ p_{1122}}$}
\put(55,122){\line(-2,3){11}}
\put(55,122){\line(2,3){11}}
\put(55,116){\circle{12}}
\put(52,113){$\mlc$}
\put(30,97){\line(-2,1){25}}
\put(30,97){\line(2,1){25}}
\put(30,91){\circle{12}}
\put(27,88){$\mld$}
\put(83,142){\scriptsize $\overline{ p_{1211}}$}
\put(108,142){\scriptsize $\overline{ p_{1212}}$}
\put(105,122){\line(-2,3){11}}
\put(105,122){\line(2,3){11}}
\put(105,116){\circle{12}}
\put(102,113){$\mlc$}
\put(133,142){\scriptsize $\overline{ p_{1221}}$}
\put(158,142){\scriptsize $\overline{ p_{1222}}$}
\put(155,122){\line(-2,3){11}}
\put(155,122){\line(2,3){11}}
\put(155,116){\circle{12}}
\put(152,113){$\mlc$}
\put(130,97){\line(-2,1){25}}
\put(130,97){\line(2,1){25}}
\put(130,91){\circle{12}}
\put(127,88){$\mld$}
\put(80,72){\line(-4,1){51}}
\put(80,72){\line(4,1){51}}
\put(80,66){\circle{12}}
\put(77,63){$\mlc$}

\put(184,142){\scriptsize $\overline{ p_{2111}}$}
\put(208,142){\scriptsize $\overline{ p_{2112}}$}
\put(205,122){\line(-2,3){11}}
\put(205,122){\line(2,3){11}}
\put(205,116){\circle{12}}
\put(202,113){$\mlc$}
\put(233,142){\scriptsize $\overline{ p_{2121}}$}
\put(258,142){\scriptsize $\overline{ p_{2122}}$}
\put(255,122){\line(-2,3){11}}
\put(255,122){\line(2,3){11}}
\put(255,116){\circle{12}}
\put(252,113){$\mlc$}
\put(230,97){\line(-2,1){25}}
\put(230,97){\line(2,1){25}}
\put(230,91){\circle{12}}
\put(227,88){$\mld$}
\put(283,142){\scriptsize $\overline{ p_{2211}}$}
\put(308,142){\scriptsize $\overline{ p_{2212}}$}
\put(305,122){\line(-2,3){11}}
\put(305,122){\line(2,3){11}}
\put(305,116){\circle{12}}
\put(302,113){$\mlc$}
\put(333,142){\scriptsize $\overline{ p_{2221}}$}
\put(358,142){\scriptsize $\overline{ p_{2222}}$}
\put(355,122){\line(-2,3){11}}
\put(355,122){\line(2,3){11}}
\put(355,116){\circle{12}}
\put(352,113){$\mlc$}
\put(330,97){\line(-2,1){25}}
\put(330,97){\line(2,1){25}}
\put(330,91){\circle{12}}
\put(327,88){$\mld$}
\put(280,72){\line(-4,1){51}}
\put(280,72){\line(4,1){51}}
\put(280,66){\circle{12}}
\put(277,63){$\mlc$}

\put(181,43){\line(-6,1){100}}
\put(181,43){\line(6,1){100}}
\put(181,37){\circle{12}}
\put(178,34){$\mld$}

\put(15,10){{\bf Figure 15:} $\sim\hspace{-3pt} \cla x\cle y\cla z/x,\hspace{-2pt}y \cle t/x,\hspace{-2pt}y \hspace{3pt} p(x,y,z,t)$ (when the universe is $\{1,2\}$)}
\end{picture}\end{center}

\noindent The above cirquents are pairwise extensionally non-identical. An interpretation separating the cirquent of Figure 13 from those of Figures 14 and 15 is one that sends  all atoms to $\twg$. And an interpretation separating the cirquent of Figure 14 from the other two cirquents (by making the former $\twg$ while the latter $\tlg$) is the one that sends the four atoms $p_{1111}$, $p_{1221}$, $p_{2112}$, $p_{2222}$ to $\twg$ and sends all other atoms to $\tlg$.  
 
The cirquent of Figure 13 can be seen to be extensionally identical to  the cirquent of figure 12. In general, the same would be the case for any pair  of cirquents that syntactically relate to each other in the same way as the cirquents of Figures 13 and 12 do, namely, where one cirquent is a cirquent  with all $\mld$-clusters in the lowest rank  and all $\mlc$-clusters in the highest rank, and the other cirquent is the result of ignoring in the first one all $\mlc$-clusters and ignoring ranking, after which it can be understood as a cirquent in the limited sense of Section \ref{s6}. 

The cirquent of Figure 14 is the exact opposite of the cirquent of Figure 13, in the sense that, under any interpretation $^*$, one is $\twg$ iff the other is $\tlg$. In general, if two cirquents $C_1$ and $C_2$ syntactically relate to each other as the cirquents of Figures 13 and 14 do, then, for any interpretation $^*$, we will have $C_{2}^{*}=\gneg C_{1}^{*}$, where $\gneg $ is computability logic's ordinary negation operation of  Definition \ref{negdef}.  As for the cirquent of Figure 15, it appears to be a less natural modification of the cirquent of Figure 13 than the cirquent of Figure 14 is. In particular, it is not clear why we, in the process of transforming the cirquent of figure 13 into the cirquent of Figure 15, not only changed the label of each node to its dual, but also swapped the ranks. Furthermore, it would not be clear how to ``swap'' ranks if we had more than two of them. So, in spite of IF logic's tradition to see $\sim$ as the primary sort of negation and treat the ``ill-behaved'' $\gneg$ as a second-class citizen, we come to the vision that it is $\gneg$ rather than $\sim$ that is truly natural and deserves the first-class status.

Descriptions \ref{dsc2} and \ref{dsc3} only generate (selectional-gate-free) cirquents with two (in normal cases) or fewer (in pathological cases) ranks, and hence these sorts of cirquents are sufficient for capturing extended IF logic. Was there then a reasonable call for also considering cirquents with greater numbers of ranks? After all, any approach in any area of mathematics may find an infinite series of generalizations, and one should simply stop somewhere. This is true but, in the process of generalizing,  one should stop only at a natural point where we have a more or less closed (in whatever sense) system. And stopping at cirquents with $\leq 2$ ranks (what extended IF logic essentially did) would not be such a natural place. For, as noted earlier, the formalism of extended IF logic is not closed under its logical operators, and we would be forced to deal with a similar sort of an artificial restriction had we limited our considerations only to cirquents with $\leq 2$ ranks.

To make the above point more clear, let us  extend the syntax of IF logic by requiring that all occurrences of quantifiers, conjunctions and disjunctions be superscripted with positive integers, satisfying the following two conditions:
\begin{enumerate}[$\bullet$]
\item Whenever $1\leq i<j$ and $j$ is the superscript of some occurrence, so is $i$.
\item Whenever $i$ is the superscript of an occurrence of $\cle$ or $\mld$, the same $i$ is not the superscript of an occurrence of $\cla$ or $\mlc$.
\end{enumerate}
Such superscripts will be understood as indications of the ranks of the clusters originating from the corresponding  occurrences of operators when turning formulas into cirquents in the style of Description \ref{dsc2}.
That is, clause 3 of Description \ref{dsc2} should now (for this new syntax of IF logic) read as follows:
\begin{quote} 
Impose ranking on the resulting cirquent, putting 
all clusters originating from occurrences of $i$-superscripted operators\footnote{That is, clusters whose gates originate from occurrences of $i$-superscripted operators.} into the $i$th rank, for any superscript $i$ occurring in $E$.
\end{quote}
We baptize this newly extended version of IF logic as {\bf ranked IF logic}. Let us call the highest superscript appearing in a formula of ranked IF logic the {\bf ranking depth} of that formula.

Of course, extended IF logic is the fragment of ranked IF logic limited to formulas of ranking depth $\leq 2$. Namely, each non-negated formula $E$ of IF logic translates into ranked IF logic as the result $F$ of adding the superscript $1$  to each occurrence of $\cle$ and $\mld$, and adding the superscript 
$2$ to each occurrence of $\cla$ and $\mlc$ --- well, unless $E$ contains no $\mld$ and $\cle$, in which case the superscript $1$ rather than $2$ should be added to the occurrences of  $\cla$ and $\mlc$. Next,  for {\em any} formula $F$ of ranked IF logic (including the cases when $F$ is obtained from $E$ as above), $\gneg F$  can be understood as an abbreviation for the result of changing in $F$ each occurrence of each literal $p$ to $\gneg p$ and vice versa, each occurrence of $\mlc$ to $\mld$ and vice versa, and each occurrence of $\cla$ to $\cle$ and vice versa, {\em without} changing any superscripts in this process. 

With $\gneg$ treated as just explained, in contrast with the situation in extended IF logic, $\gneg$  can meaningfully occur anywhere in an expression  of ranked IF logic.  For instance, we can write \[\cle u^1\gneg\cla^1 x\cle^2 y\cla^1 z\cle^2 t/x \hspace{3pt} p(x,y,z,t,u),\] which will be simply understood as an abbreviation of   \[\cle u^1\cle^1 x\cla^2 y\cle^1 z\cla^2 t/x \hspace{3pt} \gneg p(x,y,z,t,u).\]

To get a further feel for the advantages of ranked IF logic over extended IF logic,   consider the formula $\gneg \cle x\cla  y/x\hspace{-2pt}\sim\hspace{-2pt} p(x,y,z)$ of the latter  which, in ranked IF logic, will be written as $\cla ^1 x\cle^2 y/x \hspace{3pt}p(x,y,z)$.  As we are dealing with a  legal and hence semantically meaningful expression of extended IF logic, we naturally want to be able to quantify it --- say, existentially --- over $z$, and also be able to arbitrarily extend the original independences --- say, by making both quantifiers independent of $\cle z$ and vice versa. Alas, extended IF logic does not permit to apply quantification to a $\gneg$-negated compound formula. But ranked IF logic does. Namely, with a little analysis, the formula 
\begin{equation}\label{eee1}
\cle^1 z\cla ^2 x/z\cle^3 y/z,\hspace{-2pt}x \hspace{3pt}p(x,y,z)
\end{equation}
can be seen to account for the intuitions that we wanted to capture by $\cle z$-quantifying the formula and then making the new and old quantifiers independent of each other.

Figure 16 shows  formula (\ref{eee1}) as a cirquent, and Figure 17 shows two cirquents obtained from it  
by putting both existential quantifiers into the same rank in an attempt to mechanically turn (\ref{eee1}) into an equivalent formula of ranking depth $2$ (a formula of extended IF logic). Either attempt fails. Namely, let 
 $^\dagger$ be the interpretation that sends the two atoms $p_{111}$, $p_{122}$ to $\twg$  
and sends  all other atoms to $\tlg$.
Next, let  $^\ddagger$ be the interpretation that sends the two atoms $p_{111}$, $p_{222}$
to $\twg$ and sends all other atoms to $\tlg$. It can be seen that 
\[\mbox{$\bigl(\cle^1 z\cla ^2 x/z\cle^3 y/z,\hspace{-2pt}x \hspace{3pt}p(x,y,z)\bigr)^\dagger=\twg$ whereas $\bigl(\cle^1 z\cla ^2 x/z\cle^1 y/z,\hspace{-2pt}x \hspace{3pt}p(x,y,z)\bigr)^\dagger=\tlg$,}\]
 and   
\[\mbox{$\bigl(\cle^1 z\cla ^2 x/z\cle^3 y/z,\hspace{-2pt}x\hspace{3pt} p(x,y,z)\bigr)^\ddagger=\tlg$ whereas $\bigl(\cle^2 z\cla ^1 x/z\cle^2 y/z,\hspace{-2pt}x \hspace{3pt} p(x,y,z)\bigr)^\ddagger=\twg$.}\]

\begin{center} \begin{picture}(180,135)

\put(2,98){\scriptsize $3^3$}
\put(52,98){\scriptsize $3^3$}
\put(102,98){\scriptsize $3^3$}
\put(152,98){\scriptsize $3^3$}
\put(77,46){\scriptsize $1^1$}
\put(27,71){\scriptsize $2^2$}
\put(127,71){\scriptsize $2^2$}

\put(-15,112){\scriptsize $p_{111}$}
\put(8,112){\scriptsize $p_{112}$}
\put(5,92){\line(-2,3){11}}
\put(5,92){\line(2,3){11}}
\put(5,86){\circle{12}}
\put(2,83){$\mld$}
\put(35,112){\scriptsize $p_{121}$}
\put(58,112){\scriptsize $p_{122}$}
\put(55,92){\line(-2,3){11}}
\put(55,92){\line(2,3){11}}
\put(55,86){\circle{12}}
\put(52,83){$\mld$}
\put(30,67){\line(-2,1){25}}
\put(30,67){\line(2,1){25}}
\put(30,61){\circle{12}}
\put(27,58){$\mlc$}
\put(85,112){\scriptsize $p_{211}$}
\put(108,112){\scriptsize $p_{212}$}
\put(105,92){\line(-2,3){11}}
\put(105,92){\line(2,3){11}}
\put(105,86){\circle{12}}
\put(102,83){$\mld$}
\put(135,112){\scriptsize $p_{221}$}
\put(158,112){\scriptsize $p_{222}$}
\put(155,92){\line(-2,3){11}}
\put(155,92){\line(2,3){11}}
\put(155,86){\circle{12}}
\put(152,83){$\mld$}
\put(130,67){\line(-2,1){25}}
\put(130,67){\line(2,1){25}}
\put(130,61){\circle{12}}
\put(127,58){$\mlc$}
\put(80,42){\line(-4,1){51}}
\put(80,42){\line(4,1){51}}
\put(80,36){\circle{12}}
\put(77,33){$\mld$}

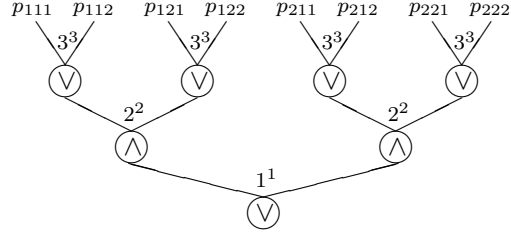
\put(-70,10){{\bf Figure 16:} $\cle^1 z\cla ^2 x/z\cle^3 y/z,\hspace{-2pt}x \hspace{3pt} p(x,y,z)$ (when the universe is $\{1,2\}$)}
\end{picture}\end{center}

\begin{center} \begin{picture}(380,135)

\put(2,98){\scriptsize $3^1$}
\put(52,98){\scriptsize $3^1$}
\put(102,98){\scriptsize $3^1$}
\put(152,98){\scriptsize $3^1$}
\put(77,46){\scriptsize $1^1$}
\put(27,71){\scriptsize $2^2$}
\put(127,71){\scriptsize $2^2$}

\put(-15,112){\scriptsize $p_{111}$}
\put(8,112){\scriptsize $p_{112}$}
\put(5,92){\line(-2,3){11}}
\put(5,92){\line(2,3){11}}
\put(5,86){\circle{12}}
\put(2,83){$\mld$}
\put(35,112){\scriptsize $p_{121}$}
\put(58,112){\scriptsize $p_{122}$}
\put(55,92){\line(-2,3){11}}
\put(55,92){\line(2,3){11}}
\put(55,86){\circle{12}}
\put(52,83){$\mld$}
\put(30,67){\line(-2,1){25}}
\put(30,67){\line(2,1){25}}
\put(30,61){\circle{12}}
\put(27,58){$\mlc$}
\put(85,112){\scriptsize $p_{211}$}
\put(108,112){\scriptsize $p_{212}$}
\put(105,92){\line(-2,3){11}}
\put(105,92){\line(2,3){11}}
\put(105,86){\circle{12}}
\put(102,83){$\mld$}
\put(135,112){\scriptsize $p_{221}$}
\put(158,112){\scriptsize $p_{222}$}
\put(155,92){\line(-2,3){11}}
\put(155,92){\line(2,3){11}}
\put(155,86){\circle{12}}
\put(152,83){$\mld$}
\put(130,67){\line(-2,1){25}}
\put(130,67){\line(2,1){25}}
\put(130,61){\circle{12}}
\put(127,58){$\mlc$}
\put(80,42){\line(-4,1){51}}
\put(80,42){\line(4,1){51}}
\put(80,36){\circle{12}}
\put(77,33){$\mld$}

\put(222,98){\scriptsize $3^2$}
\put(272,98){\scriptsize $3^2$}
\put(322,98){\scriptsize $3^2$}
\put(372,98){\scriptsize $3^2$}
\put(297,46){\scriptsize $1^2$}
\put(247,71){\scriptsize $2^1$}
\put(347,71){\scriptsize $2^1$}

\put(205,112){\scriptsize $p_{111}$}
\put(228,112){\scriptsize $p_{112}$}
\put(225,92){\line(-2,3){11}}
\put(225,92){\line(2,3){11}}
\put(225,86){\circle{12}}
\put(222,83){$\mld$}
\put(255,112){\scriptsize $p_{121}$}
\put(278,112){\scriptsize $p_{122}$}
\put(275,92){\line(-2,3){11}}
\put(275,92){\line(2,3){11}}
\put(275,86){\circle{12}}
\put(272,83){$\mld$}
\put(250,67){\line(-2,1){25}}
\put(250,67){\line(2,1){25}}
\put(250,61){\circle{12}}
\put(247,58){$\mlc$}
\put(305,112){\scriptsize $p_{211}$}
\put(328,112){\scriptsize $p_{212}$}
\put(325,92){\line(-2,3){11}}
\put(325,92){\line(2,3){11}}
\put(325,86){\circle{12}}
\put(322,83){$\mld$}
\put(355,112){\scriptsize $p_{221}$}
\put(378,112){\scriptsize $p_{222}$}
\put(375,92){\line(-2,3){11}}
\put(375,92){\line(2,3){11}}
\put(375,86){\circle{12}}
\put(372,83){$\mld$}
\put(350,67){\line(-2,1){25}}
\put(350,67){\line(2,1){25}}
\put(350,61){\circle{12}}
\put(347,58){$\mlc$}
\put(300,42){\line(-4,1){51}}
\put(300,42){\line(4,1){51}}
\put(300,36){\circle{12}}
\put(297,33){$\mld$}

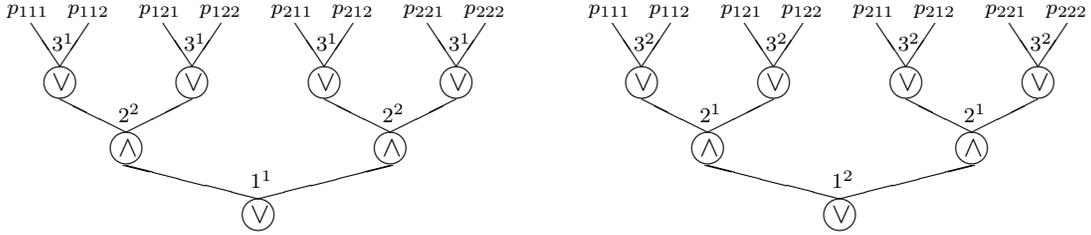
\put(30,10){{\bf Figure 17:} $\cle^1 z\cla ^2 x/z\cle^1 y/z,\hspace{-2pt}x \hspace{3pt} p(x,y,z)$ and $\cle^2 z\cla ^1 x/z\cle^2 y/z,\hspace{-2pt}x \hspace{3pt} p(x,y,z)$}
\end{picture}\end{center}

\noindent As we just had a chance to observe, ranked IF logic provides greater syntactic flexibility and convenience than extended IF logic does. This gives us reasons to  expect that the former is (not only a more direct and flexible means of expression but also) properly more expressive than the latter, in the same sense as the latter is known to be more expressive than ordinary, non-extended IF logic. A way to prove this conjecture can be expressing, through a formula $T$ of ranked IF logic, a definition of truth for formulas of ranked IF logic of ranking depth  $\leq 2$ (such formulas fully cover extended IF logic). Now, if $T$ itself was expressible as a formula of ranking depth $\leq 2$, then we would be able to produce a paradox by writing a formula of ranking depth $\leq 2$ that asserts its own not being true. 

A related and more general question, which we leave unanswered, is about whether various fragments of ranked IF logic, depending on the depth and order of ranking, yield the expressive power of various hierarchical fragments of second-order logic. In expressive power, ordinary IF logic is known to be equivalent to $\Sigma^{1}_{1}$. One can naturally expect that with higher ranks we can capture higher levels of the hierarchy. The same question can as well be asked about ranked elementary cirquents (as opposed to formulas of ranked IF logic) in general.    

As any approach, our approach allows further generalizations. For instance, our present linear orders on ranks can be relaxed to partial orders, which may give rise to an IF-logic-style approach to  independences between different \ldots ranks. But enough is enough. Things have already gone quite far, and further generalizations would be reasonable to make only if and when a clear call for them comes.   As pointed out, a call for the generalizations (of both CoL and IF logic) we have made so far in this paper was the necessity to make the formalisms reasonably complete, and neutralize certain unsettling, incompleteness-caused phenomena such as the odd status of weak negation $\gneg$ in IF logic, or the impossibility to properly develop IF logic at the purely propositional level.   

However, in our step-by-step generalization process, stopping at cirquents of the present section would not be right. There is one last necessary step remaining, which will be taken in the following section.

\section{General ports}\label{s8}

\noindent Imagine a finite cirquent $C$ in the sense of any of the previous sections. Let $p_1,\ldots,p_n$ be the atoms, listed in their lexicographic order, used (positively or negatively) in the labels of the ports of $C$. Then $C$ is, in fact, an operation that takes an $n$-tuple of elementary games (an interpretation, that is) and produces a new, not-necessarily elementary (unless $C$ only has $\mld$ and $\mlc$ gates) game. The same is the case when $C$ is infinite, only here, as an operation, $C$ may take an infinite sequence (rather than just an $n$-tuple) of elementary games. 

Cirquents thus are systematic ways to generate and express an infinite variety of operations on games. However, as just noted, as long as we only consider cirquents in the sense of the previous section(s), all such operations are limited to ones whose arguments are elementary games $\pp$ and $\oo$. These are two very special and simple cases of  games. So, a natural call comes for generalizing our approach in a way that allows cirquents to express operations with not only elementary arguments, but also with arguments that can be arbitrary interactive computational problems (static games, that is) considered in CoL. 

One way to achieve the above would be to change the concept of an interpretation $^*$ so that now it is allowed to send the atoms of a cirquent to any, not-necessarily-elementary, static games. By doing so we would certainly gain a lot, but just as much would be lost:
the class of valid cirquents would shrink, victimizing many innocent ones such as, say, 
$p\mli p\mlc p$ or $p\mld p\mli p\add p$. The point is that elementary problems (games) are meaningful and interesting in their own right, and losing the ability to differentiate them from problems in general would be too much of a sacrifice. For instance, classical logic, IF logic, or the systems of computability-logic-based arithmetic constructed in \cite{Japtowards,cla4,cla5}, are exclusively concerned with cases where atoms are interpreted as elementary problems.  

We have a better solution. It is simply allowing two sorts of atoms in the language, one for elementary problems and the other for all problems. This way, not only do we have the ability to express combinations of problems of either sort within the same formal language, but also combinations that intermix elementary problems with not-necessarily-elementary ones. 

Let us rename the objects to which we earlier refereed as (simply) ``atoms'' into {\bf elementary atoms}. In addition to elementary atoms, we fix another infinite set of alphanumeric strings, disjoint from the set of elementary atoms, and call its elements  {\bf general atoms}. We shall continue using the lowercase $p$, $q$, $r$, $s$, $p_1$, $p(3,4)$, \ldots as metavariables for elementary atoms, and we will be using the uppercase $P$, $Q$, $R$, $S$, $P_1$, $P(3,4)$, \ldots as metavariables for general atoms. As before, a {\bf literal} is $L$ or $\gneg L$, where $L$ is an atom. In the former case the literal is said to be {\bf positive}, and in the latter case {\bf negative}. Such a literal will be said to be elementary or general depending on whether the atom $L$ is elementary or general. The two literals $L$ and $\gneg L$ are said to be {\bf opposite}. It is assumed that the question on whether a  literal is elementary or general is decidable. 

A {\bf cirquent} in the sense of the present section means the same as one in the sense of the previous section, with the only difference that now not only elementary, but also general literals are allowed as labels of ports. A port is said to be {\em elementary} or {\em general}, {\em positive} or {\em negative}  depending on whether its label is so. Similarly, two ports are said to be {\em opposite} iff their labels are so. 
   
An {\bf interpretation} now is a function $^*$ that (as before) sends each elementary atom $p$ to an elementary game $p^*$, and sends each general atom $P$ to any, not-necessarily-elementary, static game $P^*$.  This function immediately extends to all literals by stipulating that, for any (elementary or general) atom $W$, $(\gneg W)^*=\gneg (W^*)$, where $\gneg$ in $\gneg (W^*)$ is the ordinary game negation operation of Definition \ref{negdef}.

For a run $\Gamma$ and a string $\alpha$, we will be using  $\Gamma^\alpha$ to denote the result of deleting in $\Gamma$ all labmoves except those that look like $\xx \alpha\beta$ (either player $\xx$ and whatever string $\beta$), and then further deleting the prefix $\alpha$ in each remaining move --- that is, replacing each $\xx\alpha\beta$ by $\xx\beta$.

Our present definition of the legal runs of the game $C^*$ represented by a cirquent $C$ under an interpretation $^*$ is similar to the earlier definition(s), with the difference that now additional moves of the form $a.\alpha$ can be made, where $a$ is (the ID of) a general port. The intuitive meaning of such a move is making the move $\alpha$ in the  copy of the game $L^*$ associated with $a$, where $L$ is the label of $a$. Accordingly, the additional condition that needs to be satisfied for a legal run $\Gamma$ is that, whenever $a,L$ are as above, $\Gamma^{a.}$ (intuitively the run of $L^*$ played in port $a$) should be a legal run of $L^*$. Below is a full definition:

\begin{defi}\label{mayanew}
Let $C$ be a cirquent,  $^*$ an interpretation, and  $\Phi$ a position. $\Phi$ is a {\bf legal position} of the game $C^*$ iff, with ``cluster'' and ``port'' below meaning those of $C$,  the following conditions are satisfied:
\begin{enumerate}[(1)]
\item Every labmove of $\Phi$ has one of the following forms: 
\begin{enumerate} 
\item $\pp c.i$, where $c$ is a $\tgd$\hspace{-2pt}-,\hspace{-2pt} $\sqd$\hspace{-2pt}-\hspace{-2pt} or $\add$-cluster and $i$ is a positive integer not exceeding  the outdegree of $c$. 
\item $\oo c.i$, where $c$ is a $\tgc$\hspace{-2pt}-,\hspace{-2pt} $\sqc$\hspace{-2pt}-\hspace{-2pt} or $\adc$-cluster and $i$ is a positive integer  not exceeding  the outdegree of $c$. 
\item $\xx a.\beta$, where $a$ is a general port, $\xx$ is either player, and $\beta$ is some string. 
\end{enumerate} 
\item Whenever $c$ is a choice  cluster, $\Phi$ contains at most one occurrence of a labmove of the form $\xx c.i$.   
\item Whenever $c$ is a sequential  cluster and $\Phi$ is of the form $\seq{\ldots, \xx c.i ,\ldots ,\xx c.j ,\ldots}$, we have $i<j$. 
\item Whenever $a$ is a general port and $L$ is its label, $\Phi^{a.}$ is a legal position of the game $L^*$.
\end{enumerate}
\end{defi}

The concept of a {\bf metaselection}, as well as the concepts {\bf unresolved}, {\bf resolved} and {\bf resolvent}, for any of the eight sorts of gates, transfer from Section \ref{s7}  to the present section without any changes. And metatruth (Definition \ref{may14ccc}) is now redefined as follows: 

 \begin{defi}\label{may14ff}
Let $C$ be a cirquent, $^*$ an interpretation,   $\Gamma$ a legal run of $C$, and $\vec{f}$ a metaselection for $C$. In this context, with ``metatrue'' to be read as ``{\bf metatrue w.r.t. $(^*,\Gamma,\vec{f})$}'', we say that: 
\begin{enumerate}[$\bullet$]
\item An  (elementary or general) $L$-port $a$  is metatrue  iff $\Gamma^{a.}$ is a $\pp$-won run of $L^*$.\footnote{Note that, if $a$ is an elementary port, then $\Gamma^{a}$ is empty, and saying that such a run is a $\pp$-won run of $L^*$ is the same as to say that $L^*=\twg$.}
\item A resolved gate (of any of the eight types) is metatrue iff so is its resolvent.
\item No unresolved disjunctive gate (of any of the four types) is metatrue.
\item Every unresolved conjunctive gate (of any of the four types) is metatrue.
\end{enumerate}
Finally, we say that $C$ is metatrue  iff so is its root.
\end{defi}

With metatruth (conservatively) redefined this way, the definition of (simply) {\bf truth} is literally the same as before (Definition \ref{may14k}).  
So is the {\bf Wn} component of the game $C^*$ represented by a cirquent $C$ under an interpretation $^*$. Namely, a legal run $\Gamma$ of $C^*$ is considered to be $\pp$-won   
iff $C$ is true w.r.t. $(^*,\Gamma)$.

In certain cases, the elementary versus general status of atoms has no effect on validity. For instance, the cirquent $\gneg p\mld p$ is valid, and so can be shown to be (whether in the weak or in the strong sense) the cirquent $\gneg P\mld P$. The same does not hold for all cirquents though. An example would be $\gneg p\mld(p\mlc p)$, which is valid but can be shown to be not so (whether in the strong or in the weak sense) with $P$ instead of $p$. At this point we can observe the resource-consciousness of computability logic even when only parallel gates/connectives are considered: while $p$ and $p\mlc p$ are ``the same'', $P$ and $P\mlc P$ (or $P\mld P$) are not so at all: $P$ stands for what will evolve as a {\em single} play of game $P^*$ (whatever interpretation $^*$ we have in mind), whereas  $P\mlc P$ stands for what will evolve as two parallel plays of $P^*$ --- that is, plays  on two boards. While the game played on either board in the latter case is the same $P^*$, the actual runs on the two boards will not necessarily be the same (unless both players are making exactly the same moves on the two boards), so that, it may well happen that the play on one board is won while on the other board is lost. Generally, winning $P\mlc P$ for $\pp$ is harder than winning $P$, and winning $P$ is harder than winning $P\mld P$.  The situation is very different from this one with elementary atoms instead of general atoms: $p$ is indeed ``the same'' as $p\mlc p$ (otherwise computability logic would not be a conservative extension of classical logic). That is because $p^*$ is an elementary game with no moves, and hence it makes no difference whether it is ``played'' on one board or two boards: all of its ``plays'' will be identical, and hence either all of them will be won or all of them will be lost. 

As an aside, the phenomenon of resource-consciousness, in a certain way, can be observed in pure IF logic as well. For instance, the two formulas $\cla x E(x)$ and $\cla x\bigl(E(x)\mld E(x)\bigr)$ can be seen to be non-equivalent there, with $E(x)$ abbreviating $\cle y/x\hspace{1pt} (x=y)$. (Hint: to see this, consider a model whose domain contains exactly two objects.) IF logic and resource logics are thus interrelated, and this fact serves as additional evidence in favor of the idea of studying them within a common general framework, such as the one elaborated in the present paper. 

Earlier we pointed out the increase in expressive power of the formalism of computability logic achieved by switching from formulas to cirquents. Such a switch has an even more dramatic impact on expressive power when, along with elementary atoms, general atoms are also allowed. As we remember, finite cirquents with only $\mlc$ and $\mld$ gates and only elementary ports are not any more expressive than formulas are. The same is no the case for cirquents with general ports though. To get a feel of this, let us compare the two cirquents of Figure 18, the only difference between which is that one (on the right) has general ports where the other has elementary ports. Here and later, following our earlier practice, node IDs are omitted in these figures. Also as agreed earlier, omitted cluster IDs indicate that clustering is trivial. i.e., all clusters are singletons. Finally, omitted rank indicators should be understood as that all $\mld$-clusters are in the lowest rank  and all $\mlc$-clusters are in the highest rank, even though this is, in fact, irrelevant: in the absence of nonsingleton clusters, ranking can be seen to be redundant, and how it is chosen has no effect on the semantics of the cirquent. 

\begin{center} \begin{picture}(310,133)

\put(24,113){$p$}
\put(74,113){$q$}
\put(124,113){$r$}

\put(27,91){\line(0,1){16}}
\put(27,91){\line(3,1){50}}
\put(27,85){\circle{12}}
\put(24,82){$\mlc$}

\put(77,91){\line(-3,1){50}}
\put(77,91){\line(3,1){50}}
\put(77,85){\circle{12}}
\put(74,82){$\mlc$}

\put(127,91){\line(0,1){16}}
\put(127,91){\line(-3,1){50}}
\put(127,85){\circle{12}}
\put(124,82){$\mlc$}

\put(77,62){\line(-3,1){50}}
\put(77,62){\line(3,1){50}}
\put(77,62){\line(0,1){17}}
\put(77,56){\circle{12}}
\put(74,53){$\mld$}

\put(200,113){$P$}
\put(250,113){$Q$}
\put(300,113){$R$}

\put(203,91){\line(0,1){16}}
\put(203,91){\line(3,1){50}}
\put(203,85){\circle{12}}
\put(200,82){$\mlc$}

\put(253,91){\line(-3,1){50}}
\put(253,91){\line(3,1){50}}
\put(253,85){\circle{12}}
\put(250,82){$\mlc$}

\put(303,91){\line(0,1){16}}
\put(303,91){\line(-3,1){50}}
\put(303,85){\circle{12}}
\put(300,82){$\mlc$}

\put(253,62){\line(-3,1){50}}
\put(253,62){\line(3,1){50}}
\put(253,62){\line(0,1){17}}
\put(253,56){\circle{12}}
\put(250,53){$\mld$}

\put(23,33){\em $(p\mlc q)\mld(p\mlc r)\mld(q\mlc r)$}
\put(226,33){\em No formula}

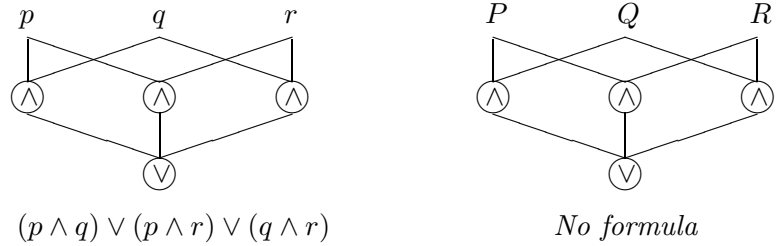
\put(55,10){{\bf Figure 18:} Elementary versus general ports}
\end{picture}\end{center}

\noindent The left cirquent of Figure 18 can be turned into an equivalent tree in the standard way (duplicate and separate shared nodes), yielding the formula $(p\mlc q)\mld(p\mlc r)\mld(q\mlc r)$. Everything is classical here, that is. The same trick fails for the right cirquent though. It represents a  game played on three boards where $\pp$, in order to win, should win on (at least) two out of the three boards. So, the cirquent that we see there is by no means equivalent to  $(P\mlc Q)\mld(P\mlc R)\mld(Q\mlc R)$: the latter represents a game on six (rather than three) boards grouped into three pairs where, in order to win, $\pp$ needs to win on both boards of at least one pair. This is a ``two out of six'' combination, which is generally easier to win than the ``two out of three'' combination represented by the cirquent under question. There is simply no tree-like cirquent (formula) extensionally identical, or equivalent in any reasonable weaker sense, to the right cirquent of Figure 18, meaning that expressing the game operation represented by it  essentially requires the ability to account for {\em sharing} --- the ability absent in formula-based languages. In our cirquent, each of the ports is shared between two conjunctive parents. What makes the formula $(P\mlc Q)\mld(P\mlc R)\mld(Q\mlc R)$ inadequate is that, for instance, it fails to indicate that the two occurrences of $P$ stand for {\em the same copy} of game $P$ rather than {\em two copies of the same game} $P$. And, as pointed out earlier, two copies of $P$ are semantically not the same as just one copy.      

Now we are done with defining ever more general concepts of cirquents and setting up a CoL semantics for them. The next natural step in this line of research would be elaborating deductive systems that adequately axiomatize the sets of valid cirquents. Of course, this is only possible for various subclasses of cirquents rather than all cirquents. A modest progress in this direction has already been made in \cite{Japdeep} where a cirquent-based sound and complete system\footnote{Soundness and completeness in \cite{Japdeep} was proven with respect to abstract resource semantics rather than the semantics of computability logic; as we are going to see later, however, these two semantics are equivalent.} was constructed. The cirquents of the language of that system have only general ports, only $\mlc$ and $\mld$ gates and, of course, no clustering and ranking. 

At the formula level, very considerable advances have already been made in the direction of axiomatizing the sets of valid principles (\cite{Japtocl1}-\cite{Japtcs},  \cite{Japjsl}-\cite{Propint}, \cite{Japseq}-\cite{Japtowards}, \cite{Japtogl}, \cite{Ver}). For instance, \cite{Japtogl} contains a sound and complete axiomatization for the propositional fragment of CoL with both elementary and general atoms, negation and all four --- parallel, choice, sequential and toggling --- sorts of conjunctions and disjunctions. Certain first-order fragments of CoL have also found sound and complete axiomatizations. The quantifiers $\ada,\ade$ turn out to be much better behaved than their classical counterparts and,   in a striking difference from the latter, yield decidable first order logics. See \cite{Japtcs}, \cite{Japtcs2}.

The present paper makes no axiomatization attempts, leaving this ambitious task for the future. Instead, we are going to show extensional equivalence between the semantics of CoL and its companion termed {\em abstract resource semantics}. This result can make the future job of axiomatizing various fragments of CoL significantly easier, as abstract resource semantics is technically much simpler and more convenient to deal with than the semantics of computability logic. 

\section{Abstract resource semantics}\label{s9}

\noindent {\bf Abstract resource semantics} ({\bf ARS}) aims to formally capture our intuitions of {\em resources} and resource management. 
 Resources are symmetric to {\em tasks}: what is a resource for the consumer, is a task for the provider. So, an equally adequate name for ARS would be ``abstract task semantics''. 

The concept of an {\em atomic resource} in ARS is taken as a basic one without any definition of its nature. This makes it open to various interpretations which ARS itself, as a general-purpose tool, does not provide. The semantics of computability logic, treating atoms as variables over static games, can be seen to be one of many such possible interpretations.  As for compound resources, technically their explication in ARS is given in terms of games. The formal language that ARS deals with is the same as that of computability logic. Precisely, in this paper, we let this (otherwise open-ended) language consist of all cirquents in the (most general) sense of Section \ref{s8}. 

There are two sorts of resources: {\em elementary} and {\em general}. Intuitively, elementary resources are ``reusable'' or ``unexhaustable'' ones, while general resources may or may not be so. Let us say in achieving a certain goal $G$ you used the fact $2+2=4$ as an (intellectual) resource. After this usage, $2+2$ will still be  $4$, so that the resource will remain equally available for future usage if needed again. On the other hand, if you also used  $\$20,000$  for achieving  $G$, you may not be able to use the same resource again later. $2+2=4$ is an elementary resource, while $\$20,000$ is not, which makes the latter a (properly) general resource. As we may guess, elementary resources will be represented in our cirquent formalism through elementary ports, and general resources through general ports.

To get some basic intuitive feel of ARS, and to see why the latter, just like CoL, naturally calls for switching from formulas to cirquents,  let us borrow a discussion from \cite{Japdeep}.   
We are talking about a vending machine that has slots for 25-cent ($25c$) coins, with each slot taking a single coin. Coins can be authentic or counterfeited. Let us instead use the more generic terms {\em true} and {\em false} here, as there are various particular situations naturally and inevitably emerging in the world of resources corresponding to those two opposite values. Below are a few examples of real-world resources/tasks and the possible meanings of the two semantical values for them:   
\begin{enumerate}[$\bullet$]
\item A financial debt, which may (true) or may not (false) be eventually paid. 
\item An electrical outlet or a battery, which may (true) or not (false) actually have sufficient power in it.
\item A standard task performed by a company's employee or an AI agent, which, eventually, may (true) or not (false) be successfully completed. 
\item A specified amount of computer memory required by a process, which may (true) or not (false) be available at a given time.
\item A promise, which may be kept (true) or broken (false).
\end{enumerate}   
See Section 8 of \cite{Cirq} for detailed elaborations of these intuitions. 

Continuing the description of our vending machine, inserting a false coin into a slot fills the slot up (so that no other coins can be inserted into it until the operation is complete), but otherwise does not fool the machine into thinking that it has received 25 cents. A candy costs 50 cents, and the machine will dispense a candy if at least two of its slots receive true coins. Pressing the ``dispense'' button while having inserted anything less than 50 cents, such as a single coin, or one true and two false coins, results in a non-recoverable loss.   

Victor has three $25c$-coins, and he knows that two of them are true while one is perhaps false (but he has no way to tell which one is false). Could he get a candy? 

The answer depends on how many slots the machine has. Consider two cases: machine $M2$ with two slots, and machine $M3$ with three slots. Victor would have no problem with $M3$: he can insert his three coins into the three slots, and the machine, having received $\geq 50c$, will dispense a candy. With $M2$, however, Victor is in trouble. He can try inserting arbitrary two of his three coins into the two slots of the machine, but there is no guarantee that one of those two coins is not false, in which case Victor will end up with no candy and only $25$ cents remaining in his pocket. 

Both $M2$ and $M3$ can be understood as resources --- resources turning coins into a candy. And note that these two resources are not the same: $M3$ is obviously stronger (``better''), as it allows Victor to get a candy whereas $M2$ does not, while, at the same time, anyone rich enough to be able to make $M2$ dispense a candy would be able to do the same with $M3$ as well.  Yet, formulas fail to capture this important difference. 
$M2$ and $M3$ can be written as
\[\mbox{$R2\mli \mbox{\em Candy}$ \ and \ $R3\mli \mbox{\em Candy}$,}\] 
respectively (with $E\mli F$, as always, abbreviating $\gneg E\mld F$): they consume a certain resource $R2$ or $R3$ and produce {\em Candy}. What makes $M3$ stronger than $M2$ is that the subresource $R3$ that it consumes is weaker (easier to supply) than the subresource $R2$ consumed by $M2$. Specifically, with one false and two true coins, Victor is able to satisfy $R3$ but not $R2$. 

The resource $R2$ can be represented as the following cirquent:
  
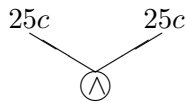
\begin{center} \begin{picture}(70,40)

\put(33,11){\line(5,3){25}}
\put(33,11){\line(-5,3){25}}

\put(0,28){$25c$}
\put(51,28){$25c$}

\put(30,3){$\mlc$}
\put(33,6){\circle{11}}
\end{picture}
\end{center}

\noindent which, due to being tree-like, can also be adequately written as the formula \[25c\mlc 25c.\] As for the resource $R3$, either one of the following two cirquents is an adequate representation of it, with one of them probably showing the relevant part of the actual physical circuitry used in $M3$: 

\begin{center} \begin{picture}(235,97)

\put(36,62){\line(-2,1){29}}
\put(36,62){\line(2,1){29}}
\put(7,62){\line(0,1){15}}
\put(7,62){\line(2,1){29}}

\put(36,36){\line(2,1){29}}
\put(36,36){\line(-2,1){29}}
\put(36,36){\line(0,1){14}}

\put(65,62){\line(-2,1){29}}
\put(65,62){\line(0,1){15}}

\put(0,80){$25c$}
\put(29,80){$25c$}
\put(58,80){$25c$}

\put(33,53){$\mlc$}
\put(36,56){\circle{11}}

\put(4,53){$\mlc$}
\put(7,56){\circle{11}}

\put(62,53){$\mlc$}
\put(65,56){\circle{11}}

\put(33,28){$\mld$}
\put(36,31){\circle{11}}
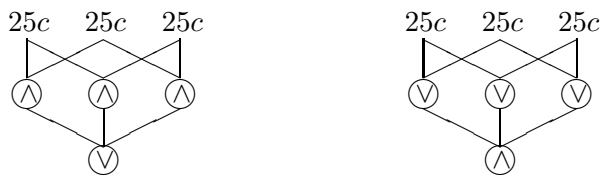
\put(-17,8){{\bf Figure 19:} Two equivalent cirquents for the resource $R3$}

\put(186,62){\line(-2,1){29}}
\put(186,62){\line(2,1){29}}
\put(157,62){\line(0,1){15}}
\put(157,62){\line(2,1){29}}

\put(186,36){\line(2,1){29}}
\put(186,36){\line(-2,1){29}}
\put(186,36){\line(0,1){14}}

\put(215,62){\line(-2,1){29}}
\put(215,62){\line(0,1){15}}

\put(150,80){$25c$}
\put(179,80){$25c$}
\put(208,80){$25c$}

\put(183,53){$\mld$}
\put(186,56){\circle{11}}

\put(154,53){$\mld$}
\put(157,56){\circle{11}}

\put(212,53){$\mld$}
\put(215,56){\circle{11}}

\put(183,28){$\mlc$}
\put(186,31){\circle{11}}
\end{picture}
\end{center}

\noindent Unlike $R2$, however, $R3$ cannot be written as a formula, for reasons similar to those that we saw when discussing Figure 18. $25c\mlc 25c$ does not fit the bill, for it represents $R2$ which, as we already agreed, is not the same as $R3$. Rewriting one of the above two cirquents --- let it be the one on the right --- into an ``equivalent'' formula in the standard way, by duplicating and separating shared nodes, results in  
\begin{equation}\label{ffff1}
(25c\mld 25c)\mlc(25c\mld 25c)\mlc (25c\mld 25c),\end{equation}
which is not any more adequate than $25c\mlc 25c$. It expresses not $R3$ but the resource consumed by 
a    machine with six coin slots grouped into three pairs, where (at least) one slot in each of the three pairs needs to receive a true coin. Such a machine thus dispenses a candy for $\geq 75$ rather than $\geq 50$ cents, which makes Victor's resources insufficient. 

The trouble here, as in the case of the right cirquent of Figure 18, is related to the inability of formulas to explicitly account for resource sharing or the absence thereof. The right cirquent of Figure 19 stands for a conjunction of three resources, each conjunct, in turn, being a disjunction of two subresources of type $25c$. However, altogether there are three rather than six $25c$-type subresources, each one being shared between two different conjuncts of the main resource. Formula (\ref{ffff1}) is inadequate because, for example, it fails to indicate that the first and the third occurrences of ``$25c$'' stand for the same resource while the second and the fifth (as well as the fourth and the sixth) occurrences stand for another resource, albeit a resource of the same $25c$-type. 
In yet another attempt to save formulas, one could try to agree that atoms always stand for individual resources rather than resource types, then give to the three ports of the right cirquent of Figure 19 three different names $P$, $Q$, $R$, and represent the cirquent as the formula 
$(P\mld Q)\mlc(P\mld R)\mlc (Q\mld R)$. 
But then a crucial piece of information would be lost, specifically the information about all inputs being of the same type $25c$,  as opposed to, say, the three different types $25c$, $10c$, $5c$. This would make it impossible to match Victor's resources with those inputs.   

Thus, any systematic attempt to develop a logic of resources would face the necessity to go beyond formulas and use a formalism that permits to account for resource sharing, as our cirquents do. And it is an absolute shame that linear logic, commonly perceived as ``the'' logic of resources, does not allow to express such simple and naturally emerging combinations of resources as the ``two out of three'' combination expressed by the cirquents of Figure 19.

The main purpose of a good semantics should be serving as a bridge between the real world and the otherwise meaningless formal expressions of logic. And, correspondingly, the value of a semantics should be judged by how successfully it achieves this purpose, which, in turn, depends on how naturally and adequately it formalizes certain basic intuitions connecting logic with the outside world. Such intuitions behind abstract resource semantics have been amply explained and illustrated in Section 8 of \cite{Cirq}. The reader is strongly recommended to get familiar with that piece of literature in order to appreciate the claim of abstract resource semantics that it is a ``real'' semantics of resources, formalizing the resource philosophy traditionally (and,  as argued in \cite{Cirq},  somewhat wrongly) associated with linear logic and its variations. In this paper we are mainly focused on just providing formal definitions, only occasionally making brief intuitive comments, and otherwise fully relying on \cite{Cirq} for extended explanations of the intuitions, motivations and philosophy underlying the semantics. Even though \cite{Cirq} dealt with only a very modest subclass of cirquents in our present sense, the basic intuitions relevant to our  treatment --- at the philosophical if not technical level --- were already given there.

From the technical point of view, ARS is  a game semantics and, as such, only differs from the semantics of CoL in the way it treats atoms. Namely, if in CoL atoms are just placeholders which (together with the entire cirquent) become games only after an interpretation $^*$ is applied to them, ARS treats each atom as an  atomic abstract resource in its own rights, and correspondingly sees the whole cirquent as a resource/task/game in its own rights, without requiring an interpretation to be applied to it. To repeat, while in CoL a cirquent $C$, by itself (without an interpretation) is just a formal expression but not a game, in ARS every cirquent $C$ is directly seen as a game, which we agree to denote by $\hat{C}$. A related difference between CoL and ARS is that, in the game $C^*$ represented by a cirquent $C$ under a given interpretation $^*$, CoL allows moves to be made within the games associated by $^*$ with the general literals of $C$. In ARS, as noted, such literals stand for atomic entities and no moves within  them can or should be made. On the other hand, ARS has an additional sort of moves by $\pp$, called (resource) {\em allocation}. Such a move looks like $(a,b)$, where $a$ is the ID of a $P$-port for some general atom $P$, and $b$ is the ID of a $\gneg P$-port. A condition here, called the {\em monogamicity condition}, is that neither $a$ nor $b$ could have been already used earlier in any allocation moves. 
As explained and illustrated in \cite{Cirq}, the intuition behind an  allocation move is that of (indeed) allocating one resource to another: a coin ($25c$) to a coin-receiving  slot ($\gneg 25c$), a memory ($100MB$) to a memory-requesting process ($\gneg 100MB$), a power source ($100w$) to a power-consuming utensil ($\gneg 100w$), an USB-interface external device ($USB$) to an USB port of a computer ($\gneg USB$), etc. And a  justification of the monogamicity condition is that if a nonelementary resource $a$ is used by (allocated to) $b$, then it cannot be also used by (allocated to) another $c\not=b$. 

Formally, an {\bf allocation} for a given cirquent $C$ is a pair $(a,b)$ --- identified with the expression ``$(a,b)$'' --- where $a$ and $b$ are (the IDs of) general ports of $C$ such that the label of $a$ is $P$ (some general atom $P$) and the label of $b$ is $\gneg P$. 

The set of legal runs of the game associated with a cirquent in ARS is defined as follows:

\begin{defi}\label{tamuna}
Let $C$ be a cirquent and $\Phi$ a position.  $\Phi$ is a {\bf legal position} of the game $\hat{C}$ (associated with $C$ in ARS)  iff,  with ``cluster'' and ``port'' below meaning those of $C$, the following conditions are satisfied:
\begin{enumerate}[(1)]
\item Every labmove of $\Phi$ has one of the following forms: 
\begin{enumerate} 
\item $\pp c.i$, where $c$ is a $\tgd$\hspace{-2pt}-,\hspace{-2pt} $\sqd$\hspace{-2pt}-\hspace{-2pt} or $\add$-cluster and $i$ is a positive integer not exceeding  the outdegree of $c$. 
\item $\oo c.i$, where $c$ is a $\tgc$\hspace{-2pt}-,\hspace{-2pt} $\sqc$\hspace{-2pt}-\hspace{-2pt} or $\adc$-cluster and $i$ is a positive integer not exceeding  the outdegree of $c$. 
\item $\pp (a,b)$, where  $(a,b)$ is an allocation for $C$. This kind of a move is said to be an {\bf allocation move}.
\end{enumerate} 
\item Whenever $c$ is a choice  cluster, $\Phi$ contains at most one occurrence of a labmove of the form $\xx c.i$.   
\item Whenever $c$ is a sequential  cluster and $\Phi$ is of the form $\seq{\ldots ,\xx c.i ,\ldots, \xx c.j ,\ldots}$, we have $i<j$. 
\item $\Phi$ does not contain any two occurrences  $\pp (a,b)$ and $\pp (a',b')$ such that $a=a'$ or $b=b'$.  
\end{enumerate}
\end{defi}\medskip
  
\noindent By an {\bf arrangement} for a cirquent $C$ we mean a set $\mathcal A$ of allocations for $C$ such that,  whenever $(a,b),(a',b')\in {\mathcal A}$, we have either $(a,b)=(a',b')$, or both $a\not=a'$ and $b\not=b'$. We call this condition (on arrangements) the {\bf monogamicity condition}.

Whenever $C$ is a cirquent and $\Gamma$ is a legal run of $\hat{C}$, by the {\bf arrangement induced by $\Gamma$} we mean the set of all allocations $(a,b)$  such that $\Gamma$ contains the labmove $\pp (a,b)$.

By a {\bf situation} $^*$ for a cirquent $C$ we mean an assignment of one of the values $\mathbb{T}$ (``{\em true}'') or $\mathbb{F}$ (``{\em false}'') to each port of $C$, satisfying the condition that, for any two elementary (but not necessarily general) ports $a$ and $b$, whenever $a$ and $b$ have the same label, they are assigned the same value, and whenever $a$ and $b$ have opposite labels, they are assigned different values. 
Any such function $^*$ is a legitimate situation, including the cases when $^*$ assigns different values to general ports that have identical labels. Intuitively this is perfectly meaningful in the world of resources because, say, one $25c$-port (slot of the vending machine) may receive a true coin while the other $25c$-port may receive a false coin or no coin at all.

Note the difference between situations and interpretations. One difference is that an interpretation associates with each port a {\em game}, while a situation simply sends 
each port to one of the values $\mathbb{T}$ or $\mathbb{F}$. Of course, in the case of elementary (but by no means general) ports, this difference is not essential, as $\mathbb{T}$ can be identified with the  game $\twg$ and $\mathbb{F}$ with the game $\tlg$. Another difference is that the domain of an interpretation is the set of {\em atoms} of a cirquent while the domain of a situation is the set of {\em ports}. This is a substantial difference, as the same atom, with or without a negation, may be the label of many different ports. But, again, this difference is not relevant if we only consider elementary ports. 
   
Let $C$ be a cirquent, $\mathcal A$ an arrangement for $C$,  and $^*$ a situation for $C$. We say that $^*$ is {\bf consistent with $\mathcal A$} iff, whenever $(a,b)\in{\mathcal A}$, the situation $^*$ assigns different values to $a$ and $b$.\footnote{In \cite{Cirq}, a weaker condition was adopted, according to which at least one (but possibly both) of the ports $a,b$ should be assigned $\mathbb{T}$. It is easy to see that either condition yields the same concept of validity, so that this difference is unimportant.}

The concept of a {\bf metaselection}, as well as the concepts {\bf unresolved}, {\bf resolved} and {\bf resolvent}, for any of the eight sorts of gates, transfer from Section \ref{s8} and hence Section \ref{s7} to the present section without any changes. And metatruth (Definition \ref{may14ff}) is now redefined as follows: 

 \begin{defi}\label{ma14ff}
Let $C$ be a cirquent, $^*$ a situation for $C$,   $\Gamma$ a legal run of $\hat{C}$, and $\vec{f}$ a metaselection for $C$. In this context, with ``metatrue'' to be read as ``{\bf metatrue w.r.t. $(^*,\Gamma,\vec{f})$}'', we say that: 
\begin{enumerate}[$\bullet$]
\item An (elementary or general) $L$-port $a$  is metatrue  iff $L^*=\mathbb{T}$.
\item A resolved gate (of any of the eight types) is metatrue iff so is its resolvent.
\item No unresolved disjunctive gate (of any of the four types) is metatrue.
\item Every unresolved conjunctive gate (of any of the four types) is metatrue.
\end{enumerate}
Finally, we say that $C$ is metatrue iff so is its root.
\end{defi}

With metatruth redefined this way, the definition of (simply) {\bf truth} is literally the same as before (Definition \ref{may14k}).  

Next, where $C$ is a cirquent  and $\Gamma$ is a legal run of $\hat{C}$, we say that   
$C$ is {\bf accomplished} w.r.t. $\Gamma$ iff, for every situation $^*$ consistent with the arrangement induced by $\Gamma$, $C$ is true w.r.t. $(^*,\Gamma)$. 

Now, where $C$ is a cirquent  and $\Gamma$ is a legal run of $\hat{C}$, $\Gamma$ is considered to be a {\bf $\pp$-won} run of $\hat{C}$ iff $C$ is accomplished w.r.t. $\Gamma$. This, together with Definition \ref{tamuna}, completes our definition of the game $\hat{C}$ associated with a cirquent $C$ in ARS.

We say that an HPM $\mathcal M$ {\bf accomplishes} a cirquent $C$ iff $\mathcal M$ wins the game $\hat{C}$.  When such an $\mathcal M$ exists, the cirquent $C$ is said to be 
{\bf accomplishable}. 

Accomplishability is the main semantical concept of ARS. In its philosophical spirit, it is an ARS-counterpart of the classical concept of truth. Technically, however, it is more a validity- (rather than truth-) style concept. Namely, it is similar to the concept of strong validity of computability logic.

 To see an example, consider the cirquent of Figure 20.

\begin{center} \begin{picture}(213,114)

\put(5,102){\footnotesize $1$}
\put(28,102){\footnotesize $2$}
\put(66,102){\footnotesize $3$}
\put(88,102){\footnotesize $4$}
\put(126,102){\footnotesize $5$}
\put(148,102){\footnotesize $6$}
\put(186,102){\footnotesize $7$}
\put(208,102){\footnotesize $8$}
\put(4,91){$P$}
\put(27,91){$P$}
\put(64,91){$P$}
\put(87,91){$P$}
\put(119,91){$\gneg P$}
\put(140,91){$\gneg P$}
\put(179,91){$\gneg P$}
\put(200,91){$\gneg P$}

\put(19,78){\line(-1,1){10}}
\put(19,78){\line(1,1){10}}
\put(19,72){\circle{12}}
\put(16,69){$\mld$}
\put(79,78){\line(-1,1){10}}
\put(79,78){\line(1,1){10}}
\put(79,72){\circle{12}}
\put(76,69){$\mld$}
\put(48,53){\circle{12}}
\put(45,50){$\mlc$}
\put(48,59){\line(-4,1){30}}
\put(48,59){\line(4,1){30}}

\put(139,78){\line(-1,1){10}}
\put(139,78){\line(1,1){10}}
\put(139,72){\circle{12}}
\put(136,69){$\mld$}
\put(199,78){\line(-1,1){10}}
\put(199,78){\line(1,1){10}}
\put(199,72){\circle{12}}
\put(196,69){$\mld$}
\put(168,53){\circle{12}}
\put(165,50){$\mlc$}

\put(168,59){\line(-4,1){30}}
\put(168,59){\line(4,1){30}}

\put(108,37){\line(-6,1){60}}
\put(108,37){\line(6,1){60}}
\put(108,31){\circle{12}}
\put(105,28){$\mld$}
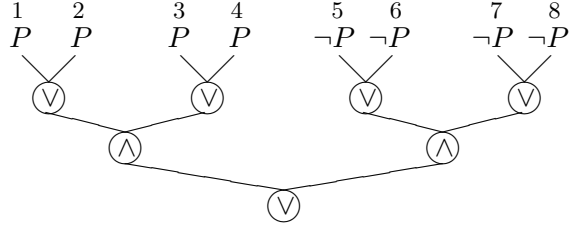
\put(18,8){{\bf Figure 20:} An accomplishable cirquent}
\end{picture}\end{center}

\noindent And consider  two HPMs ${\mathcal M}_1$ and ${\mathcal M}_2$ such that:  
\[\mbox{${\mathcal M}_1$ generates the run  $\Gamma_1=\seq{\pp (1,5), \pp (2,6), \pp (3,7),\pp (4,8)}$};\]
\[\mbox{${\mathcal M}_2$ generates the run  $\Gamma_2=\seq{\pp (1,5), \pp (2,7), \pp (3,6),\pp (4,8)}$}.\]
Then ${\mathcal M}_1$ does not accomplish the cirquent while ${\mathcal M}_2$ does. 

Indeed, the arrangement  induced by $\Gamma_1$ is  \[{\mathcal A}_1 = \{(1,5),  (2,6), (3,7),(4,8)\}.\] Let $^\dagger$ be the situation with  
\[\mbox{$1^\dagger =2^\dagger =7^\dagger= 8^\dagger =\mathbb{T}$; \ \ $3^\dagger =4^\dagger =5^\dagger =6^\dagger =\mathbb{F}$.}\]
Then $^\dagger$ is consistent with ${\mathcal A}_1$. But the cirquent is false w.r.t. $(^\dagger,\Gamma_1)$. Hence it is not accomplished w.r.t. $\Gamma_1$. Hence ${\mathcal M}_1$ does not accomplish it.
 
The above situation $^\dagger$, on the other hand, is not consistent with the arrangement \[{\mathcal A}_2=\{(1,5), (2,7), (3,6), (4,8)\}\] induced by $\Gamma_2$. Moreover, with some thought, one can see that no situation  that makes the cirquent of Figure 20 false can be consistent with ${\mathcal A}_2$. This means that ${\mathcal M}_2$, unlike ${\mathcal M}_1$, {\em does} accomplish that cirquent. 

As an aside, the cirquent of Figure 20 is tree-like and hence can be written as the formula $\bigl((P\mld P)\mlc(P\mld P)\bigr)\mld \bigl((\gneg P\mld \gneg P)\mlc(\gneg P\mld \gneg P)\bigr)$. This formula, first brought forward by Blass \cite{Bla92} in a related context, is not provable in multiplicative linear logic or even in the extension of the latter known as {\em affine logic}. The same applies to the more general principle $\bigl((P\mld Q)\mlc(R\mld S)\bigr)\mld \bigl((\gneg P\mld \gneg R)\mlc(\gneg Q\mld \gneg S)\bigr)$.
Thus, unlike the situation with classical logic or IF logic, the logic induced by ARS or by (the extensionally equivalent) computability logic is {\em not} a conservative extension of linear logic or its standard variations such as affine logic. It is this fact that makes CoL and ARS believe that linear logic is incomplete as a logic of resources.   
  
The cirquent of Figure 21 looks very ``similar'' to the cirquent of \mbox{Figure 20}. Yet, unlike the latter, it is not accomplishable. As an exercise, the reader may want to try to understand why this is so.  

\begin{center} \begin{picture}(400,123)

\put(14,107){$P$}
\put(37,107){$P$}
\put(29,94){\line(-1,1){10}}
\put(29,94){\line(1,1){10}}
\put(29,88){\circle{12}}
\put(26,85){$\mld$}

\put(75,107){$P$}
\put(98,107){$P$}
\put(90,94){\line(-1,1){10}}
\put(90,94){\line(1,1){10}}
\put(90,88){\circle{12}}
\put(87,85){$\mld$}

\put(136,107){$P$}
\put(159,107){$P$}
\put(151,94){\line(-1,1){10}}
\put(151,94){\line(1,1){10}}
\put(151,88){\circle{12}}
\put(148,85){$\mld$}

\put(90,61){\circle{12}}
\put(87,58){$\mlc$}
\put(90,67){\line(-4,1){60}}
\put(90,67){\line(4,1){60}}
\put(90,67){\line(0,1){15}}

\put(230,107){$\gneg P$}
\put(254,107){$\gneg P$}
\put(249,94){\line(-1,1){10}}
\put(249,94){\line(1,1){10}}
\put(249,88){\circle{12}}
\put(246,85){$\mld$}

\put(291,107){$\gneg P$}
\put(315,107){$\gneg P$}
\put(310,94){\line(-1,1){10}}
\put(310,94){\line(1,1){10}}
\put(310,88){\circle{12}}
\put(307,85){$\mld$}

\put(352,107){$\gneg P$}
\put(376,107){$\gneg P$}
\put(371,94){\line(-1,1){10}}
\put(371,94){\line(1,1){10}}
\put(371,88){\circle{12}}
\put(368,85){$\mld$}

\put(310,61){\circle{12}}
\put(307,58){$\mlc$}
\put(310,67){\line(-4,1){60}}
\put(310,67){\line(4,1){60}}
\put(310,67){\line(0,1){15}}

\put(200,37){\line(-6,1){110}}
\put(200,37){\line(6,1){110}}
\put(200,31){\circle{12}}
\put(197,28){$\mld$}
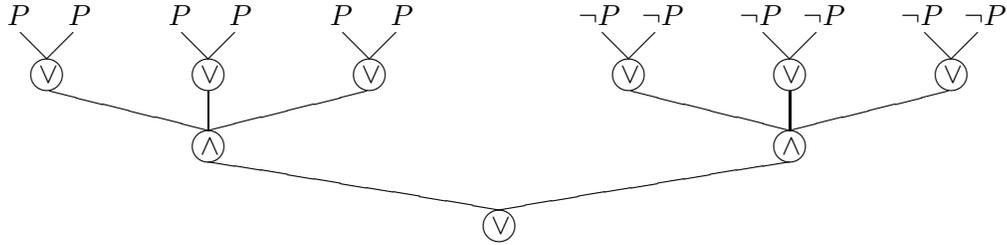
\put(106,8){{\bf Figure 21:} An unaccomplishable cirquent}
\end{picture}\end{center}

\noindent To see the difference that sharing general ports may create, compare the two cirquents of Figure 22. The left cirquent can be seen to be unaccomplishable. The right cirquent only differs from the left cirquent in that ports $6$ and $7$ are combined together into one port $8$ and shared between the two parents. This ``minor'' change makes it accomplishable. Namely, it is accomplished by an HPM that makes the three moves $(8,1)$, $(4,2)$ and $(5,3)$ in whatever order, i.e., sets up the arrangement $\{(8,1),(4,2),(5,3)\}$. 

\begin{center} \begin{picture}(350,143)

\put(6,117){$1$}
\put(31,117){$2$}
\put(54,117){$6$}
\put(77,117){$7$}
\put(100,117){$3$}
\put(123,117){$4$}
\put(146,117){$5$}

\put(206,117){$1$}
\put(231,117){$2$}
\put(266,117){$8$}
\put(300,117){$3$}
\put(323,117){$4$}
\put(346,117){$5$}

\put(0,104){$\gneg P$}
\put(25,104){$\gneg P$}
\put(52,104){$P$}
\put(44,91){\line(-1,1){10}}
\put(44,91){\line(1,1){10}}
\put(44,85){\circle{12}}
\put(41,82){$\mld$}

\put(75,104){$P$}
\put(93,104){$\gneg P$}
\put(90,91){\line(-1,1){10}}
\put(90,91){\line(1,1){10}}
\put(90,85){\circle{12}}
\put(87,82){$\mld$}

\put(121,104){$P$}
\put(144,104){$P$}
\put(136,91){\line(-1,1){10}}
\put(136,91){\line(1,1){10}}
\put(136,85){\circle{12}}
\put(133,82){$\mld$}

\put(90,50){\circle{12}}
\put(87,47){$\mlc$}
\put(90,56){\line(-2,1){45}}
\put(90,56){\line(2,1){45}}
\put(90,56){\line(0,1){22}}
\put(90,56){\line(-4,1){81}}
\put(8,101){\line(0,-1){25}}

\put(53,28){\em unaccomplishable}

\put(200,104){$\gneg P$}
\put(225,104){$\gneg P$}
\put(265,104){$P$}
\put(244,91){\line(-1,1){10}}
\put(244,91){\line(5,2){23}}
\put(244,85){\circle{12}}
\put(241,82){$\mld$}

\put(293,104){$\gneg P$}
\put(290,91){\line(-5,2){23}}
\put(290,91){\line(1,1){10}}
\put(290,85){\circle{12}}
\put(287,82){$\mld$}

\put(321,104){$P$}
\put(344,104){$P$}
\put(336,91){\line(-1,1){10}}
\put(336,91){\line(1,1){10}}
\put(336,85){\circle{12}}
\put(333,82){$\mld$}

\put(290,50){\circle{12}}
\put(287,47){$\mlc$}
\put(290,56){\line(-2,1){45}}
\put(290,56){\line(2,1){45}}
\put(290,56){\line(0,1){22}}
\put(290,56){\line(-4,1){81}}
\put(208,101){\line(0,-1){25}}

\put(257,28){\em accomplishable}
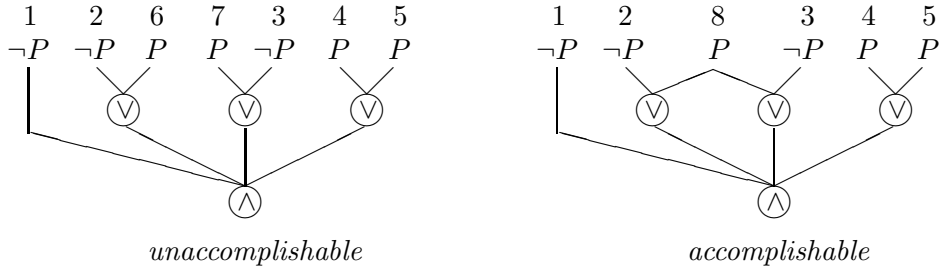
\put(80,8){{\bf Figure 22:} Unshared versus shared general ports}
\end{picture}\end{center}

\noindent Again very briefly about the intuitions behind ARS, elaborated through  several papers (\cite{Jap02,Cirq,Japdeep,Japseq}). Situations are full descriptions of the world in terms of what is true and what is false. What the ``actual'' situation is, is typically unknown to an agent (HPM) trying to accomplish a certain task (cirquent). Furthermore, the truth values of atoms may simply be indetermined before or during the process of playing the game associated with the cirquent, and can be influenced by what moves have been made. For instance, if $p$ stands for ``Victor has (or will) become a millionaire in his lifetime'', the truth value of $p$ may eventually  depend on how Victor acts towards achieving $p$ as a goal. This  value is initially indetermined, for otherwise all Victor's activities would be meaningless. It will become determined only by the time when Victor dies of when the world ends. 

Playing a game represented by a cirquent in ARS can be seen as resource or task management. The goal is to make sure that the eventual (``actual'') situation which will determine success and over which the agent only has partial control, is favorable for the agent (guarantees a win). Resource management includes allocation decisions. The effect of each such  decision/move  is narrowing down the set of all possible (and otherwise unknown) situations. For instance, allocating a coin to a slot of a vending machine rules out the situation (by making it inconsistent with the arrangement that is being set up) in which the coin is true yet the slot did not receive $25c$. Resource management also includes decision-style actions associated with selectional gates or clusters. For instance, a choice between a candy and an apple that a vending machine offers to whoever inserts $50c$ into it is a $\adc$-combination of {\em Candy} and {\em Apple} (this is so from the machine's perspective; it becomes an $\add$-combination from the user's perspective). And nature's ``decisions'' about whether Victor lives or dies is a $\sqd$\hspace{-2pt}-style combination of ``{\em Victor is alive}'' and ``{\em Victor is dead}'', where nature can switch from the former to the latter but never back, as required by the game rules associated with sequential combinations. In Section \ref{s5} we further saw resource/task intuitions associated with clustering. Remember the ``Victor and Peggy in the middle of the road'' example. 

{\bf Historical remarks}. A year before computability logic was officially born, paper \cite{Jap02} introduced an approach termed the {\em logic of tasks} which, ignoring certain unessential and flexible technical details, eventually became a conservative fragment and an ideological predecessor of both CoL and ARS. In our present terms, the language of the logic of tasks was limited to only elementary atoms (an important detail!) and formulas (tree-like cirquents) built from them using $\mlc,\mld,\adc,\add$ in both propositional and quantifier forms. For these sorts of cirquents, the concept of validity defined in \cite{Jap02} coincides with our present ARS concept of accomplishability, as well as with our present CoL concept of strong validity. The above-sketched philosophical vision of the world as a set of potential situations, with the game-playing agent trying to reduce that set to favorable ones, also was originally developed within the framework of the logic of tasks. While the logic of tasks is being further studied by a number of researchers (\cite{Ch1}, \cite{Ch2}, \cite{Ch3}), the author himself abandoned it as an approach superseded by the more general computability logic.

The idea of abstract resource semantics in the proper sense (the sense that differentiates it from the more limited logic of tasks), as well as the idea of cirquents, was born in \cite{Cirq}. Central to this idea was considering allocations as moves in their own rights and, correspondingly, considering in the language general atoms instead of --- or, rather, along with --- the elementary atoms of the logic of tasks. Cirquents that \cite{Cirq} officially dealt with were very limited, with the root always being a $\mlc$-gate, the children of the root always being $\mld$-gates, and all grandchildren of the root being general ports. The same paper, however, outlined the possibility and expediency of considering more general sorts of cirquents and correspondingly extending the associated abstract resource semantics. 

The paper \cite{Japdeep} generalized the cirquents of \cite{Cirq} and the corresponding abstract resource semantics to all finite $\mlc,\mld$-cirquents with general ports. And the paper \cite{Japseq} outlined ARS in more or less full generality, even though for formulas only. Within that outline, Mezhirov and Vereshchagin \cite{Ver} undertook a detailed study of propositional formulas in the logical signature $\{\gneg,\mlc,\mld,\adc,\add,\pst,\pcost\}$, and proved a theorem similar to our forthcoming Theorem  \ref{mth}. The main novelty in the present extension of ARS is the idea of clustering, never considered before in either CoL or ARS. In fact, as noted earlier, cirquents, in whatever form, had never been used in CoL (as opposed to ARS) as official means of expression. And ARS only had been defined for selectional-gate-free cirquents without clustering and ranking.    

Before closing this section, we want to summarize what we have already said or what the reader has probably already observed. From the technical (as opposed to, perhaps, philosophical) point of view, when cirquents with {\em only} elementary ports are considered, there is no difference between ARS and the semantics of CoL, as long as, in the latter, we limit our attention only to the concept of strong  validity. And the above-mentioned logic of tasks is a fragments of this common part of CoL and ARS. So is classical logic, IF logic or extended IF logic. ARS and the semantics of CoL start to differ only when we consider cirquents with general ports. This difference is substantial, yet only in the {\em intensional} sense. The following section shows the nontrivial fact that {\em extensionally} there is no difference --- that is, the two semantics, despite differences, still yield the same classes of valid cirquents.

\section{Accomplishability and strong validity are equivalent}\label{s10}

\noindent By a {\bf nice game} we shall mean a game $G$ such that, with $\xx$ (as always) standing for either player and $\gneg \xx$ for the other player, we have:
\begin{enumerate}[$\bullet$]
\item Every legal run of $G$ is either $\seq{}$ or $\seq{\xx m}$ or $\seq{\xx m,\gneg \xx n}$, where $m,n$ are (the decimal representations of) some positive integers. 
\item The empty run $\seq{}$ is won by $\pp$, and a legal run $\seq{\xx m}$ of $G$ is won by $\xx$.
\item A legal run  $\seq{\pp m,\oo n}$ of $G$ is $\xx$-won iff so is   $\seq{ \oo n,\pp m}$. This allows us to see legal runs of nice games as {\em sets} rather than {\em sequences} of labmoves, and write $\{\pp m,\oo n\}$ instead of $\seq{\pp m,\oo n}$.
\end{enumerate} 
Thus, different nice games differ from each other only in which runs of the form $\{\pp m,\oo n\}$  are won.

By a {\bf nice interpretation} we shall mean an interpretation that interprets each general atom as a nice game.
 
This section is entirely devoted to a proof of the following theorem. In it, as always, a ``cirquent'' means a cirquent in the most general sense defined so far,  i.e., in the sense of Section \ref{s8}. 

\begin{thm}\label{mth}
$C$ is strongly valid iff $C$ is accomplishable (any cirquent $C$).
 
Furthermore, both the ``if'' and ``only if'' parts of this theorem come in the following strong forms, respectively:\medskip 

1.\ There is an effective  procedure that takes an arbitrary HPM \ $\mathcal M$ and constructs an HPM \ $\mathcal U$ such that, if $\mathcal M$ accomplishes $C$, then $\mathcal U$ is a uniform solution for $C$.\medskip 

2.\ If $C$ is not accomplishable, then, for any HPM \ $\mathcal U$,
  there is a nice interpretation $^*$ such that $\mathcal U$ does not
  win $C^*$.
\end{thm}

Before getting down to a proof of this theorem, we need to agree on certain additional details about the (otherwise loosely defined in Section \ref{s2}) HPM model of computation. Namely, we assume that either tape of an HPM has a beginning but no end, with cells arranged in the left-to-right order. We also assume that, at any computation step, an HPM can make at most one move, whereas its environment can make any finite number of moves. When both players move, the order in which their moves are appended to the content of the run tape is that $\pp$'s move goes before $\oo$'s moves. By a  {\bf run generated} by a given HPM \ $\mathcal H$ we mean any run  that might be (depending on the environment's behavior) incrementally spelled on the run tape of $\mathcal H$ during its work.  Thus, $\mathcal H$ wins a game $A$ iff every run generated by $\mathcal H$ is a $\pp$-won run of $A$.\medskip

{\bf Proof of clause 1.}\  Consider an arbitrary cirquent $C$, and an arbitrary HPM \ $\mathcal M$.  Our goal is to construct an HPM \ $\mathcal U$ such that, as long as $\mathcal M$ accomplishes $C$, \ $\mathcal U$ wins $C^*$  for any interpretation $^*$.  

Understanding the idea behind our proof is not hard. We let $\mathcal U$ be a machine which, as far as moves associated with selectional clusters are concerned, acts the same way in $C^*$ --- i.e., makes the same selections --- as $\mathcal M$ does in the game $\hat{C}$ associated with $C$ in ARS. The machines $\mathcal U$ and $\mathcal M$ only differ from each other in their actions related to general ports. The strategy of $\mathcal U$ in general ports is that, as long as $\mathcal M$ has not allocated a given port to another one (or vice versa), $\mathcal U$ makes no moves in it. However, as soon as two general ports $a$ and $b$ are allocated to each other by $\mathcal M$, \ $\mathcal U$ starts applying copycat between the games (one being the negation of the other) associated with those ports, i.e., mimicking in either game the environment's moves made in the other game. As a result, the plays (runs) of the games associated with $a$ and $b$ are guaranteed to be symmetric. More precisely, one is a $\pp$-delay of the other. This makes it impossible that both of those plays are lost by $\pp$. We may assume that exactly one of them is won by $\pp$ (if both are won, ``even better''). Then, translating ``lost'' into the value $\mathbb{F}$ and ``won'' into the value $\mathbb{T}$, we get a situation $^\ddagger$ for $C$ consistent with the arrangement induced by the run $\Theta$ generated by $\mathcal M$. If $\mathcal M$ accomplishes $C$, the latter is true with respect to $(^\ddagger,\Theta)$.  Then the game $C^*$ can be easily seen to be won by $\mathcal U$. 

In more technical detail, $\mathcal U$ works by simulating $\mathcal M$.  
For this simulation, at any step, $\mathcal U$ maintains a record {\em Configuration} for the ``current'' configuration of $\mathcal M$. The latter contains the ``current'' state of $\mathcal M$, the locations of the two scanning heads of $\mathcal M$, and full contents of the (imaginary) work and run tapes of $\mathcal M$. Initially, the state is the start state of $\mathcal M$, the two scanning heads are looking at the leftmost cells of their tapes, and the contents of the two tapes are empty. $\mathcal U$ also maintains a variable $i$ which is initially set to $1$. 

After the above initialization step, $\mathcal U$ just acts according to the following procedure: \medskip

{\bf Procedure} LOOP:
\begin{enumerate}[(1)]
\item Using the transition function of $\mathcal M$ and based on the current value of {\em Configuration}, $\mathcal U$ computes the next state, next locations of the scanning heads, the next content of the work tape of $\mathcal M$, and correspondingly updates {\em Configuration}.
\item  If during the above transition $\mathcal M$ made a move $\omega$, \ $\mathcal U$ further updates {\em Configuration} by appending the labmove $\pp\omega$  to the content of the imaginary run tape of $\mathcal M$. In addition:
\begin{enumerate}
\item If $\omega$ is not an allocation move, $\mathcal U$ makes the same move $\omega$ in the real play.
\item Suppose now $\omega$ is an allocation move   $(a,b)$. Let $\Upsilon$ be the longest initial segment of the position spelled on  the run tape of $\mathcal U$ such that the last labmove of $\Upsilon$, as a string spelled on the tape, ends at location $j$ (i.e., the last symbol of the labmove is written in the $j$th cell) for some $j<i$.  Then $\mathcal U$ looks up, within (but not beyond) $\Upsilon$, all the labmoves $\oo a.\beta_1,\ldots,\oo a.\beta_n$ of the form $\oo a.\beta$, and makes  the moves $b.\beta_1,\ldots,b.\beta_n$ in the real play. Similarly, it looks up within $\Upsilon$ all the labmoves $\oo b.\alpha_1,\ldots,\oo b.\alpha_m$ of the form $\oo b.\alpha$, and makes  the moves $a.\alpha_1,\ldots,a.\alpha_m$. 
\end{enumerate}
 \item Now $\mathcal U$ checks its run tape  to see whether it contains a labmove $\oo\omega$ which, as a string spelled on the  tape, ends exactly at location $i$. If such an $\omega$ is found, then:
\begin{enumerate}
\item If $\omega$ looks like $c.j$ where $c$  is a selectional cluster,  $\mathcal U$ further updates {\em Configuration} by adding the labmove $\oo\omega$ to the content of the imaginary run tape of $\mathcal M$.
\item  If $\omega$ looks like $a.\delta$ where $a$ is a general port already allocated by $\mathcal M$ to a certain port $b$ (or vice versa) --- i.e.,  
the imaginary run tape of $\mathcal M$ contains the labmove $\pp (a,b)$ or $\pp (b,a)$  --- then $\mathcal U$ makes the move $b.\delta$ in the real play. 
\end{enumerate}
\item Finally, $\mathcal M$ increments $i$ to $i+1$, and repeats LOOP. 
\end{enumerate} 

Assume $\mathcal M$ accomplishes $C$.

Consider an arbitrary run $\Gamma$ generated by $\mathcal U$. To this run corresponds a run $\Theta$ generated by $\mathcal M$ --- namely, $\Theta$ is the run incrementally spelled on the imaginary run tape of $\mathcal M$ when the latter is simulated by $\mathcal U$ during playing according to the scenario of $\Gamma$.  Fix these $\Gamma$ and $\Theta$. Let us also fix $\mathcal A$ as the arrangement induced by $\Theta$. We further pick an arbitrary interpretation $^*$ and fix it, too. 

Our assumption that $\mathcal M$ accomplishes $C$ implies that $\mathcal M$ never makes an illegal move of $\hat{C}$ (unless its adversary does so first).  We may also safely assume that $\mathcal U$'s environment never makes illegal moves of $C^*$ either, for otherwise $\mathcal U$ automatically wins and we are done. 
Then a little analysis of the situation reveals that $\Gamma$ is a legal run of $C^*$ and $\Theta$ is a legal run of $\hat{C}$. We will implicitly rely on this observation in our further argument.  

Consider an arbitrary pair $(a,b)$ with $(a,b)\in{\mathcal A}$. Let $P$ be the label of $a$, and thus $\gneg P$ the label of $b$. Remember that $\Gamma^{a.}$ is the run that took place (according to the scenario of $\Gamma$) in the copy of the game $P^*$ associated with $a$ and, similarly, $\Gamma^{b.}$ is the run that took place in the copy of the game $\gneg P^*$ associated with $b$. Remember also that, for a run $\Omega$, $\gneg \Omega$ means the result of negating all labels  in $\Omega$. It is clear from our description of LOOP that $\Gamma^{a.}$ is a $\pp$-delay of $\gneg  \Gamma^{b.}$.
Assume $\Gamma^{b.}$ is a $\pp$-lost run of $\gneg P^*$. Then, by the definition of the game negation operation, $\gneg \Gamma^{b.}$ is a $\pp$-won run of $P^*$. But then, because $P^*$ is a static game and $\Gamma^{a.}$ is a $\pp$-delay of $\gneg \Gamma^{b.}$, $\Gamma^{a.}$ is a $\pp$-won run of $P^*$. 
To summarize, we have:
\begin{equation}\label{jun1a}
\begin{array}{l}
\mbox{\em Suppose $(a,b)\in{\mathcal A}$, and $P$ and $\gneg P$ are the labels of $a$ and $b$. Then}\\
\mbox{\em either $\Gamma^{a.}$ is a $\pp$-won run of $P^*$, or $\Gamma^{b.}$ is a $\pp$-won run of $\gneg P^*$, or both.}
\end{array}
\end{equation}

\noindent We define a situation $^\dagger$ for $C$ by stipulating that, for any elementary or general $L$-port $c$ of $C$, 
$c^\dagger=\mathbb{T}$ iff $\Gamma^{c.}$ is a $\pp$-won run of $L^*$.

From our description of the work of $\mathcal U$ it is clear that $\mathcal M$ and $\mathcal U$ act in exactly the same ways in the selectional clusters of $C$. That is, $\Gamma$ and $\Theta$ contain exactly the same labmoves of the form $\xx c.j$ where $c$ is a selectional cluster. From this observation and our choice of $^\dagger$, with a little thought, we can see that:
\begin{equation}\label{m31a}
\mbox{\em $C$ is true w.r.t. $(^\dagger,\Theta)$ iff $C$ is true w.r.t. $(^*,\Gamma)$.}
\end{equation}

Let $^\ddagger$ be the situation for $C$ that agrees with $^\dagger$ in all cases except that,  whenever $(a,b)\in{\mathcal A}$ and $a^\dagger=b^\dagger=\mathbb{T}$, we have $a^\ddagger=\mathbb{T}$ and $b^\ddagger=\mathbb{F}$. In view of the monotonicity of the truth conditions associated with all types of gates, it is clear that 
\begin{equation}\label{m31b}
\mbox{\em If $C$ is true w.r.t. $(^\ddagger,\Theta)$, then $C$ is also true w.r.t. $(^\dagger,\Theta)$.}
\end{equation}

Consider any $(a,b)\in{\mathcal A}$. By our choice of $^\ddagger$, it is impossible that $a^\ddagger=b^\ddagger=\mathbb{T}$. And, with (\ref{jun1a}) in mind, it is also impossible that  $a^\ddagger=b^\ddagger=\mathbb{F}$. Thus, $^\ddagger$ assigns different values to $a$ and $b$. This  means that 

\begin{equation}\label{jun1b}
\mbox{\em $^\ddagger$ is consistent with $\mathcal A$.}
\end{equation}
 
Since $\mathcal M$ accomplishes $C$,  in view of (\ref{jun1b}), $C$ is  true w.r.t. $(^\ddagger,\Theta)$. But then, by (\ref{m31b}) and (\ref{m31a}), $C$ is true w.r.t. $(^*,\Gamma)$, meaning that $\Gamma$ is a $\pp$-won run of $C^*$. But remember that $\Gamma$ was an arbitrary run generated by $\mathcal U$ and $^*$ was an arbitrary interpretation. Thus, $\mathcal U$ is a uniform solution for $C$. It remains to make the straightforward observation that, as promised in  the theorem, our construction of $\mathcal U$ from $\mathcal M$ is effective.\medskip

{\bf Proof of clause 2.}\  Our proof here is in many respects symmetric to the proof of clause 1. 
 
Consider an arbitrary cirquent $C$, and an arbitrary HPM \ $\mathcal U$ such that $\mathcal U$ wins $C^*$ for any nice interpretation $^*$.  Our goal is to construct an HPM \ $\mathcal M$ such that $\mathcal M$  accomplishes $C$. 

To understand the idea behind our proof, imagine a play of $\mathcal U$ over $C^*$ for whatever nice interpretation $^*$, on which, note, the behavior of $\mathcal U$ does not depend.  Every (legal) move in this play signifies either a move made in a selectional cluster, or a move made in a general port. Since $^*$ is nice, for each general port, either player has exactly one move that can be made there. Let us call $\mathcal U$'s environment {\em smart} if it always makes different moves in different general ports. This is the case when, say, the environment always makes the move ``$a$'' in port $a$. Let us assume that, in the play that we are considering, the environment is smart. The  best that then $\mathcal U$  can do  is to {\em match} each $P$-labeled port with one (but not more!) $\gneg P$-labeled port 
 --- match in the sense of mimicking adversary's moves so that the two games evolve in a symmetric fashion, to guarantee that at least one of them will be eventually won. Each time such a ``matching'' between ports $a$ and $b$ is detected, we let $\mathcal M$ make a move that allocates $a$ to $b$. Other than this, $\mathcal M$  plays in the same way as $\mathcal U$ does, by making the same moves (selections) in selectional clusters as $\mathcal U$ makes. We can then show that, if $\mathcal M$ does not accomplish $C$ with this strategy, any falsifying situation $^\dagger$ for $C$ translates into certain conditions on $^*$ under which  $\mathcal U$ has lost $C^*$. This, however, is impossible because, by our assumption,   $\mathcal U$ wins $C^*$ for {\em any} nice interpretation $^*$.

In more  detailed terms, $\mathcal M$ works by simulating $\mathcal U$.  
For this simulation, at any step, $\mathcal M$ maintains a record {\em Configuration} for the ``current'' configuration of $\mathcal U$. The latter contains the ``current'' state of $\mathcal U$, the locations of the two scanning heads of $\mathcal U$, and full contents of the (imaginary) work and run tapes of $\mathcal U$. Initially, the state is the start state of $\mathcal U$, the two scanning heads are looking at the leftmost cells of their tapes, and the contents of the two tapes are empty.  $\mathcal M$ also maintains a variable $i$ which is initially set to $1$. 

After the above initialization step,  $\mathcal M$ just acts according to the following procedure:\medskip

{\bf Procedure} LOOP:
\begin{enumerate}[(1)]
\item Using the transition function of $\mathcal U$ and based on the current value of {\em Configuration}, $\mathcal M$ computes the next state, next locations of the scanning heads, the next content of the work tape of $\mathcal U$, and correspondingly updates {\em Configuration}.
\item  If during the above transition $\mathcal U$ made a move $\omega$, \ $\mathcal M$ further updates {\em Configuration} by appending the labmove $\pp\omega$  to the content of the imaginary run tape of $\mathcal U$. In addition, if $\omega$ looks like $c.j$ for some selectional cluster $c$,  $\mathcal M$ makes the same move $\omega$ in the real play.

\item If $i$ is (the ID of)  a  general port, $\mathcal M$ further updates {\em Configuration} by appending the labmove $\oo i.i$ (a ``smart environment's'' move) to the content of the imaginary run tape of $\mathcal U$.

\item Next, $\mathcal M$ checks  if there is a pair $(a,b)$ of opposite general ports with $a$ positive and $b$ negative, such that the imaginary run tape of $\mathcal U$ contains the four labmoves $\oo a.a$, $\oo b.b$, $\pp b.a$, $\pp a.b$, and $\mathcal M$ has not yet made the allocation move $(a,b)$ in the real play. This is to what we earlier referred as ``detecting a match between $a$ and $b$''. If such a pair $(a,b)$ is found, $\mathcal M$ makes the move $(a,b)$ in the real play.    

 \item Now $\mathcal M$ checks its run tape  to see whether it contains a labmove $\oo\omega$ which, as a string spelled on the  tape, ends at location $i$. If such an $\omega$ is found and it looks like $c.j$ where $c$ is a selectional cluster,   
$\mathcal M$ further updates {\em Configuration} by appending $\oo \omega$ to the content of the imaginary run tape of $\mathcal U$.
\item Finally, $\mathcal M$ increments $i$ to $i+1$, and repeats  LOOP. 
\end{enumerate} 

Consider an arbitrary run $\Theta$ generated by $\mathcal M$. To this run  corresponds a run $\Gamma$ generated by $\mathcal U$ --- namely, $\Gamma$ is the run incrementally spelled on the imaginary run tape of $\mathcal U$ when the latter is simulated by $\mathcal M$ during playing according to the scenario of $\Theta$. Fix these $\Theta$ and $\Gamma$. Let us also fix $\mathcal A$ as the arrangement induced by $\Theta$. 
We further pick an arbitrary situation $^\dagger$ for $C$ consistent with $\mathcal A$ and fix it, too. 

Our assumption that $\mathcal U$ wins $C^*$ for any nice interpretation $^*$ implies that $\mathcal U$ never makes an illegal move of $C^*$ (unless its environment does so first). We may also safely assume that $\mathcal M$'s adversary never makes illegal moves of $\hat{C}$ either, for otherwise $\mathcal M$ automatically wins and we are done. 
Then, a little analysis of the situation reveals that $\Theta$ is a legal run of $\hat{C}$ and $\Gamma$ is a legal run of $C^*$ (for whatever nice interpretation $^*$). We will implicitly rely on this observation in our further argument.

We shall say that a general port $a$ of $C$ is {\bf unmatched} iff for no $b$ does $\mathcal A$ contain $(a,b)$ or $(b,a)$. Otherwise $a$ is {\bf matched}, and the port $b$ for which $\mathcal A$ contains $(a,b)$ or $(b,a)$ is said to be the {\bf match} of $a$. 

Let $^\ddagger$ be the situation for $C$ which agrees with $^\dagger$ on all elementary and matched general ports, and sends each unmatched general port to $\mathbb{F}$. Obviously $^\ddagger$, just like $^\dagger$, is consistent with $\mathcal A$. And, as $^\dagger$ sends to $\mathbb{T}$ any port that $^\ddagger$ does, in view of the monotonicity of all truth conditions associated with gates, it is clear that    

\begin{equation}\label{j2b}
\mbox{\em If $C$ is true w.r.t. $(^\ddagger,\Theta)$, then so is it  w.r.t. $(^\dagger,\Theta)$.}
\end{equation}

We now choose  a nice interpretation $^*$ such that:
\begin{enumerate}[$\bullet$]
\item For any elementary atom $p$, \ $p^*=\twg$ iff there is a $p$-labeled (resp. $\gneg p$-labeled) port $a$ such that $a^\ddagger=\mathbb{T}$ (resp. $a^\ddagger=\mathbb{F}$).
\item For any general atom $P$, \ $P^*$ is the nice game such that any legal run  $\{\oo a,\pp b\}$ of it is  $\pp$-won  iff we have one of the following:
\begin{enumerate}[(1)]
\item $a$ is a $P$-port with $a^\ddagger=\mathbb{T}$,  and  $\Gamma^{a.}=\{\oo a, \pp b\}$.
\item $b$ is a $\gneg P$-port with  $b^\ddagger=\mathbb{F}$,  and 
 $\Gamma^{b.}=\{\pp a, \oo b\}$. 
\end{enumerate} 
\end{enumerate}

We claim the following:
\begin{equation}\label{j2c}
\mbox{\em For any general $L$-port $c$ of $C$, $c^\ddagger=\mathbb{T}$  iff $\Gamma^{c.}$ is a $\pp$-won  run of $L^*$.}
\end{equation}

To verify the above claim, let us first consider the case when $L=P$ (some general atom $P$) and $c^\ddagger=\mathbb{T}$.  Since $^\ddagger$ assigns $\mathbb{T}$ only to matched general ports, $c$ is matched. Let $d$ be the match of $c$.  
From our description of the work of $\mathcal M$ it is obvious that $\Gamma^{c.}=\{\oo c, \pp d\}$. And, by clause 1 of our definition of $P^*$, such a run is a $\pp$-won run of $P^*$, as desired.

Next,   consider the case when $L=\gneg P$ and $c^\ddagger=\mathbb{T}$.  
Again, since $^\ddagger$ assigns $\mathbb{T}$ only to matched general ports, $c$ is matched. Let $d$ be its match. So, $d^\ddagger=\mathbb{F}$.
Note that  $\Gamma^{c.}=\{\pp d,\oo c\}$ and hence $\gneg\Gamma^{c.}=\{\oo d,\pp c\}$. By our definition of $P^*$, $\{\oo d,\pp c\}$  can be  a $\pp$-won run of $P^*$ only if either $d^\ddagger=\mathbb{T}$ (clause 1) or $c^\ddagger=\mathbb{F}$ (clause 2). But neither of these two is true in our case. So,  $\gneg \Gamma^{c.}$   is a $\oo$-won run of $P^*$. But then, by the definition of game negation, $\Gamma^{c.}$ is a $\pp$-won run of $\gneg P^*$, as desired.

Next,   consider the case when $L=P$ and $c^\ddagger=\mathbb{F}$. Since the smart adversary of (the simulated by $\mathcal M$) \ $\mathcal U$ makes the move $c.c$ for each general port $c$, $\Gamma^{c.}$ contains the labmove $\oo c$. If it does not contain any other labmoves, $\Gamma^{c.}$ is  a $\oo$-won run of $P^*$ because the latter is a nice game. Suppose now $\Gamma^{c.}=\{\oo c,\pp d\}$ for some $d$. So, either $c$ is unmatched, or $d$ its match. If $c$ is unmatched, then $\Gamma^{d.}$ cannot be $\{\oo d,\pp c\}$ and hence, by our definition of $P^*$, $\Gamma^{c.}$ is not a $\pp$-won run of $P^*$. Now suppose $d$ is the match of $c$. Then $d^\ddagger=\mathbb{T}$ which, again, makes it impossible that $\Gamma^{c.}$ is a $\pp$-won run of $P^*$. Thus, in all cases,  $\Gamma^{c.}$ is  a $\oo$-won run of $P^*$, as desired.

Finally,   consider the case when $L=\gneg P$ and $c^\ddagger=\mathbb{F}$. As in the previous case,  $\Gamma^{c.}$ contains the labmove $\oo c$. Hence, $\gneg \Gamma^{c.}$ contains  $\pp c$. If $\gneg \Gamma^{c.}$ does not contain any other labmoves, $\gneg\Gamma^{c.}$ is a $\pp$-won run of $P^*$ because $P^*$ is a nice game; then, $\Gamma^{c.}$ is a $\oo$-won run of $\gneg P^*$, and we are done.  Suppose now $\gneg \Gamma^{c.}= \{\oo d,\pp c\}$ for some $d$, and hence  $ \Gamma^{c.}= \{\pp d,\oo c\}$. Then, by clause 2 of our definition of $P^*$,  $\gneg \Gamma^{c.}$ is a $\pp$-won run of $P^*$, meaning that $\Gamma^{c.}$ is a $\oo$-won run of $\gneg P^*$, as desired. Claim (\ref{j2c}) is now proven.

By our choice of $^*$, claim (\ref{j2c}) is automatically true for elementary ports instead of general ports. So, we have: 
\begin{equation}\label{jj2c}
\mbox{\em For any (elementary or general) $L$-port $c$ of $C$, $c^\ddagger=\mathbb{T}$  iff $\Gamma^{c.}$ is a $\pp$-won  run of $L^*$.}
\end{equation}

From our description of the work of $\mathcal M$ one can see that $\mathcal M$ and $\mathcal U$ act in exactly the same ways in selectional clusters. More precisely, $\Theta$ and $\Gamma$ contain exactly the same labmoves of the form $\xx c.j$ where $c$ is a selectional cluster of $C$. From this observation, in conjunction with  
(\ref{jj2c}), it is obvious that $C$ is true w.r.t. $(^\ddagger,\Theta)$ iff it is true w.r.t. $(^*,\Gamma)$. But, by our assumption, $\mathcal U$ wins $C^*$ for any nice interpretation $C^*$. Thus, $C$ is true w.r.t. $(^\ddagger,\Theta)$. Then, by (\ref{j2b}), $C$ is also true w.r.t. $(^\dagger,\Theta)$. 
Now, remembering that $^\dagger$ was an arbitrary situation consistent with the arrangement induced by $\Theta$, we find that $C$ is accomplished w.r.t. $\Theta$. In other words, $\Theta$ is a $\pp$-won run of $\hat{C}$. Finally, remembering that $\Theta$ was an arbitrary run generated by $\mathcal M$, we conclude that $\mathcal M$ wins $\hat{C}$. In other words, $\mathcal M$ accomplishes $C$.

\section{Conclusion} 

\noindent We have elaborated circuit-style constructs called {\em cirquents}, and set up two  sorts of game semantics for them: the semantics of {\em computability logic}, and {\em abstract resource semantics}. The two, while substantially different, have been proven to validate the same classes of cirquents.

 Cirquents, allowing us to account for sharing substructures between different parent structures, are properly more expressive and efficient means of representing various objects of study than formulas are. This fact had already been observed in the past,  but only very limited classes of cirquents had been studied so far, and only abstract resource semantics (not the semantics of computability logic) had been defined for them. The present paper extended the earlier studied cirquents by adding three non-traditional pairs of conjunctive and disjunctive gates to them. An even more important innovation was generalizing gates to what we called {\em clusters}. 
This is a generalization naturally called for by advanced approaches to game logics and resource logics.  Clustering further increases the expressiveness and flexibility of cirquents. Among its merits is achieving the full expressive power of {\em independence friendly logic} and far beyond. 

This paper has been exclusively focused on semantics. Among the subsequent natural steps within the present line of research would be attempts to construct deductive systems for various fragments of the set of valid cirquents.

\end{document}